\documentclass[twocolumn,10pt]{asme2ej}

\pdfoutput=1
\usepackage{graphicx}

\usepackage{amsmath,amssymb}
\usepackage[colorlinks=true,linkcolor=blue,citecolor=red,urlcolor=cyan]{hyperref}

\title{Adaptive and model-based control theory applied to convectively unstable flows}

\author{Nicol\`o Fabbiane
\affiliation{Linn\`e FLOW Centre\\Department of Mechanical Engineering\\
             Royal Institute of Technology (KTH)\\S-10044 Stockholm, Sweden}}
\author{Onofrio Semeraro
\affiliation{Laboratoire d'Hydrodynamique (LadHyX)\\
			 CNRS -- Ecole Polytechnique\\91128 Palaiseau, France}}
\author{Shervin Bagheri \&  Dan S. Henningson
\affiliation{Linn\`e FLOW Centre\\Department of Mechanical Engineering\\
             Royal Institute of Technology (KTH)\\S-10044 Stockholm, Sweden}}

  \newcommand   \refsec[1]  {\S\ref{SEC:#1}}
  \newcommand   \reffig[1]  {\figurename~\ref{FIG:#1}}
  \newcommand   \refeqn[1]  {(\ref{EQN:#1})}

  \newcommand   \refeqns[2] {(\ref{EQN:#1}--\ref{EQN:#2})}

  \newcommand   \Real     {\mathbb{R}}                    
  \newcommand   \Complex  {\mathbb{C}}                    
  \newcommand   \ione     {\text{i}}                      
  \renewcommand \vec[1]   {\mathbf{#1}}                   
  \newcommand   \mat[1]   {\mathbf{#1}}                   

  \newcommand   \dd[3]    {\frac{\partial^{#1}\,#2}{\partial\,{#3}^{#1}}} 
  \newcommand   \ddt[1]   {\dot{#1}}                      
  \newcommand   \adj[1]   {{#1^H}}                        
  \newcommand   \Tr[1]    {\mbox{Tr}\left(#1\right)}      

  \newcommand   \td       {\tilde{t}}                     
  \newcommand   \xd       {\tilde{x}}                     
  \newcommand   \Lref     {\tilde{l}}                     
  \newcommand   \Uref     {\tilde{V}}                     
  \renewcommand \t        {t}                             
  \renewcommand \x        {x}                             
  
  \newcommand   \keta     {\eta}                          
  \newcommand   \knu      {\mu}                           
  \newcommand   \kudvec   {\tilde{v}}                     
  
  \newcommand   \kr       {\mathcal{R}}                   
  \newcommand   \kp       {\mathcal{P}}                   
  \newcommand   \kuvec    {v}                             

  \newcommand   \kubase   {V}                             
  \newcommand   \kuper    {v^\prime}                      
  \newcommand   \kfper    {f^\prime}                      

  \newcommand   \kb       {b}                             
  \newcommand   \kc       {c}                             
  \newcommand   \kg       {g}                             
  
  \renewcommand \u        {u}                             
  \renewcommand \d        {d}                             
  \newcommand   \n        {n}                             
  
  \newcommand   \y        {y}                             
  \newcommand   \z        {z}                             

  \newcommand   \q        {\vec{v}}                       
  \newcommand   \A        {\mat{A}}                       
  \newcommand   \B        {\mat{B}}                       
  \newcommand   \C        {\mat{C}}                       

  \newcommand   \qadj     {\vec{p}}                       
  \newcommand   \G        {\mat{G}}                       
  
  \newcommand   \Ad       {\tilde{\A}}                    
  \newcommand   \Bd       {\tilde{\B}}                    
  \newcommand   \Cd       {\tilde{\C}}                    

  \newcommand   \Pl       {\mathcal{P}}                   
  \newcommand   \pl       {\tilde{\mathcal{P}}}           
  
  \newcommand   \Sl       {\mathcal{S}}                   
  
  \newcommand   \es       {\tilde{\mathcal{E}}}           
  
  \newcommand   \Ct       {\mathcal{K}}                   
  \newcommand   \ct       {\tilde{\mathcal{K}}}           

  \newcommand   \J        {\mathcal{L}}                   
  \newcommand   \Ja       {\tilde{\J}}                    
  \newcommand   \Wq       {\mat{W}_{\q}}                  
  \newcommand   \Wu       {w_{\u}}                        
  \newcommand   \Wz       {w_{\z}}                        
  \newcommand   \K        {\mat{K}}                       
  \newcommand   \X        {\mat{X}}                       
  
  \newcommand   \Ta       {T_a}                           
  \newcommand   \F        {\mat{P}_{\zv\q}}               
  \newcommand   \Ph       {\mat{P}_{\zv\uv}}              
  \newcommand   \zv       {\vec{z}}                       
  \newcommand   \uv       {\vec{u}}                       
  \newcommand   \Wuv      {\mat{W}_{\uv}}                 
  \newcommand   \Wzv      {\mat{W}_{\zv}}                 
  \newcommand   \Wzq      {\mat{W}_{\q}}                  

  \newcommand   \R        {\mathcal{R}}                   
  \newcommand   \var      {R}                             
  
  \newcommand   \e        {\vec{e}}                       
  \newcommand   \N        {\mathcal{N}}                   
  \newcommand   \Na       {\tilde{\N}}                    
  \newcommand   \Wd       {\var_{\d}}                     
  \newcommand   \Wn       {\var_{\n}}                     
  \renewcommand \L        {\mat{L}}                       
  \newcommand   \Y        {\mat{\var}_{\e}}               

  \newcommand   \err      {e}                             
  \newcommand   \lam      {\lambda}                       
  \newcommand   \step     {\mu}                           
  
\begin{document}

\maketitle

\begin{abstract}
Research on active control for the delay of laminar-turbulent transition in boundary layers has made a significant progress in the last two decades, but the employed strategies have been many and dispersed. Using one framework, we review  model-based techniques, such as linear-quadratic regulators,  and model-free adaptive methods, such as least-mean square filters. The former  are supported by a elegant and powerful theoretical basis, whereas the latter may provide a  more practical approach in the presence of complex disturbance environments, that are difficult to model.  We compare the methods with a particular focus on efficiency, practicability and robustness to uncertainties. Each step  is exemplified on the one-dimensional  linearized Kuramoto-Sivashinsky equation, that shows many similarities with the initial linear stages of the transition process of  the flow over a  flat plate. Also, the source code for the examples are provided.
\end{abstract}

\section{Introduction}              \label{SEC:introduction}
The key motivation in research on drag reduction is to develop new technology that will result in the design of vehicles with a significantly lower fuel consumption. The field is broad, ranging from passive methods, such as coating surfaces with materials  that are super-hydrophobic or non-smooth \cite{Bushnell1991Drag-reduction-},  to active methods, such as applying wall suction or using measurement-based  closed-loop control \cite{arfm2007-kim-bewley}. This work positions itself in the  field of active control methods for skin-friction drag.  In general, the mean skin friction of a turbulent boundary layer on a flat plate is  an order of magnitude larger compared to a laminar boundary layer. One strategy to reduce skin-friction drag is thus to push  the  laminar-turbulent transition on a flat plate downstream \cite{schlichting}. Different transition scenarios may occur in a boundary layer flows, depending on the intensity of the external disturbances acting on the flow, \cite{arfm2002-saric-et-al}. Under low levels of free-stream turbulence and sufficiently far downstream, the transition process is initiated by the linear growth of small perturbations called Tollmien-Schlichting (TS) waves \cite{schlichting}. Eventually, these perturbations reach finite amplitudes and breakdown to smaller scales via nonlinear mechanisms \cite{2001STSF-schmid-henningson}. However, in presence of stronger free-stream disturbances, the exponential growth of TS waves are {bypassed} and transition may be directly triggered by the algebraic growth of stream-wise elongated structures, called streaks \cite{arfm2002-saric-et-al}. One may delay transition by damping the growth of TS waves and/or streaks, and thus postpone their nonlinear breakdown. This strategy enables the use of linear theory for control design.

Fluid dynamists noticed in the early 90's, that many of the emerging concepts in  hydrodynamic stability theory already existed in linear systems theory \cite{jovanovic, arfm2007-schmid}. For example, the analysis of a system forced by harmonic excitations is referred to as signalling problem by fluid dynamicists, while control theorists analyze the problem by constructing a Bode diagram, \cite{2000CT-glad-ljung}; similarly, a large transient growth of a fluid system corresponds to large norm of a transfer function and matrix with stable eigenvalues can be called either globally stable or Hurtwitz, \cite{2001STSF-schmid-henningson,arfm1990-huerre-monkewitz}.

However, the systems theoretical approach had taken one step further, by ``closing the loop'', i.e~providing rigorous conditions and tools to modify the linear system at hand. It was realized by fluid dynamists  that the extension of hydrodynamic stability theory to include tools and concepts from linear control theory was natural \cite{joshi, bewley98, cortelezzi98}. A long series of numerical investigations addressing the various aspects of closed-loop control of  transitional \cite{hogberg:bewley:henning:03:a, chevalier:hoepffner:akervik:henning:07, monokrousos2008dns} and turbulent flows \cite{lee01,hogberg:bewley:henning:03:b, chevalier:hoepffner:bewley:henning:06} followed in the wake of these initial contributions.

At the same time, research on active control for transition delay has been advanced from a more practical approach using system identification methods \cite{ljung1999system} and active wave-cancellation techniques \cite{Elliott1993Active-noise-co}. Most work (but not all) is experimental, which due to feasibility constraints, has favoured an engineering and occasionally  {\it ad hoc} methods. One of the  first examples of this approach is the control of TS waves in the experiments by \cite{milling1981tollmien} using a wave-cancellation control;  the propagating waves are cancelled by generating perturbations with opposite phase. This work was followed by number of  successful experimental investigations \cite{98:jacobson:reynolds,ijhff2003-sturzebecher-nitsche, 2003:jfm:Rathnasingham:breuer, 2007:jfm:lundell:control} of transition delay using more sophisticated system identification techniques.

Whereas both numerical and experimental approaches have pushed forward  flow control research, they have in a large extent evolved  disconnected from each other; the systems control theoretical approach has provided very important insights into physical mechanisms and constraints that has to be addressed in order to design active control that is optimal and robust, but most work has stayed at a proof-of-concept level and have not yet been fully implemented in practical applications. Although, there are  exceptions \cite{McKeon2013Experimental-ma, Goldin2013Laminar-flow-co}, the majority of experimental active control has essentially suffered from the opposite; most controllers are developed directly in the experimental setting on a trial-and-error basis, with many tuning parameters, that have to be chosen for each particular set-up.  

This review aims at presenting  model-based and model-free techniques that are appropriate for the control of TS waves in a flat-plate boundary layer. We compare and link the two approaches using a linear model, that similar to the linearized Navier-Stokes equations, exhibits a large transient amplification behaviour and time delays. 
This presentation is unavoidably  influenced by the authors background and previous work; complementary reviews on flow control can be found in \cite{arfm2007-kim-bewley,sipp2010dynamics,phtr2011-bagheri-henningson}, where the linear approach is analyzed, and in the reviews by \cite{amr2009-bagheri-et-al,sipp2013closed}, focussed on the identification of reduced-order models for the linear control design. Finally, we refer to \cite{gad96, Bewley01, Collis2004Issues-in-activ} for a broader prospective.

\subsection{The control problem}\label{SEC:control-problem}
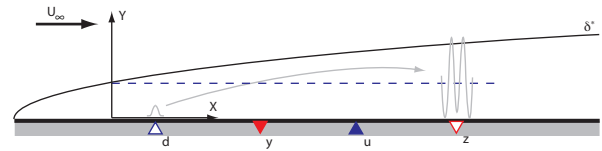
\begin{figure}
 \centering
 \includegraphics[width=.45\textwidth]{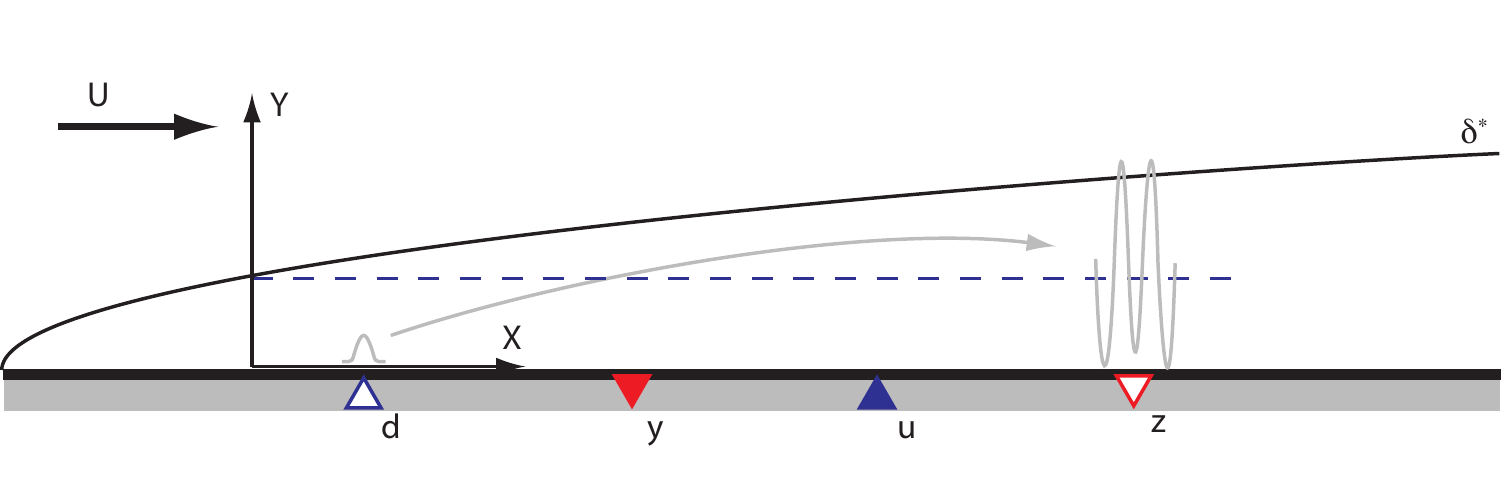}
 \setlength{\unitlength}{\textwidth} \begin{picture}(0,0) \put(-.428, .115){\tiny$\infty$}\end{picture}\hspace{-6pt}
 \caption{Scheme of a Blasius boundary-layer flow developing over a flat plate. A disturbance modelled by $\d$ grows exponentially while convected downstream. The actuator $\u$ is used to attenuate the disturbance before it triggers transition to turbulence; the actuation signal is computed based on the measurements provided by the sensor $\y$. The output $\z$, located downstream of the actuator, estimates the efficiency of the control action.}
 \label{FIG:bl-sketch}
\end{figure}

\begin{figure}
 \centering
 \includegraphics[width=.45\textwidth]{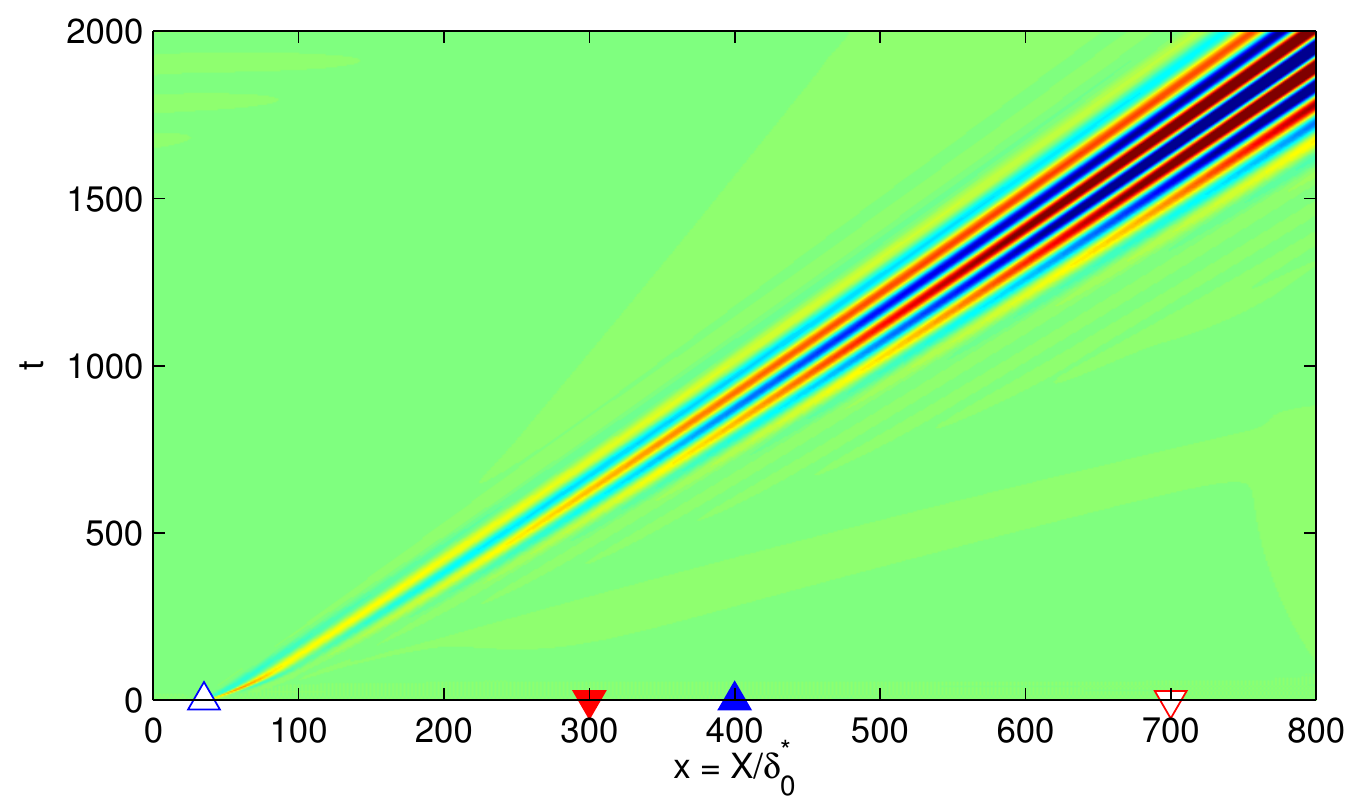}
 \caption{Response to a small, localized initial condition in a Blasius boundary-layer flow. A Tollmien-Schlichting wave-packet emerges and grows exponentially while propagating downstream. Contours of the streamwise component of the velocity are shown as a function of the streamwise direction (\emph{x}) and time (\emph{t}). The location along the normal-direction $y$ is chosen in the vicinity of the wall.}
 \label{FIG:xt-and-signals_DNS-2DBL}
\end{figure}

Consider a steady uniform flow $U_{\infty}$ over a thin flat plate of length $L$ and infinite width. Inside the two-dimensional (2D) (Blasius) boundary layer that develops over the plate, we place a small localized disturbance (denoted by $\d$ in \reffig{bl-sketch}) of simple Gaussian shape; {the set-up is the same as in \cite{bagheri2009input} and the simulation is performed using a spectral code \cite{2007SIMSON-chavalier}}. \reffig{xt-and-signals_DNS-2DBL} summarizes the spatio-temporal evolution of the  disturbance. It shows a contour plot of the stream-wise component of the perturbation velocity at a wall normal position $Y = \delta^*(0)$, where $\delta^*(X)$ is the displacement thickness of the boundary layer. The temporal growth of this disturbance is determined by classical linear stability theory (i.e.\ eigenvalue analysis of the linearized Navier-Stokes equations). Such an analysis reveals that asymptotically  a compact wave-packet emerges -- a TS wave-packet -- that grows in time at an exponential rate while travelling downstream at group velocity of approximately $U_{\infty}/3$. This disturbance behaviour is observed as long as the amplitude is below a critical value (usually a few percent of $U_{\infty}$)  \cite{2001STSF-schmid-henningson}. Above the critical value, nonlinear effects have to be taken into account; they eventually result in  a break down of the disturbance to smaller scales and finally to transition from a laminar to a turbulent flow\cite{2001STSF-schmid-henningson}. However, the key point -- that enables the use of linear  theory for transition control -- is that the disturbance may grow several orders of magnitude before it breaks down.

Using a spatially localized forcing (denoted by $\u$ in  \reffig{bl-sketch}) downstream of the disturbance, one may modify the conditions in order to  reduce the amplitude of the wave-packet and  thus  delay the transition to  turbulence. Physically this forcing is provided by devices called \emph{actuators}. An example of an actuator is a loudspeaker that generates short pulses  through a small orifice in the plate. The volume of the loudspeaker and the shape of the orifice determines the type of actuation. Another example is plasma actuators, where a plasma arch is used to induce a forcing on the flow \cite{ef2008-grundmann-tropea}.

In closed-loop control, a sensor (denoted by $\y$ in \reffig{bl-sketch}) is used to measure the disturbance that is meant to be cancelled by the actuator $(\u)$: based on these measurements one computes the actuator action in order to effectively reduce the amplitude of the perturbation. Examples of sensors include pressure measurements  using a small microphone membrane mounted flush to the wall, velocity measurements using hot-wire anemometry near the wall or shear-stress measurements using thermal sensors (wall wires). Finally, we place a second sensor (denoted by $\z$ in  \reffig{bl-sketch}) downstream of the actuator to measure the amplitude of the perturbation after the actuator action. The minimization of this output signal may serve as an objective of our control design, but the measurements also provide a means to assess the performance of the controller.

Having introduced the inputs and outputs, the control problem can be formulated as the following: given the measurement $\y(\t)$, compute the modulation signal $\u(\t)$ in order to minimize a cost function based on $\z(\t)$. The system that when given the measurement $\y(\t)$, provides the control signal $\u(\t)$ is referred to as the \emph{compensator}. The design of the compensator  has to take into account competing aspects such as robustness, performance and practical feasibility.

\bigskip
The objective of this review is to guide the reader through the steps of compensator design process. We will exemplify the theory and the associated methods on a one-dimensional (1D) model based on the linearized Kuramoto-Sivashinsky (KS) equation (presented in \refsec{system}).  The model  reproduces the most important stability properties of the flat-plate boundary layer, but it  avoids the problem of high-dimensionality and thus the high numerical costs. In \refsec{control} full-information control problem is addressed via optimal control theory; linear quadratic regulator (LQR) and model-predictive controller (MPC) strategies are derived and compared. The disturbance estimation problem is addressed in \refsec{estimation}, where classical Kalman estimation theory and least-mean-square techniques will be introduced and compared. The techniques of sections \refsec{control} and \refsec{estimation}, will be combined in order to design the compensator in \refsec{compensator}. This section also contains adaptive algorithms  that enhance the robustness of the compensator. The review finalizes with a discussion \refsec{reduction} about some important features characterizing the control problem when applied to three-dimensional (3D) fluid flows and conclusions \refsec{conclusions}.

\section{Framework}       \label{SEC:system}
We first introduce our choice of model KS equation, inputs (actuators/disturbances) and sensors. This is followed by a presentation of concepts pertinent to our work, namely  the state-space formulation (\refsec{state-space}), transfer functions and finite-impulse response (\refsec{io}), controllability and observability (\refsec{gramians}), closed-loop system (\refsec{cloop}) and robustness (\refsec{robustness}). This chapter contains  the mathematical ingredients that will be used in the following sections. 

\subsection{Kuramoto-Sivashinsky model}   
In this paper, we focus our attention on flows dominated by convection/advection, where disturbances have negligible upstream influence and are quickly swept downstream with the flow. We make use of a particular variant of the KS equation to model a linear and convection-dominated flow. Originally, the KS equation was developed to describe the flame front flutter in laminar flames, \cite{ptp1976-kuramoto-tsuzuki,aa1977-sivashinsky}. This model exhibits in its space-periodic form a spatio-temporal chaotic behaviour, with some similarities to turbulence \cite{manneville1995dissipative}. The standard KS equation reads
\begin{equation}
 \dd{}{\kudvec}{\td} + \kudvec\dd{}{\kudvec}{\xd} = - \keta\,\dd{2}{\kudvec}{\xd} - \knu\dd{4}{\kudvec}{\xd},\label{EQN:kuramoto-equation-dim}
\end{equation}
where $\td$ is the time, $\xd\in[0,\tilde{L})$ the spatial coordinate and $\kudvec = \kudvec(\xd,\td)$ the velocity. The boundary conditions accompanying \refeqn{kuramoto-equation-dim} are periodic in $\xd$. The second term on the left side in \refeqn{kuramoto-equation-dim} is the nonlinear convection term, while on the right side two viscosity terms appear. The two latter terms may be associated to the production and dissipation of energy at different spatial scales. In particular, the second-order derivative term is related to the production of the energy via the variable $\keta$, called \textit{anti-viscosity}, while the dissipation of the energy is connected to the fourth-order derivative term, multiplied by the \textit{hyper-viscosity} $\knu$, \cite{2012CCQ4-cvitanovic-et-al}.

Equation \refeqn{kuramoto-equation-dim} can be rewritten such that it is parametrized by a Reynolds-number-like coefficient. Introducing a reference length $\Lref$ and a reference velocity $\Uref$, define the non-dimensional position $\x$, velocity $\kuvec$ and time $\t$ by 
\begin{equation}
 \x = \frac{\xd}{\Lref}, \qquad \kuvec = \frac{\kudvec}{\Uref}, \qquad \t = \frac{\Uref}{\Lref}\,\td.
\end{equation}
Applying the transformation to \refeqn{kuramoto-equation-dim}, the KS equation in dimensionless form becomes
\begin{equation}
 \dd{}{\kuvec}{\t} + \kuvec\dd{}{\kuvec}{\x} = - \frac{1}{\kr} \left(\kp\,\dd{2}{\kuvec}{\x} + \dd{4}{\kuvec}{x}\right), \label{EQN:kuramoto-equation}
\end{equation}
where $\x\in[0,L)$. The parameters $\kr$ and $\kp$ are defined as
\begin{equation}
 \kr = \frac{\Uref\Lref^3}{\knu}, \qquad \kp = \frac{\keta}{\knu}\Lref^2,
\end{equation}
where $\kr$ takes the role of the Reynolds number $Re_{\delta^*}$, and $\kp$ regulates the balance between energy production and dissipation.

We  assume that the system is sufficiently close to a steady solution $\kubase(\x) = \kubase$. Then, it is possible to describe the dynamics of perturbations using the linearized KS equation. For the chosen parameters, the steady solution is stable, but an external perturbation may be amplified by an order-of-magnitude before it dies out (this requires non-periodic boundary conditions in the streamwise direction as we impose below). Introduce the perturbation $\kuper(\x,\t)$
\begin{equation}
 \kuvec(\x,\t) = \kubase + \epsilon\,\kuper(\x,\t), \label{EQN:decomp-velocity}
\end{equation} 
where $\epsilon\ll 1$.  By inserting this decomposition into \refeqn{kuramoto-equation} and neglecting the terms of order $\epsilon^2$ and higher, the linearized KS  equation is obtained
\begin{equation}
  \dd{}{\kuper}{\t} = -\kubase\dd{}{\kuper}{\x} -\frac{1}{\kr} \left(\kp\,\dd{2}{\kuper}{\x} + \dd{4}{\kuper}{\x}\right). \label{EQN:lin-kuramoto}
\end{equation}
It is the convective and amplifying properties of this non-normal system that makes it a good model of the 2D Blasius boundary layer flow.  Following  \cite{2011HI-charru}, we analyze the stability properties of \refeqn{lin-kuramoto}, by assuming travelling wave-like solutions:
\begin{equation}
 \kuper = \hat{\kuvec}\;e^{\ione\left( \alpha\x - \omega\t \right)}, \label{EQN:wave-solution}
\end{equation}
where $\alpha\in\Real$ and $\omega = \omega_r +\ione\omega_i \in\Complex$. Substituting \refeqn{wave-solution} in \refeqn{lin-kuramoto}, a dispersion relation between the spatial wave-number $\alpha$ and the temporal frequency $\omega$  is obtained
\begin{figure}
 \centering
 \includegraphics[width=.45\textwidth]{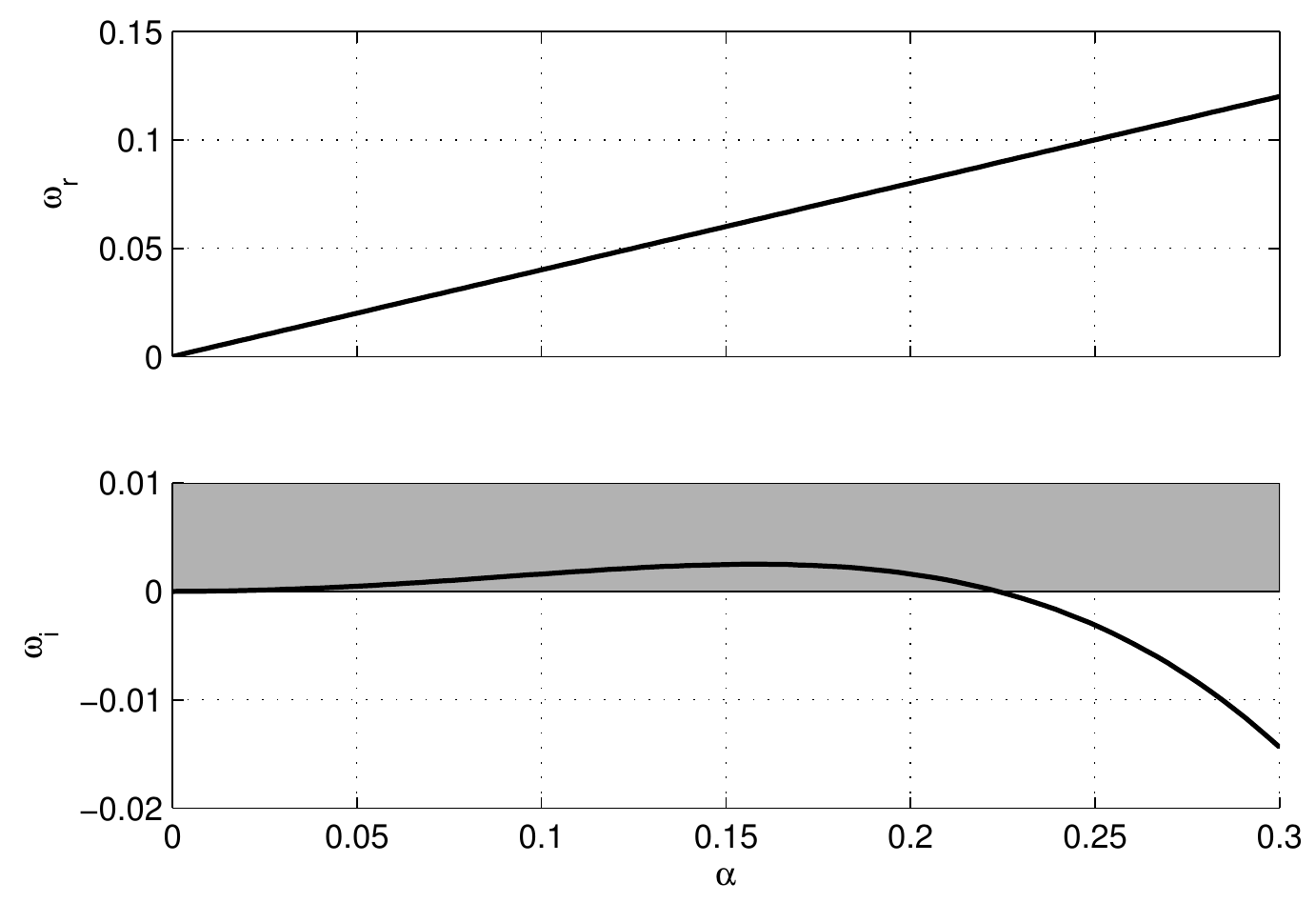}
 \setlength{\unitlength}{\textwidth} \begin{picture}(0,0) \put(-.015, .287){\scriptsize(a)}
                                                          \put(-.015, .132){\scriptsize(b)}\end{picture}\hspace{-6pt}
 \caption{The real frequency $\omega_r$ and its imaginary part $\omega_i$ are shown as a function of the spatial frequency $\alpha$, in (a) and (b), respectively. The relation among the spatial and temporal frequencies is given by the dispersion relation \refeqn{dispersion-relation-kuramoto}. Positive values of $\omega_i$ characterize unstable waves (grey region).}
 \label{FIG:dispersion-relation}
\end{figure}
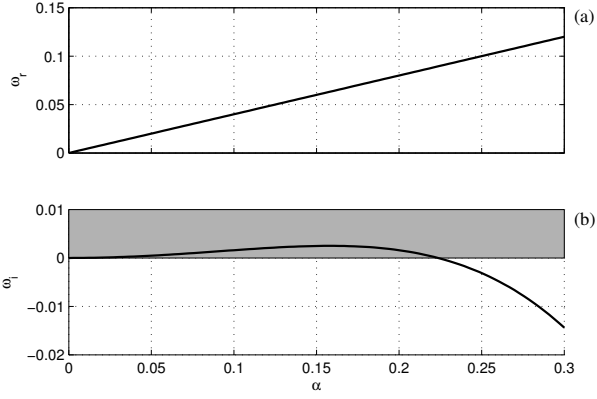
\begin{equation}
 \omega = \kubase\,\alpha + \ione\left( \frac{\kp}{\kr}\,\alpha^2 - \frac{1}{\kr}\,\alpha^4 \right). \label{EQN:dispersion-relation-kuramoto}
\end{equation}
This relation is shown in \reffig{dispersion-relation} for $\kr = 0.25$, $\kp = 0.05$ and $V = 0.4$. The parameters  are chosen to closely model the Blasius boundary layer at $Re_{\delta^*}=1000$.  The imaginary part of the frequency $\omega_i$ is the exponential temporal growth rate of a wave with wave-number $\alpha$. In  \refeqn{dispersion-relation-kuramoto} it can be observed that the term in $\alpha^2$ (associated to the production parameter $\kp$), is providing a positive contribution to $\omega_i$, while the $\alpha^4$ term (related to the dissipation parameter $\kr$),  has a stabilizing effect. The competition between these two terms determines stability of the considered wave. From \reffig{dispersion-relation}, it can be observed that for an interval of wave-numbers $\alpha$, $\omega_i>0$, i.e. the wave is unstable. 
The real part $\omega_r$  determines the phase speed of the wave in the $\x$ direction,
\begin{equation}
 c \triangleq \frac{\omega_r}{\alpha} = \kubase. \label{EQN:phase-speed}
\end{equation}
Note that the phase speed $c$ is independent of $\alpha$, in contrast to the
boundary-layer flow, which is dispersive \cite{2001STSF-schmid-henningson}.

\subsection{Outflow boundary condition}
\begin{figure}
 \centering
 \includegraphics[width=.45\textwidth]{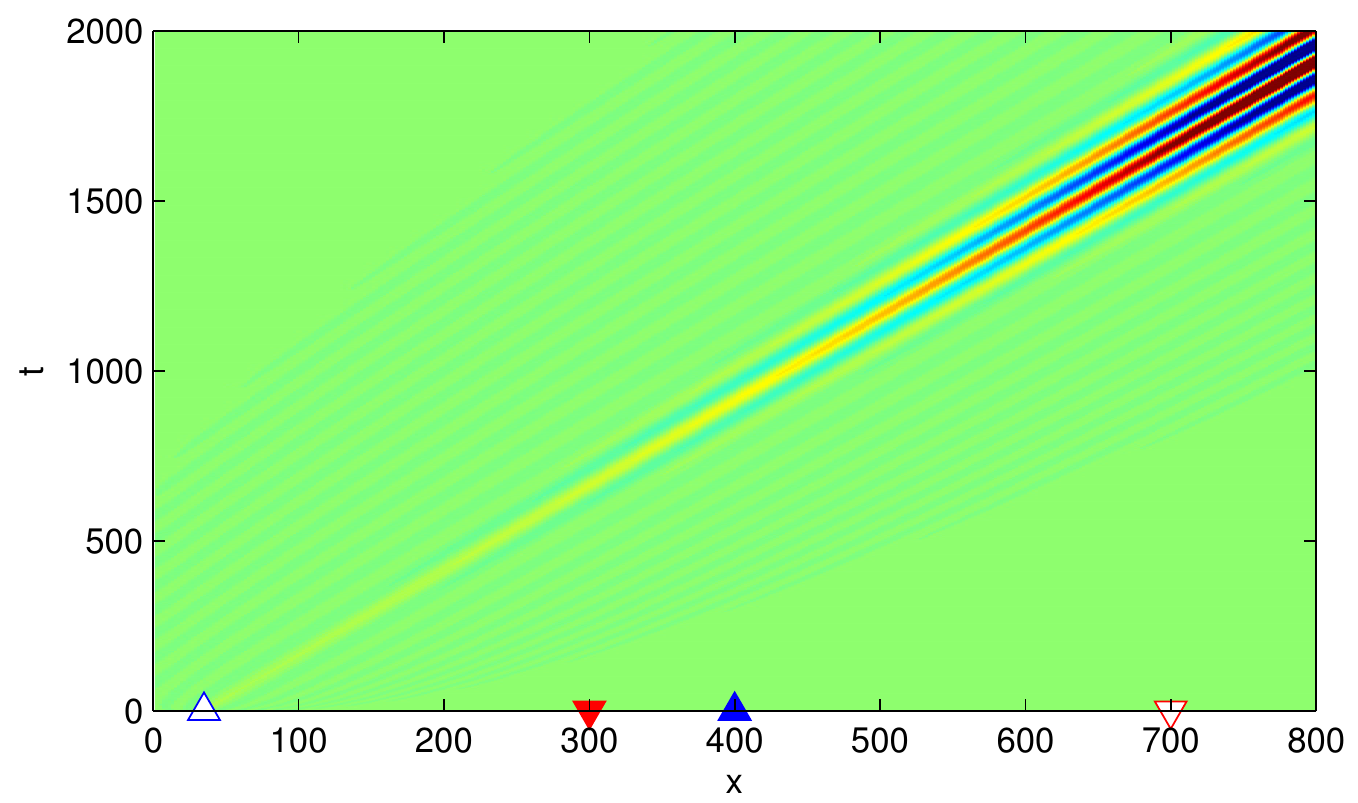}
 \caption{Response to a small, localized initial condition in a 1D KS flow \refeqn{lin-kuramoto} with $\kr = 0.25$, $\kp = 0.05$ and $V = 0.4$. The contours are shown as a function of the streamwise direction ($x$) and the time ($t$). The initial condition triggers a growing and travelling wave-packet, similar to the 2D boundary-layer flow shown in  \reffig{xt-and-signals_DNS-2DBL}. {\small\texttt{[script00.m]}}.}
 \label{FIG:xt-and-signals_Kuramoto}
\end{figure}
So far in our analysis we have assumed periodic boundary conditions for the KS equation. As we are interested in modelling the amplification of a propagating wave-packet near a stable steady solution (as observed in the case of boundary-layer flow), it is appropriate to change the boundary conditions to an outflow condition on the right side of the domain
\begin{equation}
  \left.\dd{3}{\kuper}{\x}\right|_{\x=L} = 0,\quad \left.\dd{}{\kuper}{\x}\right|_{\x=L} = 0,  \label{EQN:bc-L}
\end{equation}
while on the left side of the domain, at the inlet, an unperturbed boundary condition is considered
\begin{equation}
  \left.\kuper\right|_{\x=0} = 0,\quad \left.\dd{}{\kuper}{\x}\right|_{\x=0} = 0. \label{EQN:bc-0}
\end{equation}
With an outflow boundary condition, a localized initial perturbation in the upstream region of the domain travels in the downstream direction while growing exponentially in amplitude until it leaves the domain. This is the signature of a convectively unstable flow. Note the this choice of boundary conditions is the main variant with respect of the original KS equation, characterized by periodic boundaries. \reffig{xt-and-signals_Kuramoto} shows the spatio-temporal response to a localized initial condition of KS equation with outflow boundary condition. The set of parameters $\kr$, $\kp$ and $V$ has been chosen to mimic the response of the 2D boundary-layer flow, shown in \reffig{xt-and-signals_DNS-2DBL}. 
However,  note that in the KS model the wave crests travel parallel to each other with the same speed of the wave-packet, whereas in the boundary layer, they travel faster than the wave-packet which they form. Indeed the system is not dispersive, i.e. the phase speed $c$ equals the group speed $c_g$ as shown by \refeqn{phase-speed}; conversely, as already noticed, the 2D BL is dispersive.

\subsection{Introducing inputs and outputs} \label{SEC:system-state-space}
\begin{figure}
 \centering
 \includegraphics[width=.45\textwidth]{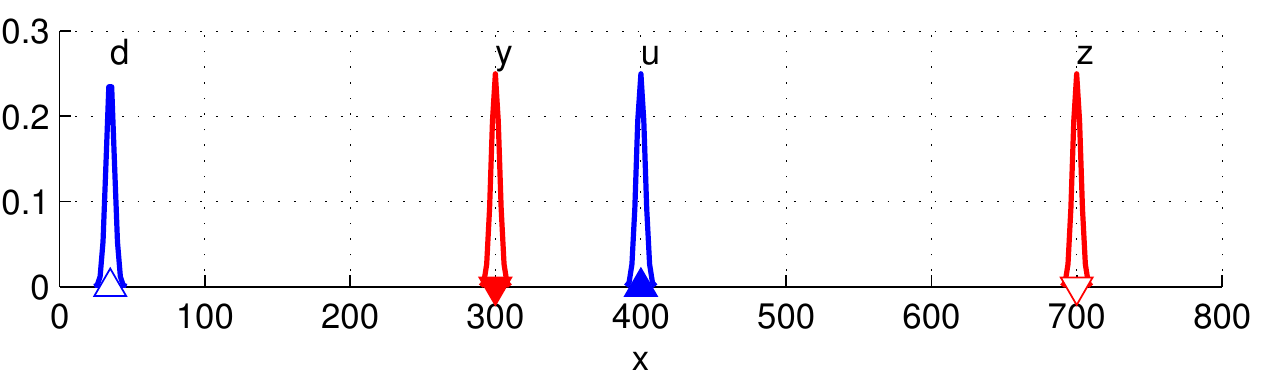}
 \caption{Spatial support of the inputs and outputs along the streamwise direction. All the elements are modelled as a Gaussian function \refeqn{kuramoto-b-structure}, with $\sigma_\d = \sigma_\u = \sigma_\y = \sigma_\z = 4$.}
 \label{FIG:inout}
\end{figure}
Having presented the dynamics of the linear system, we now proceed with a more systematic analysis of the inputs (actuators/disturbances) and sensor outputs described in \S\ref{SEC:control-problem}. Consider the linearized KS equation in \refeqn{lin-kuramoto}
\begin{equation}
  \dd{}{\kuper}{\t} = -\kubase\dd{}{\kuper}{x} -\frac{1}{\kr} \left(\kp\,\dd{2}{\kuper}{x} + \dd{4}{\kuper}{x}\right) + \kfper(\x,\t), \label{EQN:LKS}
\end{equation}
where the forcing term $\kfper(\x,\t)$ now appears on the right-hand side. This term is decomposed into two parts,
\begin{equation}
 \kfper(\x,\t) = \kb_{\d}(\x)\,\d(\t) + \kb_{\u}(\x)\,\u(\t) \label{EQN:forcing-definition}.
\end{equation}
The temporal signal of the incoming external disturbance and of the actuator are denoted by $\d(t)$ and $\u(t)$, respectively, while the corresponding spatial distribution is described by $\kb_{\d}$ and $\kb_{\u}$.  In this work, the time-independent spatial distribution of the inputs is described by the Gaussian function,
\begin{equation}
 \kg(\x;\,\hat{x},\sigma) = \frac{1}{\sigma} \exp\left [-\left( \frac{\x-\hat{\x}}{\sigma} \right)^2\right ]
 \label{EQN:kuramoto-b-structure}.
\end{equation}
The scalar parameter $\sigma$ determines the width of the Gaussian distribution, whereas $\hat{x}$ determines the centre of the Gaussian. The two forcing distributions in \refeqn{forcing-definition} are 
\begin{equation}\label{EQN:bds}
 \kb_{\d}(\x)= \kg(\x;\,\hat{x}_\d,\sigma_\d),  \qquad   \kb_{\u}(\x)= \kg(\x;\,\hat{x}_\u,\sigma_\u).
\end{equation}
The disturbance $\d$ is positioned in the beginning of the domain at $\hat{\x}_{\d} = 35$, while the actuator $\u$ in the middle of the domain at $\hat{\x}_{\u} = 400$ (see \reffig{inout}).  In the presentation above, the particular shape $\kb_{\d}(\x)$ of the disturbance $\d$ is part of the modelling process. However, note that the introduction of the upstream disturbance using a localized and well defined shape $\kb_{\d}(\x)$ is a model. In practice, due to the receptivity processes, the distribution and the appearance of the incoming disturbance is not known \emph{a-priori}, and thus difficult to predict using -- for instance -- a low-order model. 

A similar issue may arise for the model of the actuator $\kb_{\u}(\x)$, where the forcing distribution can even be time varying. For example the spatial force that a plasma actuator induces in the flow depends on the supplied voltage, e.g. modulated by the amplitude $u(t)$ \cite{ef2008-grundmann-tropea}. As we will discuss in the following sections, one may design a controller without knowing $ \kb_{\d}(\x)$ and $\kb_{\u}(\x)$, but for the sake of presentation we may assume in this section, that such models exist.

By using \refeqn{kuramoto-b-structure} as integration weights, we define two outputs of the system as
\begin{align}
 \y(\t) &= \int_{0}^{L} \kc_{\y}(\x)\,\kuper(\x,\t)\;d\x + \n(\t) \label{EQN:y-pert}, \\
 \z(\t) &= \int_{0}^{L} \kc_{\z}(\x)\,\kuper(\x,\t)\;d\x          \label{EQN:z-pert},
\end{align}
where $L$ is the length of the domain defined earlier and
\begin{equation*}
 \kc_{\y}(\x)= \kg(\x;\,\hat{x}_\y,\sigma_\y),  \qquad   \kc_{\z}(\x)= \kg(\x;\,\hat{x}_\z,\sigma_\z).
\end{equation*}
The output $\y$ provides a measurement of an observable physical quantity -- for example shear-stress, a velocity component or pressure near the wall -- averaged with the Gaussian weight. In realistic conditions, this measured quantity is subject to some form of noise, that may arise from calibration drifting, truncation errors and/or incomplete cable shielding, etc.  This is taken into account by the forcing term $\n(\t)$. It is often modelled as random noise with Gaussian  distribution of zero-mean and variance $\alpha$, {and can be regarded as an input of the system}. The second output $\z(t)$, located far downstream, represents the \textit{objective} of the controller: assuming that the flow has been already modified due to the action of the controller, this \emph{controlled} output is the quantity that we aim to keep as small as possible. 

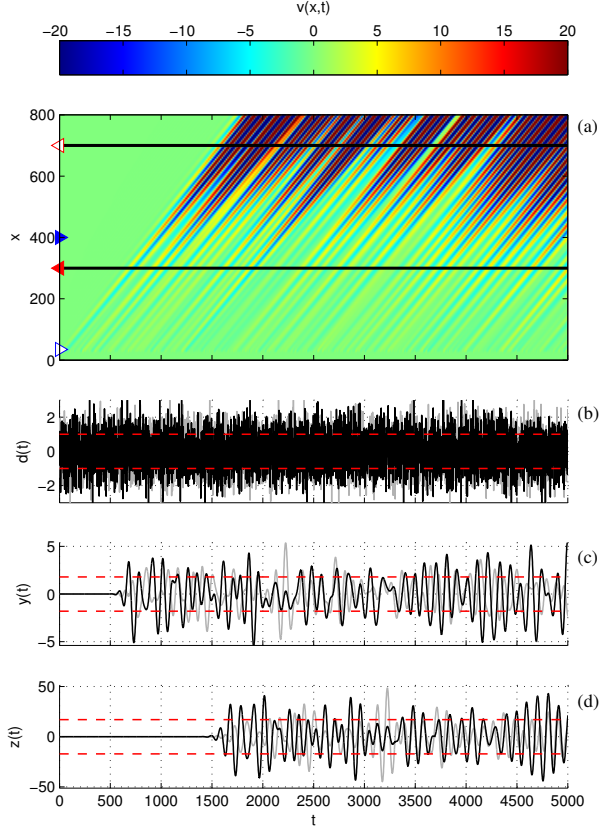
\begin{figure}
 \centering
 \includegraphics[width=.45\textwidth]{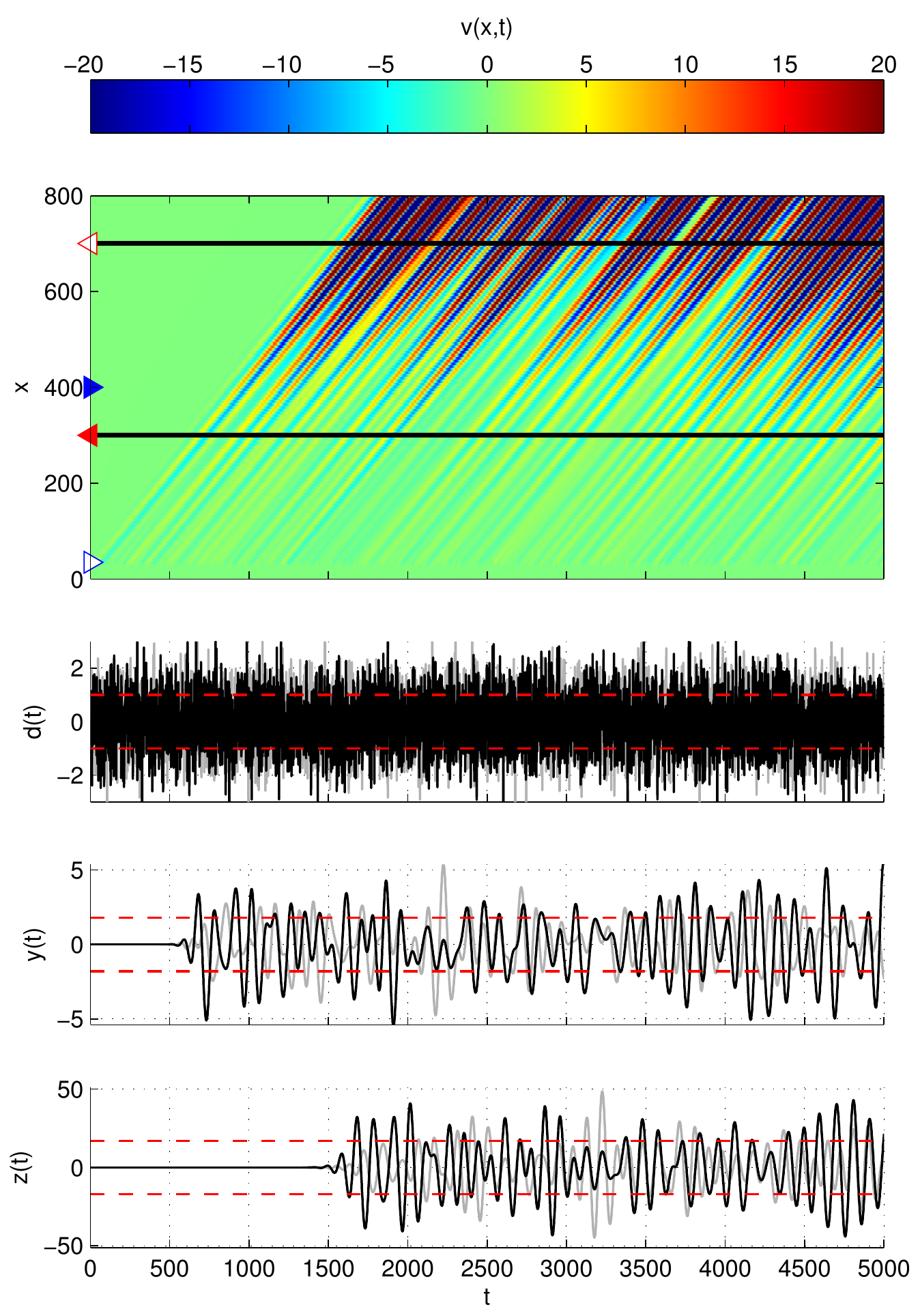}
 \setlength{\unitlength}{\textwidth} \begin{picture}(0,0) \put(-.020, .535){\scriptsize(a)}
                                                          \put(-.020, .315){\scriptsize(b)}
                                                          \put(-.020, .205){\scriptsize(c)}
                                                          \put(-.020, .095){\scriptsize(d)}\end{picture}\hspace{-6pt}
 \caption{Top frame (a) shows the spatio-temporal response to white noise $\d(\t)$, (b). The velocity contours are shown as a function of the streamwise direction ($x$) and time ($t$). The signals $\y(\t)$ and $\z(\t)$ are shown for two different realizations (black and grey lines) in (c) and (d), respectively. Red dashed lines indicate the standard deviation of the signals. {\small\texttt{[script01.m]}}}
 \label{FIG:xt-and-signals-noise_Kuramoto}
\end{figure}

In  \reffig{xt-and-signals-noise_Kuramoto}, we show the response of our system to a Gaussian white noise in $d(t)$ with a unit variance, where all  temporal frequencies are excited. Via the dispersion relation \refeqn{dispersion-relation-kuramoto}, each temporal frequency $\omega_r$  is related to a spatial frequency $\alpha = \kubase\,\omega_r$. The input signal $\d(\t)$ is thus filtered by the system, where after a short transient, only the unstable spatial wavelengths are present in the state $\q(\t)$, \reffig{xt-and-signals-noise_Kuramoto}(a), and the two output signals $\y(\t)$ and $\z(\t)$, \reffig{xt-and-signals-noise_Kuramoto}(c-d). The variance of the output $\z(\t)$ is higher than the variance of $\y(\t)$ by a factor 10, independently by the realization; this is because the wave-packets generated by $\d$ is growing in amplitude while convected downstream. We note that each realization will generate a different time evolution of the system but with the same statistical properties (black and grey lines in \reffig{xt-and-signals-noise_Kuramoto}(b-d)).

\subsection{State-space formulation} \label{SEC:state-space}
We discretize the spatial part of \refeqn{LKS}  by a finite-difference scheme. As further detailed in \refsec{matlab}, the solution is approximated by 
\begin{equation*}
  \kuper_i(\t) = \kuper(\x_i,\t) \qquad i = 1,2,...,n_\q
\end{equation*}
defined on the equispaced nodes $\x_i = i L/n_\q$, where  $n_\q = 400$.
The spatial derivatives  are approximated by a finite difference scheme based on five-points stencils. Boundary conditions in \refeqns{bc-0}{bc-L} are imposed using four ghost nodes $i = -1, 0$ and $i = n_\q + 1, n_\q + 2$. 
The resulting finite-dimensional state-space system (called \emph{plant}) is
\begin{align}
 \ddt{\q}(t) &= \A~    \,\q(t) + \B_{\d~} \,\d(t) + \B_{\u~} \,\u(t)   \label{EQN:space-state-detailed},\\
       \y(t) &= \C_{\y}\,\q(t) + \n(t)                                 \label{EQN:y-detailed},\\
       \z(t) &= \C_{\z}\,\q(t)                                         \label{EQN:z-detailed},
\end{align}
where $\q\in\Real^{n_\q}$ represents the nodal values $\kuper_i$. The output matrices $\C_{\y}$ and $\C_{\z}$ approximate the integrals in \refeqns{y-pert}{z-pert} via the trapezoidal rule, while the input matrices $\B_{\d}$ and $\B_{\u}$ are given by the evaluation of \refeqn{bds}  at the nodes.

Some of the control algorithms that we will describe are preferably formulated in a  time-discrete setting. The time-discrete variable corresponding to $a(\t)$ is
\begin{equation}
 a(k) = a(k\Delta\t), \qquad k = 1,2,... \label{EQN:time-discrete-variable}
\end{equation}
where $\Delta\t$ is the sampling time. Accordingly, the time-discrete state-space system is defined as:
\begin{align}
  \q(k+1) &= \Ad\,\q(k) + \Bd_d\,\d(k) +  \Bd_u\,\u(k)\label{EQN:state-time-discrete-general},\\
 \y(k) &= \Cd_y\,\q(k) + n(k) \label{EQN:output-time-discrete-general},\\
  \z(k) &= \Cd_z\,\q(k)  \label{EQN:output-time-discrete-genera:2l},
\end{align}
where $\Ad = \exp\left (\A\,\Delta\t\right ),\Bd = \Delta\t\,\B$ and $\Cd = \C$. For more details, the interested reader can refer to any control book (see e.g. \cite{2000CT-glad-ljung}).

\subsection{Transfer functions and Finite-impulse responses}\label{SEC:io}
Given a measurement signal $\y(t)$, our aim is to design an actuator signal $\u(t)$. The relation between input and output signals is of primary importance. Since we are interested in the effect of the control signal $\u(\t)$ on the system, we assume the disturbance signal $\d(\t)$ to be zero. Thus, given an input signal $\u(t)$ and a zero initial condition of the state, the output $\z(t)$ of \refeqns{space-state-detailed}{z-detailed} may formally be written as
\begin{align}
 \z(\t) &= \int_{0}^{\t}\Pl_{\z\u}(\t)\;\u(\t-\tau)\,d\tau,\label{EQN:time-conv-Hzu}
\end{align}
where the kernel is defined by
\begin{align}
 \Pl_{\z\u}(\t) &\triangleq \C_{\z}\,e^{\A\t}\,\B_{\u} ,\quad\t\geq 0 \label{EQN:Hzu}.
\end{align}
Note that the description of the input-output (I/O) behaviour between $\u(t)$ and $\z(t)$ does not require the knowledge of the full dynamics of the state but only a representation of the impulse response between the input $\u$ and the output $\z$, here represented by \refeqn{Hzu}. A Laplace transform results in a transfer function
\begin{equation*}
\hat{\z}(s) =  \hat\Pl_{\z\u}(s)\hat{\u} (s) = (\C_\z (s I -\A)^{-1}\B_\u)\hat{\u}(s)
\end{equation*}
with $s\in\mathbb{C}$.  Henceforth the $hat$ on the transformed quantities is omitted since related by a linear transformation to the corresponding quantities in time-domain. One may formulate a similar expression for the other input-output relations, which for our case with  three inputs and two outputs, induces $6$ transfer functions, i.e.
\begin{align}
\left[\begin{array}{c}
\z(s)\\
\y(s)
\end{array}\right] 
= 
\left[\begin{array}{ccc}
\Pl_{\z\d}(s)\,\,\,
\Pl_{\z\u}(s)\,\,\,
\Pl_{\z\n}(s)\\ 
\Pl_{\y\d}(s)\,\,\,
\Pl_{\y\u}(s)\,\,\,
\Pl_{\y\n}(s) \end{array}\right]
\left[\begin{array}{c}
\d(s)\\
\u(s)\\
\n(s)
\end{array}\right]
\label{EQN:tf-system}.
\end{align}

\begin{figure}
 \centering
 \includegraphics[width=.45\textwidth]{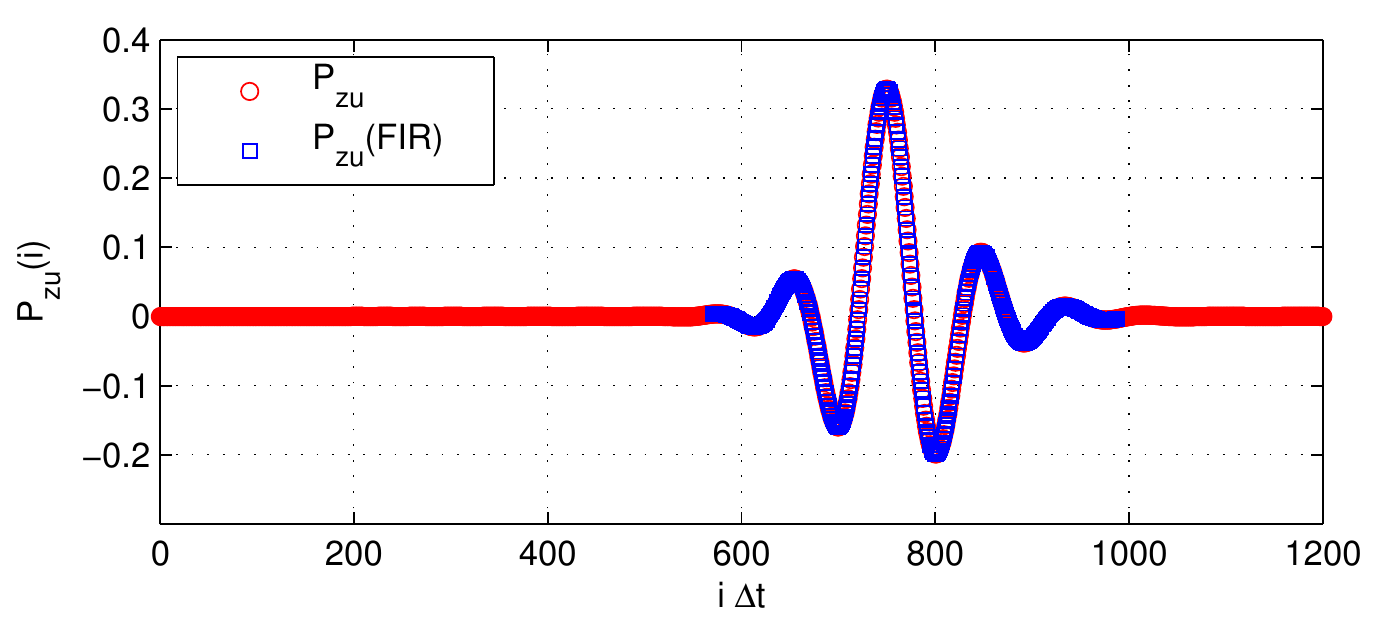}
 \caption{Time discrete impulse response ($\color{red}\circ$) between the input $u$ to the output $z$; due to the presence of strong time-delays in the system, a lag of $t\approx550$ is observed. The relevant part of the kernel is reconstructed via a FIR filter ($\color{blue}\square$). {\small \texttt{[script02.m]}}}
 \label{FIG:hzu-fir}
\end{figure}

I/O relations similar to \refeqn{time-conv-Hzu} can be found for the time-discrete system. The response $z(k)$ of the system (with $\q_0=0$) to an input $\u(k)$ is
\begin{equation}
\begin{split}
 \z(k)   &=  \sum_{i=1}^{k} \pl_{\z\u}(i)\;\u(k-i),
\end{split} \label{EQN:hzq-hzu}
\end{equation}
where
\begin{equation}
 \pl_{\z\u}(k) \triangleq \Cd_z\,\Ad^{k-1}\,\Bd_{\u},\quad k=1,2,...
\end{equation}
This procedure is usually referred to as \emph{z-transform}; for more details, we refer to \cite{2000CT-glad-ljung,skogestad05}. In the limit  of $k\rightarrow\infty$, it is possible to truncate \refeqn{hzq-hzu}, since the propagating wave-packet that is generated by an impulse in $\u$ will be detected by the output $\z$ after a time-delay (this can be observed in \reffig{hzu-fir}, where the impulse response is depicted). Thus, $\pl_{\z\u}(i)$  is non-zero only in a short time interval
and one may truncate the sum to a finite number of time steps, $N_{zu,\,f}$. Due to the strong time-delay, the initial part of the sum is also zero and the lower limit of the sum can start from $N_{zu,\,i}$. This results in a sum 
\begin{equation}
 \z(k) \approx \sum_{i=N_{zu,\,i}}^{N_{zu,\,f}} \pl_{\z\u}(i)\;\u(k-i),
\end{equation} 
which is called the Finite Impulse Response (FIR), \cite{1995AC-astrom-wittenmark}.
Note that the presence of time delays in the system is a limiting factor of the control performance. In general, a disturbance with a time scale smaller than the time delay that affects the system is difficult to control \cite{2000CT-glad-ljung}. In particular, while the compensator could still be able to damp those disturbances, it may lack  robustness, \refsec{robustness}.

\subsection{Controllability and observability}\label{SEC:gramians}
The choice of sensors and actuators is particular relevant for the control design; indeed, the measurement of the sensor $\y$ enables to compute the control signal $\u(t)$, that feeds the actuator. Thus, it is important to know: (i) if the system can be affected by the actuator $\u$; (ii) if the system can be detected by the sensor $\y$. In other words, we aim at identify the states of the system that are \emph{controllable} and/or \emph{observable}. These two properties of the I/O system are referred to as \emph{observability} and \emph{controllability}, \cite{2000CT-glad-ljung,amr2009-bagheri-et-al} and can be analyzed introducing the corresponding Gramians $\G_{o}$ and $\G_{c}$
\begin{align}
 \G_{o} &\triangleq \int_0^{\infty} e^{\A^H\t}\,\C^H\C\,e^{\A\t}\;d\t, \label{EQN:obser-gramian}\\
 \G_{c} &\triangleq \int_0^{\infty} e^{\A \t}\,\B\,\B^H\,e^{\A^H\t}\;d\t. \label{EQN:contr-gramian}
\end{align}
By construction, the Gramians ($\G_{o}$,$\G_{c}$) are positive semi-definite matrices in $\Real^{n_\q\times n_\q}$ and can be computed for each or all the outputs/inputs. It can be proved that the two Gramians are solutions of the Lyapunov equations, \cite{2000CT-glad-ljung}
\begin{align}
 \A^H\,\G_{o} + \G_{o}\,\A + \C^H\,\C = \mat{0},\label{EQN:obser-lyap}\\
 \A\,\G_{c} + \G_{c}\,\A^H + \B\,\B^H = \mat{0}.\label{EQN:contr-lyap}
\end{align}
\begin{figure}
 \centering
 \includegraphics[width=.45\textwidth]{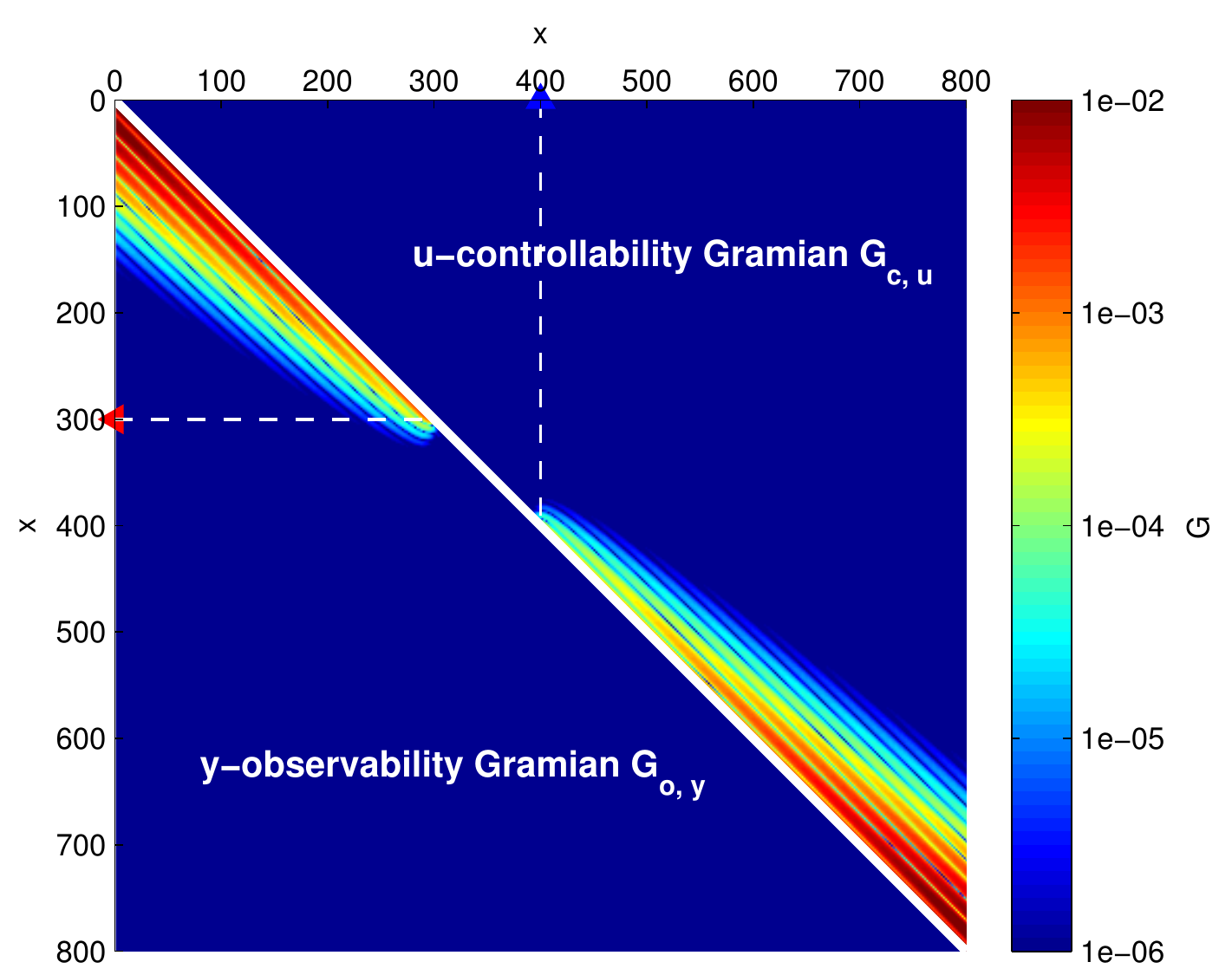}
 \caption{Controllability ($\G_{c,\,\u}$) and observability ($\G_{o,\,\y}$) Gramians, normalized by their trace; the absolute values are reported in logarithmic scale as a function of the streamwise direction ($x$). Due to the symmetry, only the upper/lower triangular part of each Gramian is shown. {\small\texttt{[script03.m]}}}
 \label{FIG:gramians}
\end{figure}
The spatial information related to the Gramians can be analyzed by diagonalizing them; the corresponding decompositions allow to identify and rank the most controllable/observable structures \cite{amr2009-bagheri-et-al}. On the other hand, for systems characterized by a small number of degrees of freedom, it is possible to directly identify the regions where the flow is observable and/or controllable. \reffig{gramians} shows the controllability Gramian related to the actuator $\u$ $(\G_{c,\,\u})$ and the observability Gramian related to the sensor $\y$ $(\G_{o,\,\y})$ for our system. The region downstream of the actuator is influenced by its action, due to the strong convection of the flow. The observability Gramian $\G_{o,\,\y}$ indicates the region where a propagating perturbation can be observed by the sensor $\y$. Note that the two regions do not overlap, thus wave-packets generated at the location $\u$ are not detected by a sensor $\y$, when is placed upstream of the actuator. This feature has important consequences on the closed-loop analysis, as introduced in the next section.

\subsection{Closed-loop system}\label{SEC:cloop}
\begin{figure}
 \centering
 \includegraphics[width=.45\textwidth]{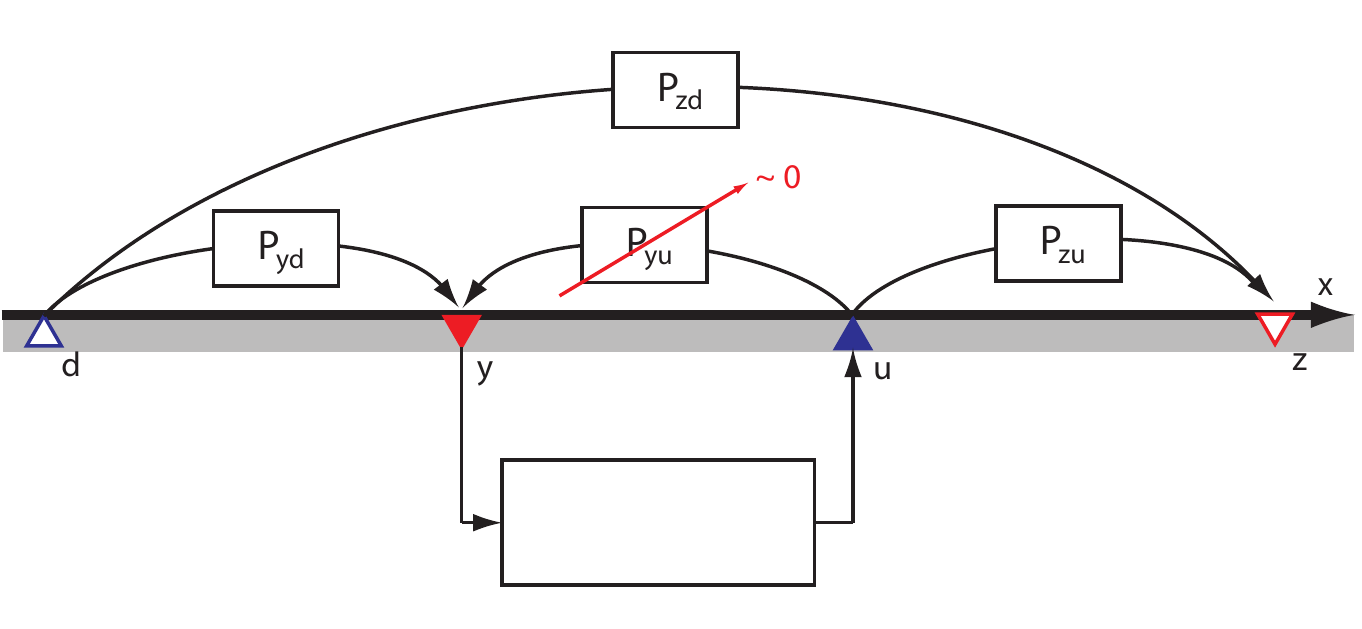}
 \setlength{\unitlength}{\textwidth} \begin{picture}(0,0) \put(-.282, .033){\scriptsize\sffamily compensator}\end{picture}\hspace{-6pt}
 \caption{{Schematic figure showing the 5 transfer functions defining the closed-loop system \refeqn{icl-system}. The transfer functions $\Pl_{\y\d}$, $\Pl_{\z\d}$ describe the input/output behaviour between the disturbance $\d$ and the outputs $\y$ and $\z$, respectively; $\Pl_{\y\u}$ and $\Pl_{\z\u}$ relate the actuator $\u$ to the two outputs $\y$ and $\z$, respectively, while $\Ct_{\u\y}$ is the compensator transfer-function. Because of the convectively unstable nature of the flow, $\Pl_{\y\u}$ is negligible for the chosen sensor/actuator locations; thus it does not allow any feedback.}}
 \label{FIG:transfer-function}
\end{figure}

The aim of the control design is to identify a second linear system $\Ct_{\u\y}$, called \emph{compensator}, that provides a mapping between the measurements $\y(t)$ and the control-input $\u(t)$, i.e.
\begin{equation*}
  \u(t)= \Ct_{{\u}{\y}}(t)\y(t)
\end{equation*}
The chosen compensator is also called \emph{output feedback controller}
\cite{doyle89, zhou:doyle:glover:02}. This definition underlines the
dependency of the control input $\u(t)$ from the measurements $\y(t)$. By considering the relation in frequency domain and inserting it into the plant \refeqn{tf-system}, the \emph{closed-loop} system between $\d(s)$ and $\z(s)$ is obtained in the form, 
\begin{equation}
\z(s) =
\left[\Pl_{\z\d}(s)\, +\,
\dfrac{\Pl_{\z\u}(s)\,\Ct_{\u\y}(s)\,\Pl_{\y\d}(s)}
{{1}-\Pl_{\y\u}(s)\,\Ct_{\u\y}(s)}\right]\d(s).
\label{EQN:icl-system}
\end{equation}
By choosing an appropriate $\Ct_{\u\y}(s)$, we may modify the system dynamics. The  graphical representation of the closed-loop system is shown in \reffig{transfer-function}. The transfer function $\Pl_{\y\u}(s)$ describes the signal dynamics from the actuator $\u$ to the sensor $\y$. By definition, a feedback configuration is obtained when $\Pl_{\y\u}(s)\neq 0$, i.e. when the sensor can measure the effect of the actuation. On the other hand, if $\Pl_{\y\u}(s)$ is zero (or very small), the closed-loop system reduces to a disturbance feedforward configuration\cite{doyle89,zhou:doyle:glover:02}. In this special case, from the dynamical point of view such a system behaves as an open-loop system despite the closed-loop design \cite{skogestad05}. Due to this inherent ambivalence within the framework of the output feedback control, sometimes the definition of \emph{reactive control} is used for indicating all the cases where the control signal is computed based on measurements of the system; thus, the definition of closed-loop system more properly applies to a system where the reactive controller is characterized by feedback \cite{gad2007flow}.

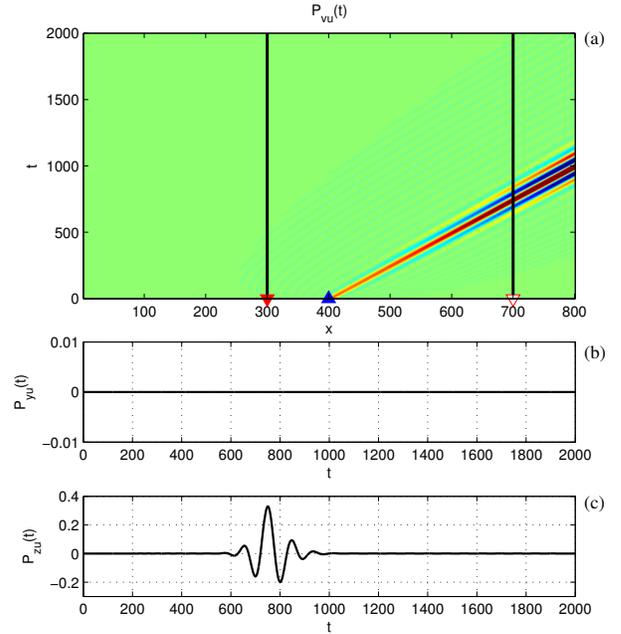
\begin{figure}
 \centering
 \includegraphics[width=.45\textwidth]{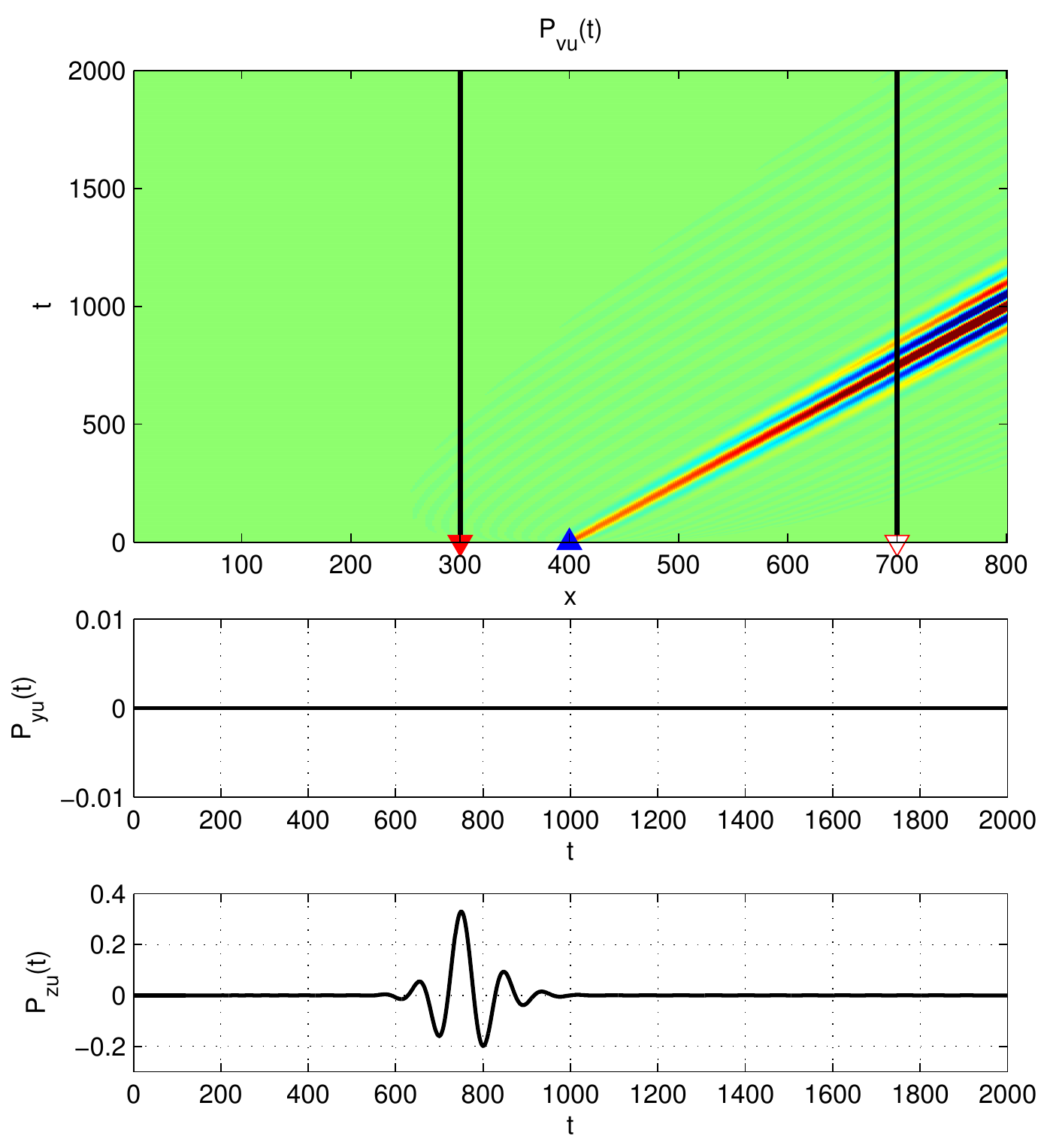}
 \setlength{\unitlength}{\textwidth} \begin{picture}(0,0) \put(-.020, .455){\scriptsize(a)}
                                                          \put(-.020, .215){\scriptsize(b)}
                                                          \put(-.020, .100){\scriptsize(c)}\end{picture}\hspace{-6pt}
 \caption{The disturbance generated by the impulse response of the system at the actuator location $\u$ in (a) is shown as a function of the streamwise direction ($x$) and time ($t$). The wave-packet is detected only by the output $\z$ (c); due to the convective nature of the flow, the sensor placed upstream of the actuator can not detect the propagating disturbance, and the resulting signal is practically null ($b$). {\small\texttt{[script02.m]}}}
 \label{FIG:Hqu-Hzu}
\end{figure}

In a convection-dominated system, the sensor should be placed upstream of the actuator, in order to detect the upcoming wave-packet before it reaches the actuator (see also \reffig{gramians}); if it is placed downstream, the actuator has no possibility to influence the propagating disturbance once it has reached the sensor. \reffig{Hqu-Hzu} shows the state and signal responses of the KS system to impulse in $\u$, where it is clear that the actuator's action is not detected by the sensor $\y$, in practice $\Pl_{\y\u}(s)\approx 0$. Note that no assumptions about the compensator has been made; the {feedback} or {feedforward} setting is determined by the choice of sensor and actuator placement.

\subsection{Robustness} \label{SEC:robustness}
In practice, model uncertainties are unavoidable and it is important to estimate how much the error arising from the mismatch between the physical system and the model affects the stability and performance of the closed-loop system. In general, one wishes to have a controller that does not amplify un-modelled errors over a range of off-design conditions: a robustness analysis aims at identify this range. A useful quantity in this context, is the sensitivity transfer function, which is defined as the denominator in the second term on the right-hand side of \refeqn{icl-system}, i.e.
\begin{align}
\Sl(s) &= \dfrac{1}{{1}-\Pl_{\y\u}(s)\Ct_{\u\y}(s)}.\label{EQN:sen-system}
\end{align}
Robustness can be quantified as the infinity norm of $\mathcal{S}(s)$. Good stability margins are guaranteed when this norm is bounded, typically $\|\Sl\|_{\infty}<2.0$, see \cite{skogestad05}. A second measure is the phase margin, that represents the maximum amount of allowable phase error before the instability of the closed-loop occurs. Indeed, the gain margin and the phase margin are the upper limit of amplification and phase error, respectively, that guarantee marginal stability of the closed-loop system. 

Note that the internal stability functions are characterized by a proper dynamics. In the \emph{loop-shaping} approach, the controller is designed by shaping the behaviour of the internal transfer function \cite{skogestad05}. Unfortunately, this methodology is difficult to be applied in complex system. A systematic approach for the robust design is represented by the optimal, robust  $\mathcal{H}_{\infty}$  (see \cite{zhou:doyle:glover:02}), where the sensitivity margins can be optimized. A more computationally demanding alternative is represented by the controllers based on numerical optimization running on-line, such as the model-predictive control (MPC) (\refsec{MPC}) or adaptive controllers  (\refsec{compensator-FXLMS}).

Thus, feedback controllers may be designed to have small sensitivity. In that regard robustness is a non-issue in a pure feedforward configuration; indeed, $\Pl_{\y\u}(s)\approx0$ and $\|\Sl\|_{\infty}\approx1$. However, a feedforward controller is highly affected by unknown disturbances and model uncertainty, that drastically reduce the overall performance of the device. Moreover, a feedforward controller is not capable in modifying the dynamics of an unstable plant; thus, feedback controllers are required for globally unstable flows \cite{sipp2013closed}.

The studies performed by \cite{jfm2013-julliet-schmid-huerre} and \cite{pof2013-belson-semeraro-et-al} show that in convectively unstable flows a feedback configuration allows the possibility of robust-control design but it does not guarantee  \emph{optimal} performances in terms of amplitude reduction. In this review, we adopt a feedforward configuration in order to achieve optimal performances.  As we will show in  \refsec{compensator-FXLMS}, robustness may be addressed to some extent using adaptive control techniques.

\section{Model-based control}       \label{SEC:control}
 In this section, we assume the full knowledge of the state $\q(t)$ for the computation of the control signal $\u(t)$. This signal is  fed back into the system   in order to minimize the energy of the output $\z(t)$.  For linear systems, it is possible to identify a  \emph{feedback gain} $\K(t)$,  relating the control signal to the state, i.e.
\begin{equation}
\u(t) = \K(t)\q(t).\label{EQN:feedback-law-K}
\end{equation}
The aim of the section is to compare and link the classical LQR problem \cite{1995OC-lewis} to the  more general MPC approach\cite{2001-bewley-moin-temam, arfm2007-kim-bewley}. In the former approach, one assumes an infinite time horizon ($\t\rightarrow\infty$), allowing the computation of the feedback gain by solving a Riccati equation (see \refsec{control-optimal-LQR}). In the latter approach,  the  optimization is performed  with a final time $T$ that is receding, i.e.~it slides forward in time as the system evolves. In \refsec{MPC-lincon}, we introduce this technique for the control of a linear system with constraints on the actuator signal, while in \refsec{MPC-wlincon} the close connection between the unconstrained MPC and the LQR is shown.
Finally, note that the framework introduced in this section makes use of a system's model. Model-free methods based on adaptive strategies are introduced in \refsec{compensator}.

\subsection{Optimal control}\label{SEC:control-optimal}
The aim of the controller is to compute a control signal $\u(t)$ in order to
minimize the norm of the fictitious output
\begin{equation}
{\z}'(t) = \left[\begin{array}{c}  \z(t) \\ \u(t) \end{array}\right] =
\left[\begin{array}{c} \C_{\z} \\ \mat{0} \end{array}\right]\,\q(t) +
\left[\begin{array}{c} 0
    \\ 1 \end{array}\right]\,\u(\t) \label{EQN:control-cost-signal-def}, 
\end{equation}
where now the control signal is also included. We define a \emph{cost function} of the system  
\begin{equation}
\begin{split}
 \J\left(\q(\u),\u\right) = \frac{1}{2} \int_0^T \left[\begin{array}{c}  \z \\ \u \end{array}\right]^H
                                                 \left[\begin{array}{cc} \Wz & 0   \\
                                                                         0   & \Wu \end{array} \right]
                                                 \left[\begin{array}{c}  \z \\ \u \end{array}\right]\,d\t. \label{EQN:control-cost-function-def}
\end{split}
\end{equation}
This cost function is quadratic and includes the constant matrices $\Wz\ge 0$ and $\Wu > 0$. The matrix $\Wz$ is used to normalize the cost output, specially when multiple $\z(\t)$ are used, while the weight $\Wu$ determines the amount of penalty  on control effort \cite{1995OC-lewis}. Using \refeqn{control-cost-signal-def}, \refeqn{control-cost-function-def} is rewritten as
\begin{equation}
\begin{split}
\J\left(\q(\u),\u\right) &= \frac{1}{2} \int_0^T \left( \q^H\,\left(\C_{\z}^H \Wz \C_{\z}\right)\,\q + \u^H\,\Wu\,\u \right)d\t = \\
                         &= \frac{1}{2} \int_0^T \left( \q^H\,\Wq\,\q + \u^H\,\Wu\,\u \right)d\t\label{EQN:control-cost-signal}
\end{split}
\end{equation}
where $\Wq=\C_{\z}^H\Wz\C_{\z}$. We recall from \refsec{system-state-space}  that  the sensor $\C_{\z}$ is placed far downstream in the domain, so we are minimizing the energy in localized region. We seek a control signal $\u(t)$ that  minimizes the cost function $\J\left(\q(\u),\u\right) $ in some time interval $t\in[0,T]$ subject to the dynamic constraint
\begin{equation}
 \ddt{\q}(t) = \A~\,\q(t) + \B_{\u} \,\u(t) \label{EQN:control-space-state}.
\end{equation}
Note that we do not consider the disturbance $\d(t)$ for the solution of the optimal control problem. In a variational approach, one defines a Lagrangian
\begin{equation}
\begin{split}
\Ja\left(\q(\u),\u\right) = \frac{1}{2}&\int_0^T \left( \q^H\,\Wzq\,\q + \u^H\,\Wu\,\u \right)d\t + \\
                      +&\int_0^T \qadj^H\left(\ddt{\q} -\A~\,\q -\B_{\u}\u\right)d\t,\label{EQN:augmented-control-cost}
\end{split}
\end{equation}
where the term $\qadj(t)$ acts as a Lagrangian multiplier \cite{gunzburger2003perspectives},  (also called the adjoint state). The expression in the last term is obtained via integration by parts. Instead of minimizing $\J$ with a constraint \refeqn{control-space-state} one may minimize $\Ja$ without any constraints.

The dynamics of the adjoint state $\qadj(t)$ is obtained by requiring ${\partial{\Ja}} /\partial{\q}=\vec{0}$, which leads to
 \begin{equation}
\begin{split}
-\ddt{\qadj}(t) &=\adj{\A}\,\qadj(t) +\Wq\,\q(t),\\ 
     \vec{0} &=\qadj(T).             \label{EQN:control-lagrangian-adjstate}
\end{split}
\end{equation}
 The adjoint field $\qadj(t)$ is computed by marching backwards in time this
 equation, from $t=T$ to $t=0$. The optimality condition is obtained by the gradient
\begin{equation}
\dd{}{\Ja}{\u} = \adj{\B_{\u}}\,\qadj + \Wu\,\u \label{EQN:control-lagrangian-u}.
\end{equation}
The resulting equations' system can be solved iteratively as follows:
\begin{enumerate}
\item The state $\q(t)$ is computed by marching forward in time  \refeqn{control-space-state} in $t\in[0,T]$. At the first iteration step, $k=1$, an initial guess is taken for the control signal $\u(t)$.

\item The adjoint state $\qadj(t)$ is evaluated marching \refeqn{control-lagrangian-adjstate} backward in time, from $t=T$  to $t=0$. The initial condition $\qadj(T)$ is taken to be zero.

\item Once the adjoint state $\qadj(t)$ is available, it is possible to compute the gradient via \refeqn{control-lagrangian-u} and apply it for the update of the control signal using a gradient-based method;  one may for example apply directly the negative gradient $\Delta \u_k = -\dfrac{\partial \Ja_k}{\partial \u}$, such that the update of the control signal at each iteration is given by
\begin{equation*}
\u_{k+1} = \u_k +\step_k\Delta \u_k.
\end{equation*}
The scalar-valued parameter $\step_k$ is the step-length for the optimization, properly chosen by applying backtracking or exact line search \cite{boyd2004convex}. An alternative choice to the steepest descent algorithm is a conjugate gradient method \cite{2007NR-press-et-al}.
\end{enumerate}
The iteration stops when the difference of the cost function $\J$ estimated at two successive iteration steps is below a certain tolerance or the gradient value $\partial \Ja / \partial \u \rightarrow 0$. We refer to \cite{gunzburger2003perspectives} for more details and to \cite{corbett2001optimal} for an application in flow optimization.

\subsubsection{Linear-quadratic regulator (LQR)} \label{SEC:control-optimal-LQR}
\begin{figure}
 \centering
 \includegraphics[width=.45\textwidth]{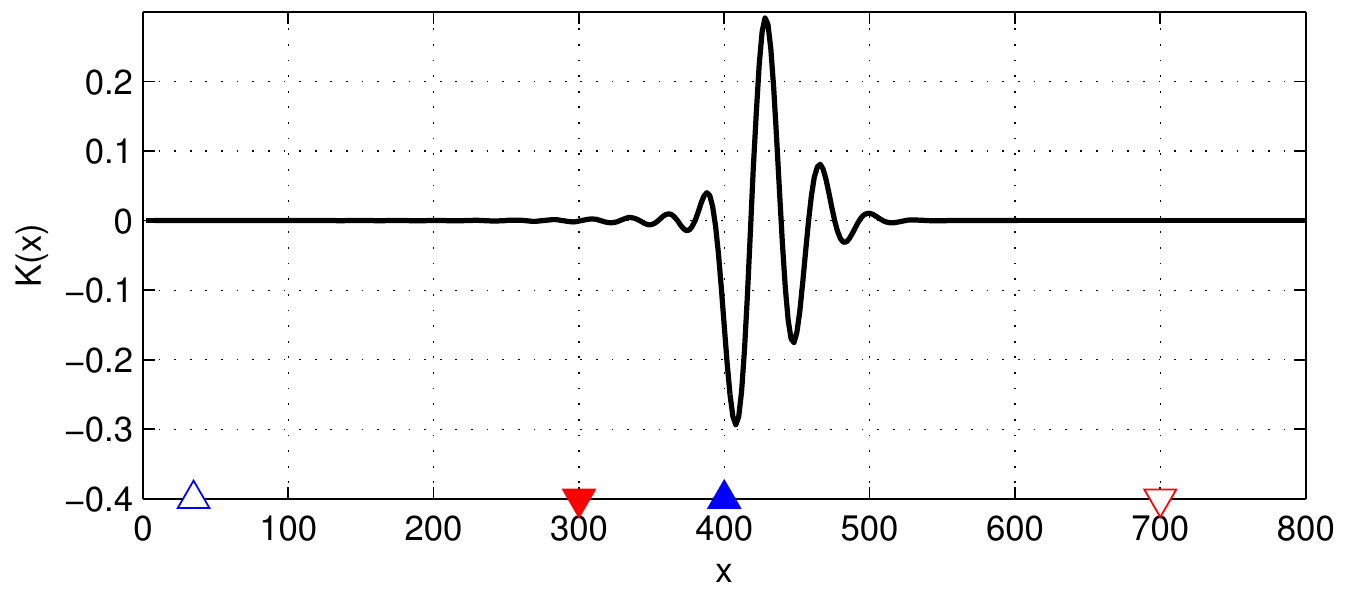}
 \caption{Control gain $\K$ computed using the LQR technique for $\Wz = 1$ and $\Wu = 1$, (see \S\ref{SEC:control-optimal-LQR}). {\small\texttt{[script04.m]}}}
 \label{FIG:K-controller}
\end{figure}
The framework outlined in the previous section is rather general and it can be applied for the computation of the control signal $\u(t)$ also when nonlinear systems or receding finite-time horizons are considered. However, a drawback of the procedure is the necessity of running an optimization on-line, next to the main flow simulation/experiment. When a linear time-invariant system is considered, a classic way to proceed is to directly use the optimal condition \refeqn{control-lagrangian-u} in order to identify the optimal control signal $\u(t)$
\begin{equation}
 \u(t) = -\Wu^{-1} \adj{\B_{\u}}\,\qadj(t).
\end{equation}
The computed control signal $u(\t)$ is \emph{optimal} as it minimizes the cost function $\J\left(\q(\u),\u\right)$ previously defined.
Assuming a linear relation between the adjoint state and the direct state, $\qadj(t)=\X(t)\q(t)$, the feedback gain is given by
\begin{equation}
 \K(\t) = - \Wu^{-1} \adj{\B_{\u}}\,\X(\t). \label{EQN:control-optimal-gain}
\end{equation}
It can be shown that the matrix $\X(t)$ is the solution of a differential Riccati equation \cite{1995OC-lewis}. When $\A$ is stable, $\X(t)$ reaches a steady state as $T\rightarrow\infty$, which is a solution of the algebraic Riccati equation
\begin{equation}
\mat{0} = \adj{\A}\X + \X\A - \X\,\B_{\u}\Wu^{-1}\B_{\u}^H\,\X + \Wzq. \label{EQN:control-Riccati}
\end{equation}
The advantage of this procedure is that $\K$ is a constant and needs to be computed only once. The spatial distribution of the control gain $\K$ is shown in \reffig{K-controller} for the KS system analysed in \refsec{system}, where the actuator is located at $x=400$ and the objective output at $x=700$.  From \reffig{K-controller} one can see that the gain is a compact structure  between the elements $\B_{\u}$ and $\C_{\z}$.  The control gain is independent on the shape of external disturbance $\B_{\d}$.

For low-dimensional systems ($n_{\q}<10^3$), solvers for the Riccati equations \refeqn{control-Riccati} are available in standard software packages \cite{arnold1984riccati}. For larger systems $n_{\q}>10^3$, as the ones investigated in flow control, direct methods are not computationally feasible. Indeed, the  solution of \refeqn{control-Riccati}  is a full matrix, whose storage requirement is at least of order $O(n_{\q}^2)$. The computational complexity is of order $O(n_{\q}^3)$ regardless the structure of the system matrix $\A$ \cite{nlaa2008-benner-li-penz}.  Alternative techniques include the Chandrasekhar method \cite{siamjco1991-bank-ito}, Krylov subspace methods \cite{ieeecs2004-benn}, decentralized techniques based on Fourier transforms for spatially invariant system \cite{bamieh2002, 2000:hogberg:bewley, hogberg:bewley:henning:03:a} and finally iterative algorithms \cite{2010:akht:borg:miro:ziet, maartensson2011synthesis, iutam2009-pralits-luchini , semeraro2013riccati}. Yet, a different approach consists of reducing $n_{\q}$ before the control techniques are applied. In practice, we seek a low-order surrogate system, typically of  $O(n_{\q,\,r})\approx 10-10^2$, whose dynamics reproduces the main features of the original, full-order system. Once the low-order model is identified, the controller is designed and fed into the full-order system; such an approach enables the application of a controller next to real experiments, using small (and fast) real-time computations. The \emph{model-reduction} problem is an important aspect of  control design for flow control; we refer to \refsec{discussion} for a brief overview.

\subsection{Model-predictive control (MPC)} \label{SEC:MPC}
\begin{figure}
 \centering
 \includegraphics[width=.45\textwidth]{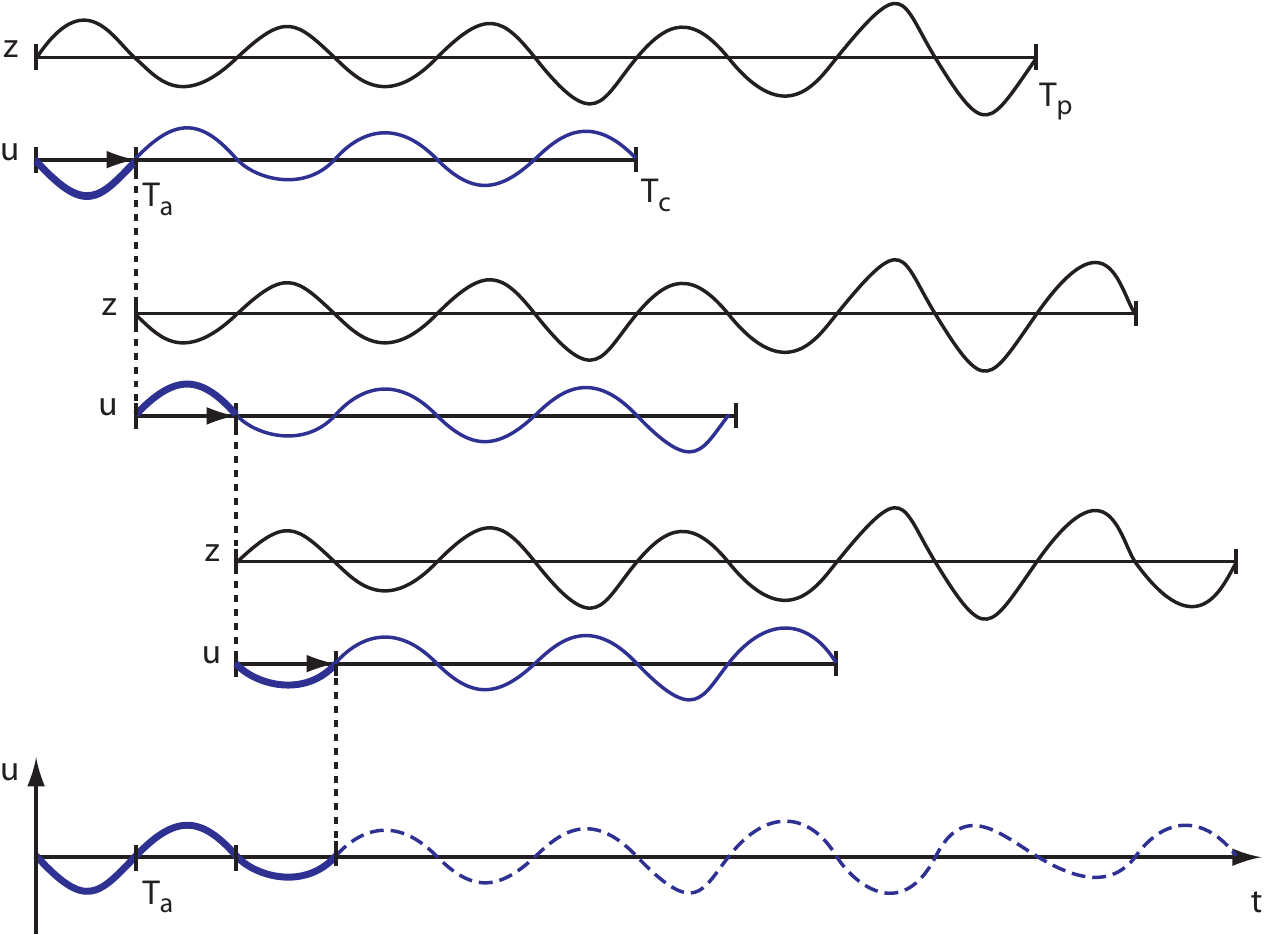}
 \caption{MPC strategy: the controller is computed over a finite time-horizon $T_c$, based on the a predicted time-horizon $T_p$. Once the solution is available, the control signal is applied on a shorter time windows $T_a$. In the successive step, the time-window slides forward in time and the optimization is performed again, starting from a new initial condition at $t=T_a$. The procedures is iterated while proceeding forward in time.}
 \label{FIG:K-strategy-control}
\end{figure}
MPC controllers make use of an identified model to predict the behaviour of the system over a finite-time horizon (see \cite{garcia1989model}, \cite{qin2003survey} and \cite{noack2011reduced} for an overview on the technique). In contrast with the optimal controllers presented in the previous section, the iterative procedure is characterized by a receding finite horizon of optimization. This strategy is illustrated in \reffig{K-strategy-control}; at time $t_0$, a control signal is computed for a short window in time $[t_0,t_0+T_c]$ by minimising a cost function (not necessarily quadratic); $T_c$ is the final time of optimization for the control problem. The minimization is performed on-line, based on the prediction of the future trajectories emanating from the current state at $t_0$ over a window of time $[t_0,t_0+T_p]$, such that $T_p\ge T_c$. In other words, the control signal is computed over an horizon $T_c$ in order to minimize the predicted deviations from the reference trajectory evaluated on a (generally) longer time of prediction $T_p$. Once the calculation is performed, only the first step $T_a$ is actually used for controlling the system. After this step, the plant is sampled again and the procedure is repeated at time $t=t_0+T_a$, starting from the new initial state.  

The MPC approach is applicable to nonlinear  models as well as all nonlinear constraints (for example an upper maximum amplitude for the actuator signals). We present an example of the latter case in the following section. 

\subsubsection{MPC for linear systems with constraints} \label{SEC:MPC-lincon}
Although it is possible to define MPC in continuous-time formulation (see for instance \cite{garcia1989model}, \cite{2001-bewley-moin-temam}), we make use of the more convenient discrete-time formulation. Let $M = T_p/\Delta t$ and $N = T_c/\Delta t$, where the parameter $\Delta t$ is the sampling time. Since $T_p\ge T_c$, we have $M\ge N$. Augmenting the expression \refeqn{hzq-hzu} with a term representing an initial state $\q(k)$ at time $k$, we get
\begin{equation}
\begin{split}
 \z(k+j|k) &= \Cd_{\z}\Ad^j\;\q(k) +\sum_{i=1}^{\min(j,N)}\Cd_{\z}\Ad^{i-1}\Bd_{\u}\;\u(k+j-i) = \\
          &= \pl_{\z\q}(j)\;\q(k) +\sum_{i=1}^{\min(j,N)}\pl_{\z\u}(i)            \;\u(k+j-i),
\end{split}
\end{equation}
where $j = 1,2,\ldots, M$. The state equation can be written in matrix form by recursive iteration, resulting in the matrix-relation
\begin{equation}
\zv_\vec{p}(k)= \F \q(k) + \Ph \uv_\vec{p}(k) \label{EQN:mpc-signal}.
\end{equation}
The matrix $\F$ appearing in \refeqn{mpc-signal} is the observability matrix of the discrete-time system
\begin{equation}
\F  =\left[\begin{array}{c} \pl_{\z\q}(1) \\
                             \pl_{\z\q}(2) \\
                             \vdots        \\
                             \pl_{\z\q}(M) \end{array}\right] = 
      \left[\begin{array}{c} \Cd_{\z}\Ad   \\
                             \Cd_{\z}\Ad^2 \\
                             \vdots        \\
                             \Cd_{\z}\Ad^M \end{array}\right],\label{EQN:obsemat-mpc}
\end{equation}
while the matrix $\Ph$, related to the convolution operator, reads
\begin{equation}
\begin{split}
\Ph =&\left[\begin{array}{cccc} \pl_{\z\u}(1) &                 &        &              \\
                                \pl_{\z\u}(2) & \pl_{\z\u}(1)   &        &              \\
                                \vdots        & \vdots          & \ddots &              \\
                                \pl_{\z\u}(N) & \pl_{\z\u}(N-1) & \cdots & \pl_{\z\u}(1)\\
                                \vdots        & \vdots          &        & \vdots       \\
                                \pl_{\z\u}(M) & \pl_{\z\u}(M-1) & \cdots & \pl_{\z\u}(M-N+1)\end{array}\right] = \\
 &\left[\begin{array}{cccc} \Cd_{\z}\Bd_{\u}          &                           & & \\
                            \Cd_{\z}\Ad\Bd_{\u}       & \Cd_{\z}\Bd_{\u}          & & \\
                            \vdots                    & \vdots                    & \ddots &             \\
                            \Cd_{\z}\Ad^{N-1}\Bd_{\u} & \Cd_{\z}\Ad^{N-2}\Bd_{\u} & \cdots & \Cd_{\z}\Bd_{\u}\\
                            \vdots                    & \vdots                    &        & \vdots       \\
                            \Cd_{\z}\Ad^{M-1}\Bd_{\u} & \Cd_{\z}\Ad^{M-2}\Bd_{\u} & \cdots & \Cd_{\z}\Ad^{M-N}\Bd_{\u}\end{array}\right].\label{EQN:dynamicmat-mpc}
\end{split}
\end{equation}
In literature, the matrix $\Ph$ is also referred to as dynamic matrix, because it takes into account the current and future input changes of the system. Note that the entries of the observability matrix \refeqn{obsemat-mpc} are directly obtained from the model realization, while the entries of the dynamic matrix \refeqn{dynamicmat-mpc} are represented by the time-discrete impulse response between the actuator $\u$ and the sensor $\z$. The input vector $\zv_\vec{p}(k)$ and output vector $\uv_\vec{p}(k)$ are defined collecting the corresponding time-signals at each discrete step
\begin{equation}
\begin{split}
\zv_\vec{p}(k) =\left[\begin{array}{c}
\z(k+1|k)   \\
\z(k+2|k)   \\
\vdots     \\
\z(k+M|k)   \\
\end{array}\right], \qquad
\uv_\vec{p}(k) = \left[\begin{array}{c}
\u(k|k)    \\
\u(k+1|k)  \\
\vdots    \\
\u(k+N-1|k)\\
\end{array}\right].
\end{split}
\end{equation}
Thus, the matrix relation \refeqn{mpc-signal} provides a linear relation between the state $\q(k)$ and the output $\zv_\vec{p}(k)$ when the system is forced by the control input $\uv_\vec{p}(k)$. The evaluation of the future output vector $\zv_\vec{p}(k)$ represents the \emph{prediction} step of the procedure; indeed, assuming that the control signal contained in the vector $\uv_\vec{p}(k)$ is known, we aim at computing the future output $\zv_\vec{p}(k)$, related to the trajectory emanating from the initial condition $\q(k)$.

By following the same rationale already adopted in the optimal control problem, a cost function $\J(k)$  that minimizes the output $\z(t)$ while limiting the control expense is defined,
\begin{equation}
\begin{split}
\J(k) &= \sum_{i=1}^M    \z^H(k+i|k)\,\Wz\,\z(k+i|k)  \\
      &+ \sum_{i=0}^{N-1}\u^H(k+i|k)\, \Wu\,\u(k+i|k) =\\
      &= \zv_\vec{p}(k)^H\,\Wzv\,\zv_\vec{p}(k) + \uv_\vec{p}(k)^H \,\Wuv\,\uv_\vec{p}(k). \label{EQN:mpc-cost}
\end{split}
\end{equation}
The parameters $\Wzv$ and $\Wuv$ are represented by block diagonal matrices
containing the weights $w_z$ and $w_u$. One may also have non-quadratic costs functions in    MPC; examples are given by \cite{2001-bewley-moin-temam} for the control of a turbulent channel. In our case, we choose a quadratic cost function in order to compare performance with the LQR controller. By combining the cost function \refeqn{mpc-cost} and the state equation \refeqn{mpc-signal}, we get
\begin{equation}
\begin{split}
\J(k)&= \zv_\vec{p}(k)^H\,\Wzv\,\zv_\vec{p}(k) + \uv_\vec{p}(k)^H\,\Wuv\,\uv_\vec{p}(k) =\\
     &= \left[\F \q(k) + \Ph \uv_\vec{p}(k)\right]^H\,\Wzv\,\left[\F \q(k) + \Ph \uv_\vec{p}(k)\right] + \\
     &\quad+\uv_\vec{p}(k)^H\,\Wuv\,\uv_\vec{p}(k).\label{EQN:mpc-cost-aug}
\end{split}
\end{equation}
Note that this manipulation is analogous to the definition of Lagrangian already shown for the LQR problem \refeqn{augmented-control-cost}. The minimization of $\J(k)$ with respect of $\uv_\vec{p}(k)$ reads
\begin{equation}
\min_{\uv_\vec{p}(k)}\left\{\frac{1}{2}\uv_\vec{p}^H(k)\,\mat{H}\,\uv_\vec{p}(k)+\vec{c}(k)\,\uv_\vec{p}(k):\; {C}\uv_\vec{p}(k) \leqslant D \right\}\label{EQN:mpc-minimi}
\end{equation}
where
\begin{equation}
\begin{split}
\mat{H}    &= 2\left(\Ph^H\Wzv\Ph+\Wuv\right) \\
\vec{c}(k) &= 2\,\q^H(k)\F^H\Wzv\Ph
\end{split}\label{EQN:mpc-minimiB}
\end{equation}
and $C\uv_\vec{p}(k) \leqslant  D$ is a constraint \cite{siamjo1999-bryd-et-al}, which we have not specified yet.  Once this minimization problem is solved, the control signal is applied for one time step, corresponding to $\Delta T = T_a$,  followed by a new iteration at step $k+1$.

\subsubsection{\bf Actuator saturation as constraint}\label{SEC:saturation}
The need of introducing constraints in the optimization process usually arises when we consider real actuators characterized by nonlinear behaviour, due for instance to saturation effects. For example, the body force generated by plasma actuators \cite{ef2008-grundmann-tropea,corke2010}  -- usually approximated by considering the macroscopic effects on a flow -- is often modelled as a nonlinear function of the voltage  \cite{suzen2005numerical, 2011:Kriegseis}.
\begin{figure}
 \centering
 \includegraphics[width=.45\textwidth]{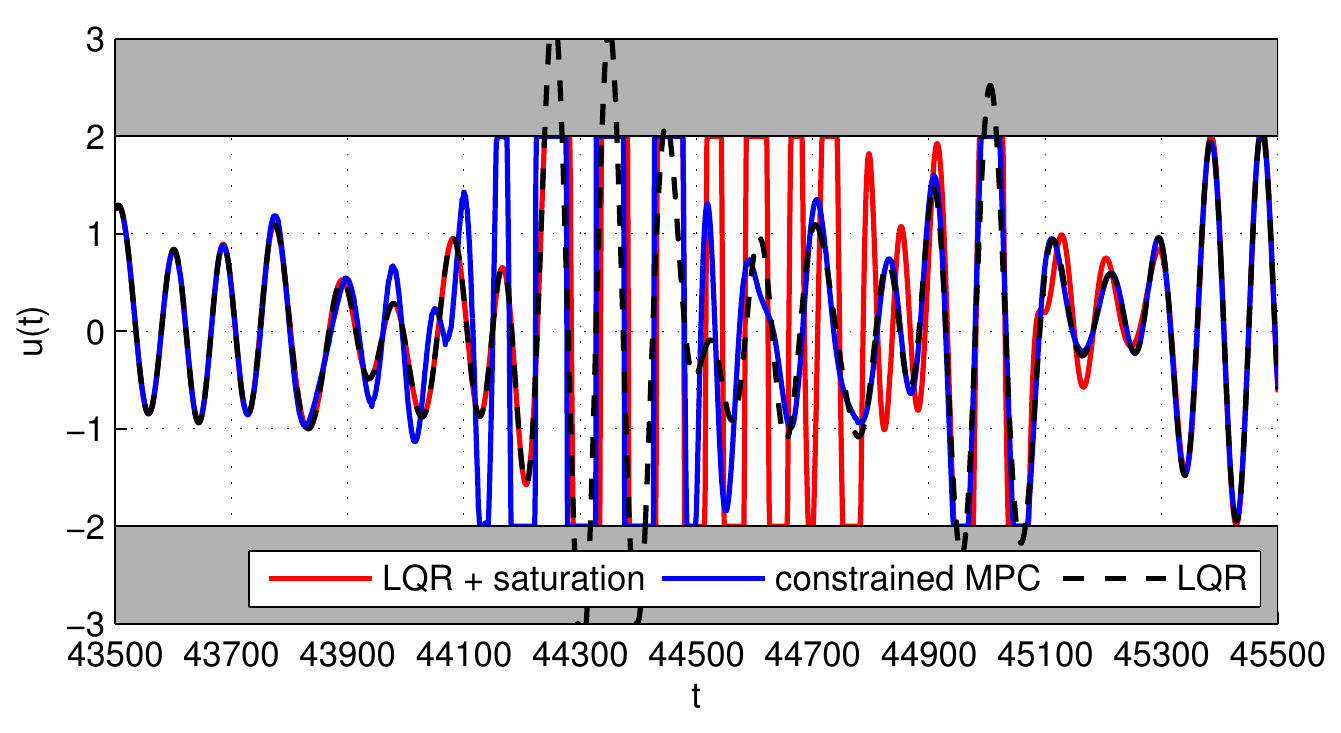}
 \caption{Control design in presence of constraints: the grey regions indicate the limits imposed to the amplitude of the control signal $\u(t)$. The control $\u(t)$ is designed following two different strategies: LQR with a saturation function ($\color{red}-$) and constrained MPC ($\color{blue}-$), see \S\ref{SEC:saturation}. The LQR solution ($\color{black}-~-$) is introduced as reference. The performances of the controllers are shown in terms of $\textstyle{rms}$-velocity reduction in \reffig{rms-saturationcontroller}.} 
 \label{FIG:u-saturationrms-controller}
\end{figure}

Consider now a control signal, whose amplitude is required to be bounded in the interval $-u_{\max} \leqslant \u \leqslant u_{\max}$. We thus minimize
\begin{equation}
  \min_{\uv_\vec{p}(k)}\,\left\{\frac{1}{2} \uv_\vec{p}^H(k)\,\mat{H}\,\uv_\vec{p}(k) + \vec{c}(k)\,\uv_\vec{p}(k):\;
                     \bar{\uv}_{min} \leqslant \uv_\vec{p}(k) \leqslant \bar{\uv}_{max}\right\},
\end{equation}
where $\mat{H}$ and $\vec{c}$ are given by \refeqn{mpc-minimiB}. One may solve this constrained MPC using nonlinear programming \cite{boyd2004convex}. Since the function to be minimized is a quadratic function, we have used a reflective Newton method suggested by \cite{siamjo1996-coleman-li}; this method is implemented in the MATLAB\textregistered~routine \texttt{quadprog.m}.
\begin{figure}
 \centering
 \includegraphics[width=.45\textwidth]{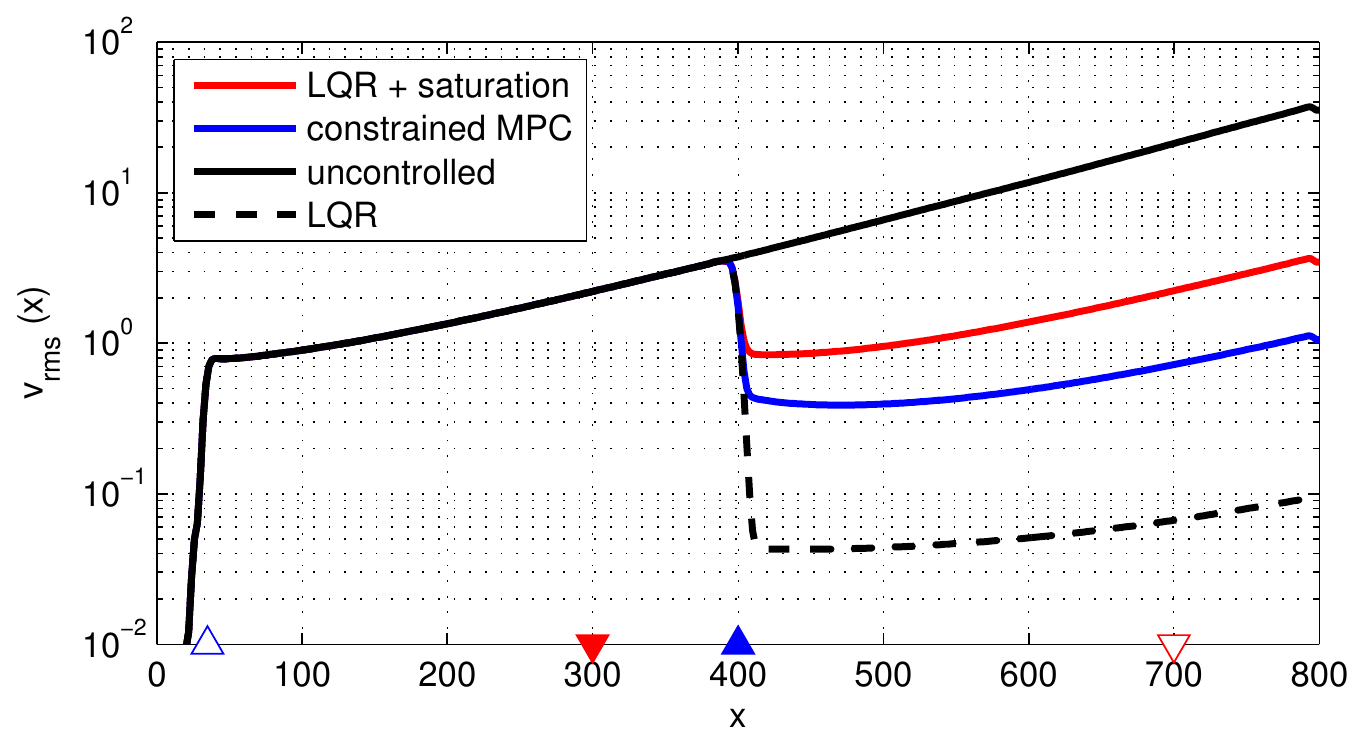}
 \caption{Control of the KS equation. The $\textstyle{rms}$ velocity as a function of the $x$ direction is analyzed; the uncontrolled configuration ($\color{black}{-}$) is compared to three diffrent control strategies already considered in \reffig{u-saturationrms-controller} (same legend).}
 \label{FIG:rms-saturationcontroller}
\end{figure}

We proceed by comparing the performance of the MPC controller with the LQR solution discussed in \refsec{control-optimal-LQR}. For a direct comparison, we apply an {\it ad hoc} saturation function to the LQR control signal, i.e.
\begin{equation}
 \u_{LQR} = \left\{\begin{array}{lllrcl} &\u_{LQR} &\qquad\mbox{if~} &\bar{\u}_{min} <  &\u_{LQR} &< \bar{\u}_{max} \\
                                     &\bar{\u}_{min} &\qquad\mbox{if~} & \bar{\u}_{min}   \geqslant &\u_{LQR} & \\
                                     &\bar{\u}_{max} &\qquad\mbox{if~} & \bar{\u}_{max}    \leqslant &\u_{LQR} &
               \end{array}\right..\label{EQN:saturation}
\end{equation}
As shown in \reffig{u-saturationrms-controller}, the control signal computed by the MPC (blue solid line) closely follows the LQR solution (dashed black line), except in the intervals where the value is larger or smaller than the imposed constraint. By simply applying the saturation function in \refeqn{saturation} to the LQR  signal, the controller becomes suboptimal; the resulting solution deviates from the optimal one and settles back on it after $t\approx300$ time units. Simply cutting off the actuator signal of LQR results in a significant reduction of performance, which in terms of root-mean-square ($\textstyle{rms}$) is almost one order of magnitude (shown in \reffig{rms-saturationcontroller}). The main drawback of the constrained MPC is the computational time required by the on-line optimization, that can be prohibitive in experimental settings.

\subsubsection{MPC for linear systems without constraints}  \label{SEC:MPC-wlincon}
For a linear system with the quadratic cost function \refeqn{control-cost-signal} but without constraints, a prediction/actuation time sufficiently long allows to approximate the solution of the LQR. This is not obvious from the mere comparison of the continuous-time LQR-objective function, \refeqn{control-cost-signal} and \refeqn{augmented-control-cost}, and the discrete-time MPC-objective function, \refeqn{mpc-cost} and \refeqn{mpc-cost-aug}. For a detailed discussion, we refer to \cite{anderson:moore:90},  where the equivalence is demonstrated analytically. In the following, the equivalence is exemplified using the KS equation.
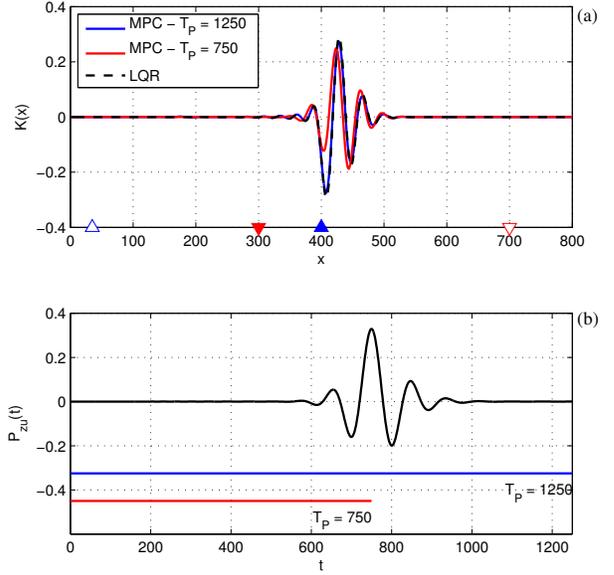
\begin{figure}
 \centering
 \includegraphics[width=.45\textwidth]{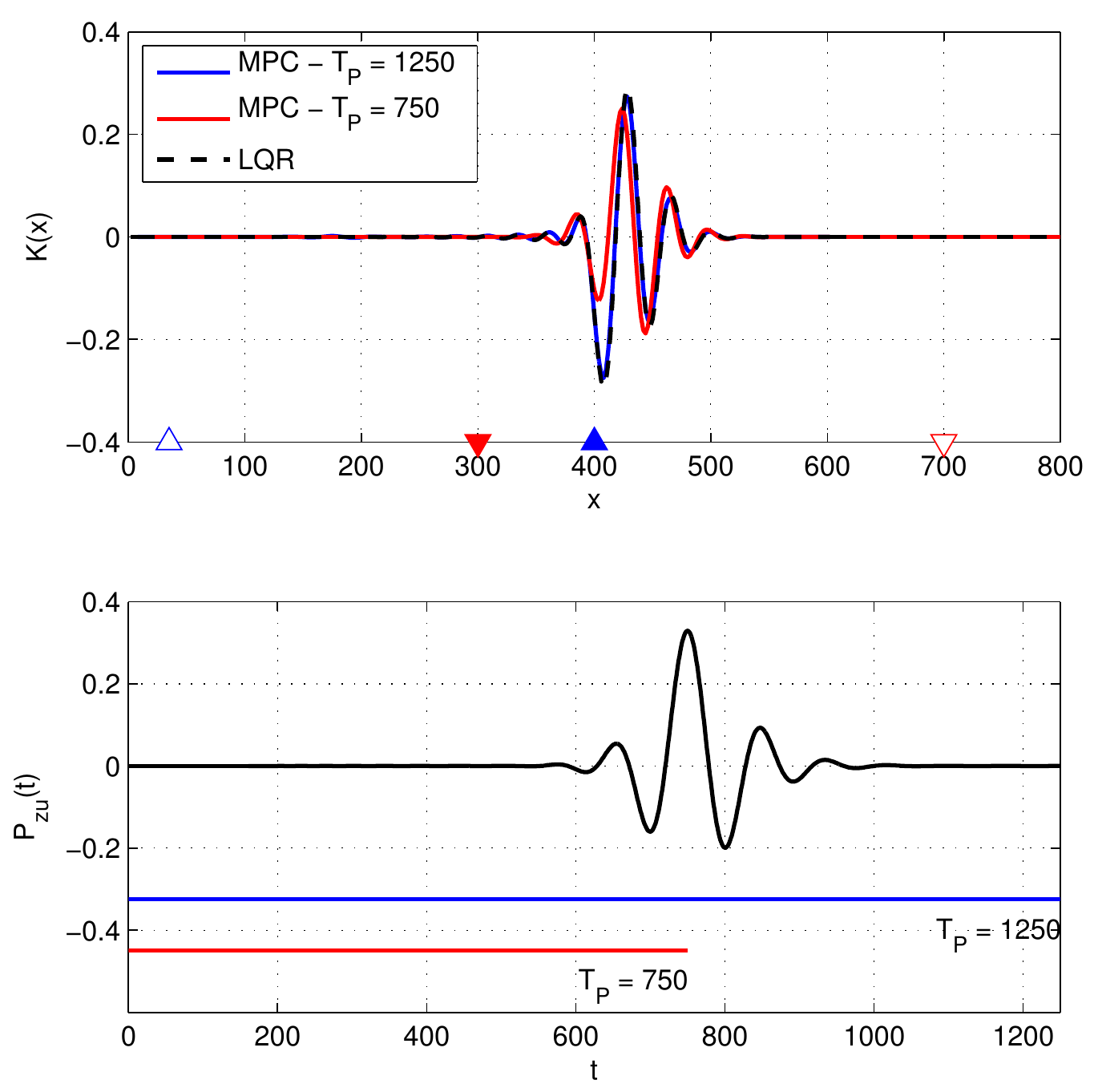}
 \setlength{\unitlength}{\textwidth} \begin{picture}(0,0) \put(-.020, .425){\scriptsize(a)}
                                                          \put(-.020, .193){\scriptsize(b)}\end{picture}\hspace{-6pt}
 \caption{In (a) the LQR  solution (\S\ref{SEC:control-optimal-LQR}) is compared to the MPC gains computed for two different times of optimization $T_p$ without constraints, see \S\ref{SEC:MPC-wlincon}. The optimization times are compared to the impulse response $\Pl_{zu}(\t)$ (b). Note that for longer time $T_p$, covering the main dynamics of the impulse response $\Pl_{zu}(\t)$, the MPC and LQR solutions are equivalent.}
 \label{FIG:K-controller-compare}
\end{figure}

When there are not imposed constraints, the optimization problem in \refeqn{mpc-minimi} corresponds to a Quadratic Program \cite{boyd2004convex}; by taking the derivative of $\J(k)$ with respect of $\uv_\vec{p}(k)$, we may obtain $\uv_\vec{p}(k)$ as solution of the following least-square problem 
\begin{equation}
\begin{split}
\uv_\vec{p}(k) &= -\mat{H}^{\dagger}\vec{c}^H =\\ 
       &= -\left(\Ph^H\Wzv\Ph+\Wuv\right)^{\dagger}\Ph^H\Wzv\F\q(k) =\\ 
       &=  \left[\begin{array}{c}\K_0   \\
                                 \K_1   \\
                                 \vdots \\
                                 \K_{N-1}  \end{array}\right]\,\q(k), 
\end{split}
\end{equation}
where $(\cdot)^{\dagger}$ indicates the Moore-Penrose generalized inverse matrix, \cite{mpcps1955-penrose}. Note that this is a least square problem (in general, $M\ge N$). If we assume an actuation time-horizon $\Ta = \Delta\t$, at each time step the control signal $\u(k)$ reads 
\begin{equation}
 \u(k) = \K_0\,\q(k).
\end{equation}
In \reffig{K-controller-compare}(a), the solid dashed line corresponds to the LQR gain obtained by solving a Riccati equation, while the coloured lines correspond to the unconstrained MPC solution for different final time of prediction $T_p$. For a shorter time of optimization ($T_p=750$, red solid line) only a portion of the dynamics of $\pl_{\z\u}(i)$ (see \reffig{K-controller-compare}(b)) is contained in the MPC gain. For longer times ($T_p=1250$, blue solid line) the MPC converges to the infinite-time horizon LQR solution.

\section{Estimation}                \label{SEC:estimation}
In this section, we assume that the only information we can extract from the system is the measurement $\y(\t)$. This signal is used to provide an estimation $\hat{\q}(\t)$ of the state  such that the error given by
\begin{equation}
 \e(\t) = \q(\t) - \hat{\q}(\t) \label{EQN:estimation-error-def},
\end{equation}
is kept as small as possible. We first derive the classical Kalman Filter, where in addition to $\y(\t)$, one requires a state-space model of the physical system. Then we discuss the least-mean square (LMS) technique, which only relies on the  measurement $\y(\t)$.

\subsection{Luenberger observer and Kalman filter} \label{SEC:estimation-kalman}
The observer is a system in the following form
\begin{align}
 \ddt{\hat{\q}}(\t) &= \A~    \,\hat{\q}(\t) + \B_{\u~} \,\u(\t) - \L\,\left(\y(\t) - \hat{\y}(\t)\right) \label{EQN:estimation-space-state},\\ 
       \hat{\y}(\t) &= \C_{\y}\,\hat{\q}(\t)                                                  \label{EQN:estimation-y},\\
       \hat{\z}(\t) &= \C_{\z}\,\hat{\q}(\t)                                                  \label{EQN:estimation-z}.
\end{align}
This formulation was proposed for the first time by Luenberger in \cite{1979IDS-luenberger}, from whom it takes the name.
Comparing this system with \refeqn{space-state-detailed}, it can be noticed that it takes into account the actuator signal $\u(\t)$ but it ignores the unmeasurable inputs -- the disturbance $\d(\t)$ and the measurement error $\n(\t)$. In order to compensate this lack of information, a correction term based on the estimation $\hat{\y}(\t)$ of the measurement $\y(\t)$ is introduced, filtered by the gain matrix $\L$.

The aim is to design $\L$ in order to minimize the magnitude of the error between the real and the estimated state, i.e. expression defined in \refeqn{estimation-error-def}. Taking the difference term by term between \refeqn{space-state-detailed} and \refeqn{estimation-space-state}, an evolution equation for the $\e(\t)$ is obtained,
\begin{equation}
 \ddt{\e}(\t) = \left(\A + \L\C\right)\,\e(\t) + \B_{\d}\,\d(\t) - \L\,\n(\t) \label{EQN:estimation-error-space-state}.
\end{equation}
It can be seen that the error is forced by the disturbance $\d(t)$ and the measurement error $\n(t)$, i.e. precisely the unknown inputs of the system.

\subsubsection{Kalman filter}\label{SEC:kalman}
In the Kalman filter approach both the disturbance $\d(t)$ and the measurement error $\n(\t)$ are modelled by white noise,  requiring a statistical description of the signals. The auto-correlation of the disturbance signal is given by 
\begin{equation}
  \R_{\d}(\tau) \triangleq \int_{-\infty}^{+\infty} \d(t)\,\d^H(\t-\tau)\;dt.
\end{equation}
This function tells us how much a signal is correlated to itself after a shift $\tau$ in time. For a white noise signal this function is non-zero only when a zero shifting ($\tau = 0$) in time is considered and its value is the variance of the signal.  Hence, the correlation functions for the considered inputs signal $\d(\t)$ and $\n(\t)$ are
\begin{equation}
  \R_{\d}(\tau) = \var_{\d}\,\delta(\tau) \quad\mbox{and}\quad \R_{\n}(\tau) = \var_{\n}\,\delta(\tau),
\end{equation}
where $\var_{\d}$ and $\var_{\n}$ are the variances of the two signals and $\delta(\tau)$ is the continuous Dirac delta function. 
When a system is forced by random signals, also the state becomes a random process and it has to be described via its statistical properties. Generally the calculation of these statistics requires a long time history of the response of the system to the random inputs.  But for the linear system \refeqn{estimation-error-space-state}, it is possible to calculate the variance of the state $\Y\in\mathbb{R}^{n_\q\times n_\q}$ by solving the following Lyapunov equation, \cite{amr2009-bagheri-et-al}
\begin{equation}
 \left(\A+\L\C_{\y}\right)^H\Y + \Y\left(\A+\L\C_{\y}\right) + \B_{\d}\,\Wd\,\B^H_{\d} + \L\,\Wn\,\L^H = \mat{0}. \label{EQN:Lyapunov-error}
\end{equation}
The trace of $\Y$ is a measure of how much the mean value of the error $\e(\t)$ differs from zero during its time evolution. One may thus define the following cost function for the design of $\L$
\begin{equation}
 \N =  \Tr{\Y} = \lim_{T\rightarrow\infty}\,\frac{1}{2T}\,\int_{-T}^{T}\, \e^H(t)\,\e(\t)\;dt,
\end{equation}
where $\Tr{\cdot}$ indicates the trace operator. 
\begin{figure}
 \centering
 \includegraphics[width=.45\textwidth]{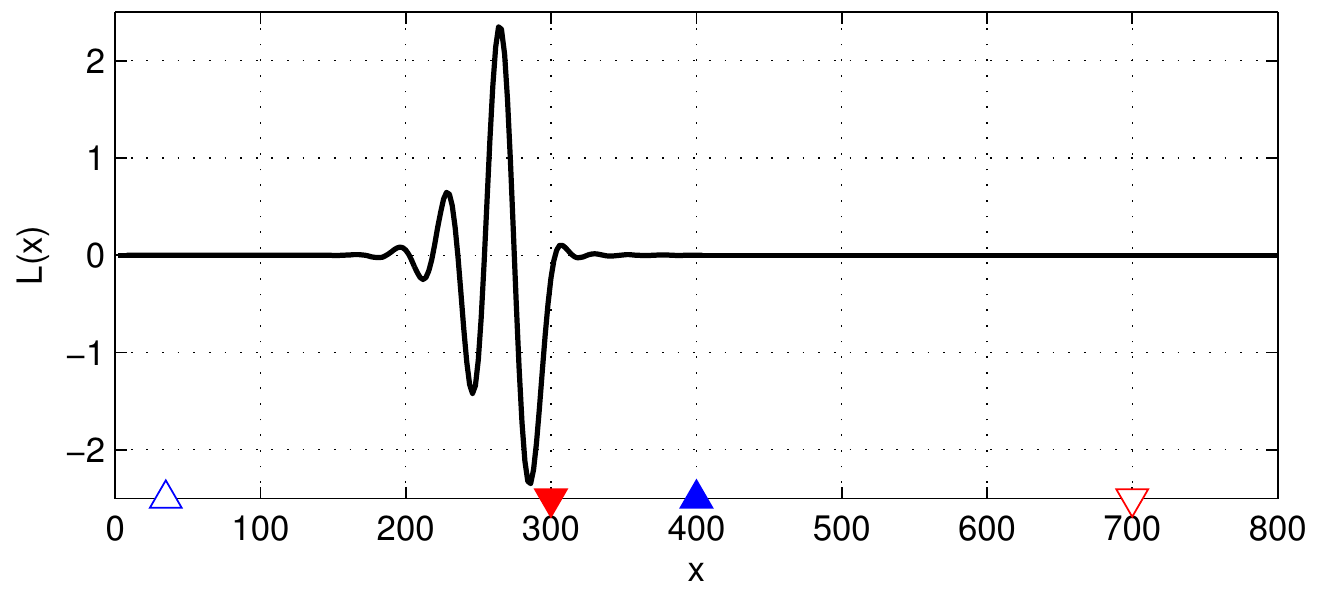}
 \caption{Kalman estimation gain $\L$ computed for $\Wd = 1$ and $\Wn = 0.1$, (see \S\ref{SEC:kalman}). {\small\texttt{[script06.m]}}}
 \label{FIG:L-estimator}
\end{figure}

With a similar approach as in \refsec{control-optimal}, we define a Lagrangian:
\begin{equation}
\begin{split}
 \Na = \mbox{Tr}\left\{\Y + \pmb{\Lambda}\left[\left(\A+\L\C_{\y}\right)^H\Y + \Y\left(\A+\L\C_{\y}\right) +\right.\right. \quad&\\
                                           \left.\left. + \B_{\d}\,\Wd\,\B^H_{\d} + \L\,\Wn\,\L^H\right]\right\}&
\end{split}
\end{equation}
where the Lagrangian multiplier $\pmb{\Lambda}$ enforce the constraint given by \refeqn{Lyapunov-error}. The solution of the minimization is obtained by the imposing the solution to be stationary respect the three parameters $\L$, $\Y$ and $\pmb{\Lambda}$. The zero-gradient condition for $\L$ gives us the expression for the estimation gain,
\begin{equation}
\L=-\Wn^{-1}\C_{\y}\,\Y \label{EQN:estimation-gain}.
\end{equation}
The zero-gradient condition for the Lagrangian multiplier $\pmb{\Lambda}$ returns the Lyapunov equation in \refeqn{Lyapunov-error}: combining this equation with \refeqn{estimation-gain}, a Riccati equation is obtained  for $\Y$:
\begin{equation}
 \A^H\Y + \Y\A - \Y\,\C_{\y}^H\Wn^{-1}\C_{\y}\,\Y + \B_{\d} \Wd \B_{\d}^H = \mat{0}. \label{EQN:estimation-Riccati}
\end{equation}
In \reffig{L-estimator} the estimation gain $\L$ is shown, where it can be observed that the spatial support is localized in the region immediately upstream of the sensor $\y$. In this region the amplitude of the forcing term in the estimator is the largest to suppress estimation error.
\begin{figure}
 \centering
 \includegraphics[width=.45\textwidth]{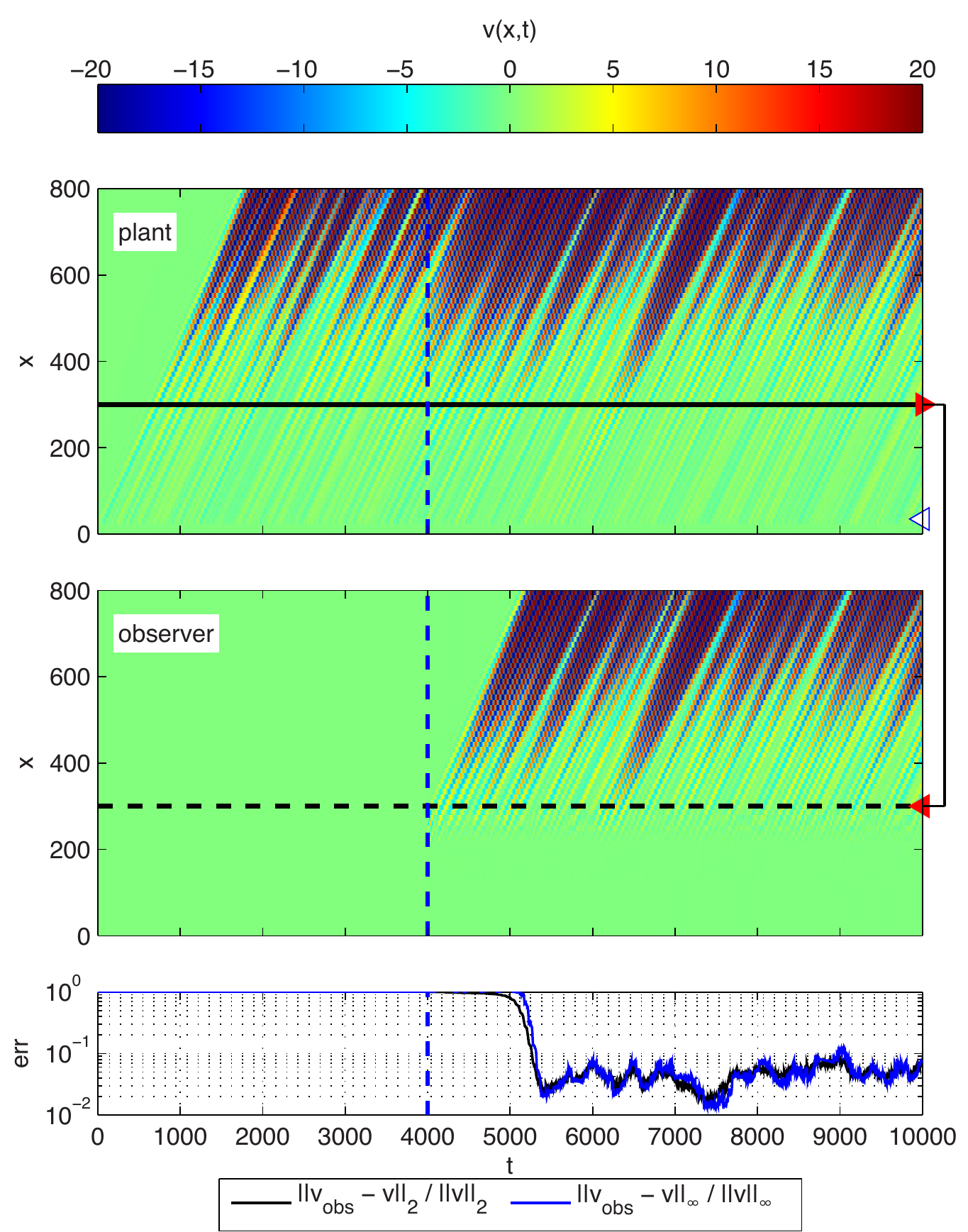}
 \setlength{\unitlength}{\textwidth} \begin{picture}(0,0) \put(-.025, .475){\scriptsize(a)}
                                                          \put(-.025, .140){\scriptsize(b)}
                                                          \put(-.025, .102){\scriptsize(c)}\end{picture} \hspace{-6pt}
 \caption{Spatio-temporal evolution of the response of the system to a disturbance $\d(t)$ (a), compared to the estimated full-order state, using a Kalman filter (b); the contours are shown as a function of the streamwise direction ($x$) and time ($t$). The error-norm between the original state and the estimated state is shown in (c). The vertical blue, dashed line indicates when the estimator is turned on. {\small \texttt{[script06.m]}}}
 \label{FIG:kalman-state}
\end{figure}
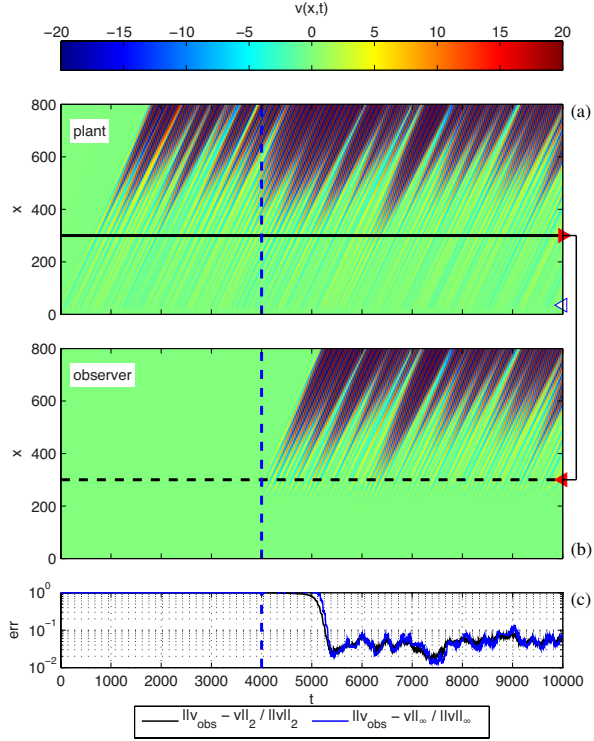
In \reffig{kalman-state} we compare the full state (a) to the estimated state (b) when the system is forced by a noise signal $\d(\t)$. As a result of strong convection, we observe that an estimation is possible only after the disturbance has reached the sensor at $x=300$,  since upstream of this point there are no measurements.  For control design it is important that $\q(\t)$ is well estimated in the region where the actuators are placed; hence, the actuators have to be placed downstream of the sensors \cite{pof2013-belson-semeraro-et-al,jfm2013-julliet-schmid-huerre}.

\subsection{Estimation based on linear filters}
A significant drawback of the Kalman filter, is that it requires a model of the disturbance $\B_{\d}$ for the solution of the Riccati equation \refeqn{estimation-Riccati}.  One may circumvent this issue by using FIR to formulate the estimation problem. In analogue to the formulations based LQR (model based) and on MPC (FIR based), we will compare and link the Kalman filter to a system identification technique called the Least-Square-Mean filter (LMS).  Many other system identification technique exists, the most common being the AutoRegressive-Moving-Average  with eXogenous inputs (ARMAX) employed in the work of \cite{jfm2012-herve-sipp-schmid-samuelides}.

From \refeqns{estimation-space-state}{estimation-z}, we observe that the estimator-input is the measurement $\y(k)$, while the output is given by the estimated values of $\z(k)$. The associated FIR of this system is
\begin{equation}
 \hat{\z}(k) = \sum_{i=N_{i,\,\z\y}}^{N_{f,\,\z\y}} \left(-\C_{\z}\,\hat{\Ad}^{i-1}\;\Delta\t\,\L\right)\;\y(k-i) 
             = \sum_{i=N_{i,\,\z\y}}^{N_{f,\,\z\y}}\es_{\z\y}(i)\;\y(k-i) \label{EQN:fir-hzy}
\end{equation}
where $\hat{\Ad} = e^{(\A+\L\C_y)\Delta\t}$ and $\es_{\z\y}(i)$ denotes the impulse response from the measurement $\y(k)$ to the output $\z(k)$. Note that, since we are considering a convectively unstable system, the sum in \refeqn{fir-hzy} is truncated using appropriate limits $N_{i,\,\z\y}$ and $N_{f,\,\z\y}$, \cite{1995AC-astrom-wittenmark}.  Next, we present a method where $\es_{\z\y}(i)$ is approximated directly from measurements, instead of its construction using the state-space model.

\subsubsection{Least-mean-square (LMS) filter}\label{SEC:estimation-sign-proc-lms}
\begin{figure}
 \centering
 \includegraphics[width=.45\textwidth]{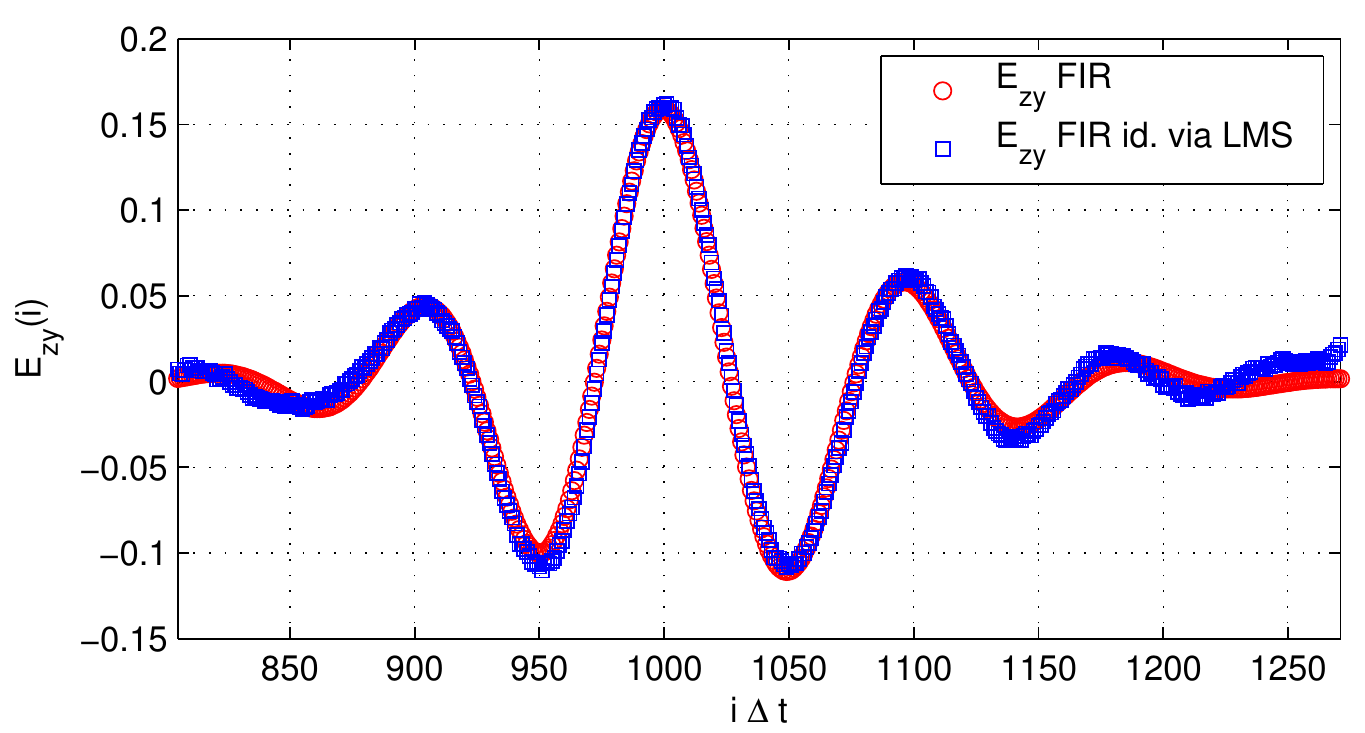}
 \caption{Impulse responses ($\y\rightarrow\z$) of the estimator as a function of the discrete-time. Red circles ({$\color{red}\circ$}) correspond to the FIR time-discrete Kalman-filter-based kernel ${\es}_{\z\y}(i)$ and the blue squares ({\tiny$\color{blue}\square$}) to the one identified by the LMS algorithm. {\small\texttt{[script07.m]}}}
 \label{FIG:hzy}\label{FIG:hzy-lms}
\end{figure}
The main idea is to identify an estimated output $\hat{\z}(k)$ for the system, by minimizing the error
\begin{equation}
 \err(k) = \hat{\z}(k) - \z(k) = \left(\sum_{i=N_{i,\,\z\y}}^{N_{f,\,\z\y}} \es_{\z\y}(i)\;\y(k-i)\right) - z(k),
\end{equation}
where $\z(k)$ is the reference measurement. The unknown of the problem is the time-discrete kernel ${\es}_{\z\y}(i)$. Thus, we aim at adapt the kernel ${\es}_{\z\y}(i)$ such that at each time step the error $\err(k)$ is minimized, i.e.
\begin{equation}
 \min_{\es_{\z\y}} \err^2(k).\label{EQN:lms-err}
\end{equation}
The minimization can be performed using a steepest descent algorithm \cite{1986AFT-haykin}; thus, starting from an initial guess at $k=0$ for  $\hat{\z}(k)$, $\es_{\z\y}$ is updated at each iteration as
\begin{equation}
 \es_{\z\y}(i|k+1) = \es_{\z\y}(i|k) + \step(k)\,\lam(i|k),\label{EQN:lms-ker}
\end{equation}
where $\lam({i|k})$ is the direction of the update and $\step(k)$ is the step-length. Note that each iteration corresponds to one time step. The direction can be obtained from the local gradient, which is given by,
\begin{equation}
 \lam(i|k) = -\dd{}{\err^2(k)}{\es_{\z\y}(i)} = -2\:\err(k)\;\y(k-i).\label{EQN:lms-lam}
\end{equation}
This expression was obtained by forming the gradient of the error $\err(k)$ with respect to $\es_{\z\y}(i)$  and making use of the estimated output $\hat{\z}(k)$ \refeqn{fir-hzy}.

The second variable that needs to be computed in \refeqn{lms-ker} is the  step-length $\step(k)$. Consider the error at time-step $k$ computed with the updated kernel $\es_{\z\y}(i|k+1)$
\begin{equation}
\begin{split}
 \tilde{\err}(k) &= \left(\sum_{i=N_{i,\,\z\y}}^{N_{f,\,\z\y}} \es_{\z\y}(i|k+1)\;\y(k-i)\right) - \z(k) = \\
 				 &= \err(k) + \step(k)\,\left(\sum_{i=N_{i,\,\z\y}}^{N_{f,\,\z\y}} \lam(i|k)\;\y(k-i) \right),
\end{split}\label{EQN:lms-errtilde}
\end{equation}
where \refeqn{lms-err} and \refeqn{lms-ker} have been used. The step-length $\step(k)$ is calculated at each time step in order to fulfil
\begin{equation}
 \min_{\step(k)} \tilde{\err}(k)^2
\end{equation}
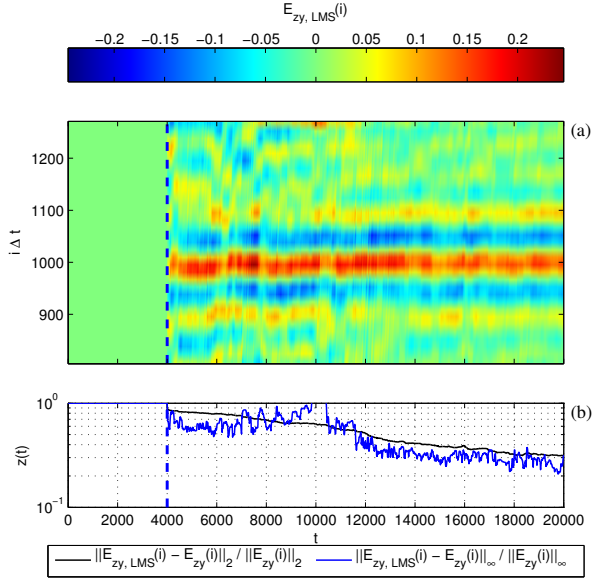
\begin{figure}
 \centering
 \includegraphics[width=.45\textwidth]{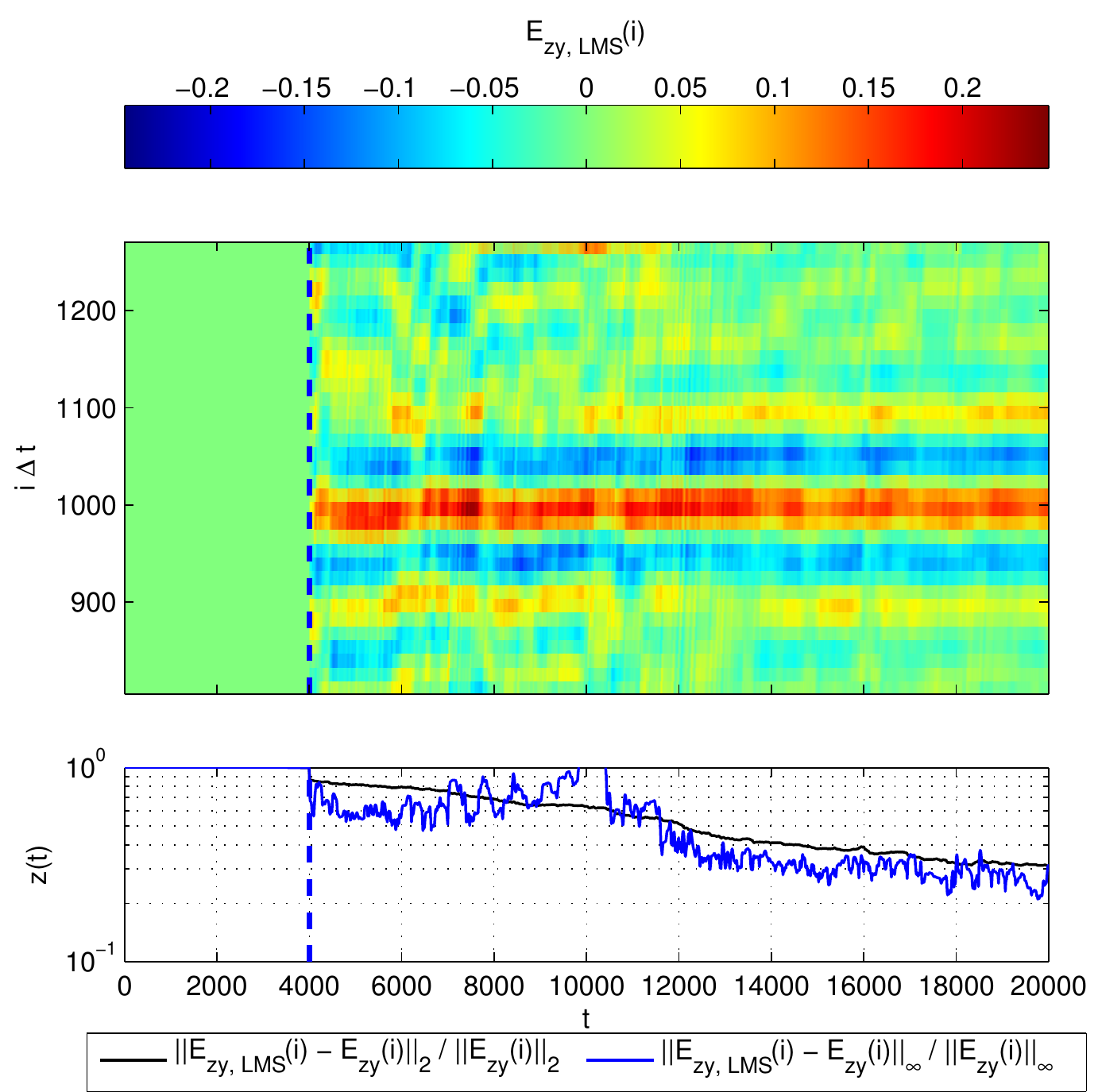}
 \setlength{\unitlength}{\textwidth} \begin{picture}(0,0) \put(-.025, .337){\scriptsize(a)}
                                                          \put(-.025, .122){\scriptsize(b)}\end{picture} \hspace{-6pt}
 \caption{In (a) the evolution of $\es_{\z\y}(i)$ is calculated by an adaptive LMS filter and shown as a function of the discrete-time ($i\Delta t$). The estimation starts at $t=4000$, as indicated by a blue dashed line ({\color{blue}{-~-}}). As the iteration progresses, the error-norm constantly reduces (c).{\small\texttt{[script07.m]}}}
 \label{FIG:hzy-lms-evolution}
\end{figure}
by imposing a zero-derivative condition with respect to $\step(k)$,
\begin{equation}
 \dd{}{\tilde{\err}(k)^2}{\step(k)} = 2\,\tilde{\err}(k)\;\left( \sum_{i=N_{i,\,\z\y}}^{N_{f,\,\z\y}} \lam(i|k)\:\y(k-i) \right)=0.
\end{equation}
Assuming that 
\begin{equation}
\sum_{i=N_{i,\,\z\y}}^{N_{f,\,\z\y}} \lam(i|k)\:\y(k-i) \neq 0
\end{equation}
and considering \refeqn{lms-errtilde}, the optimal step length becomes
\begin{equation}
 \step(k) = -\frac{\err(k)}{\sum_i \lam(i|k)\y(k-i)}.\label{EQN:lms-ste}
\end{equation}

In \reffig{hzy-lms-evolution}(a), the LMS-identified kernel $\es_{\z\y}(i)$ is shown as a function of time $t = k\Delta t$. When the LMS filter is turned on at $t=4000$, the filter starts to compute the kernel, which progressively adapts. While the iteration proceeds, the error decreases as shown in \reffig{hzy-lms-evolution}(b). In the limit of $T\rightarrow\infty$, when a steady solution can be assumed, the kernel computed by the LMS filter converges to the kernel $\es_{\z\y}$ obtained by the Kalman filter (see \reffig{hzy-lms}).  
 
The main drawback of the LMS approach is that the method is susceptible to a numerical stability, \cite{1986AFT-haykin}. A usual way for improving the stability is to bound the the step-length $\step(k)$ by introducing an upper limit. In particular, it can be proven that in order to ensure the convergence of the algorithm, the following condition has to be satisfied
\begin{equation}
 0 < \step(k) < \bar{\step} = \frac{2}{\var_{\y}},
\end{equation}
where the upper-bound $\bar{\step}$ is defined by the variance $\var_{\y}$ of the measurement $\y$, i.e. the input signal to LMS filter.

\section{Compensator}               \label{SEC:compensator}
\begin{figure}
 \centering
 \includegraphics[width=.45\textwidth]{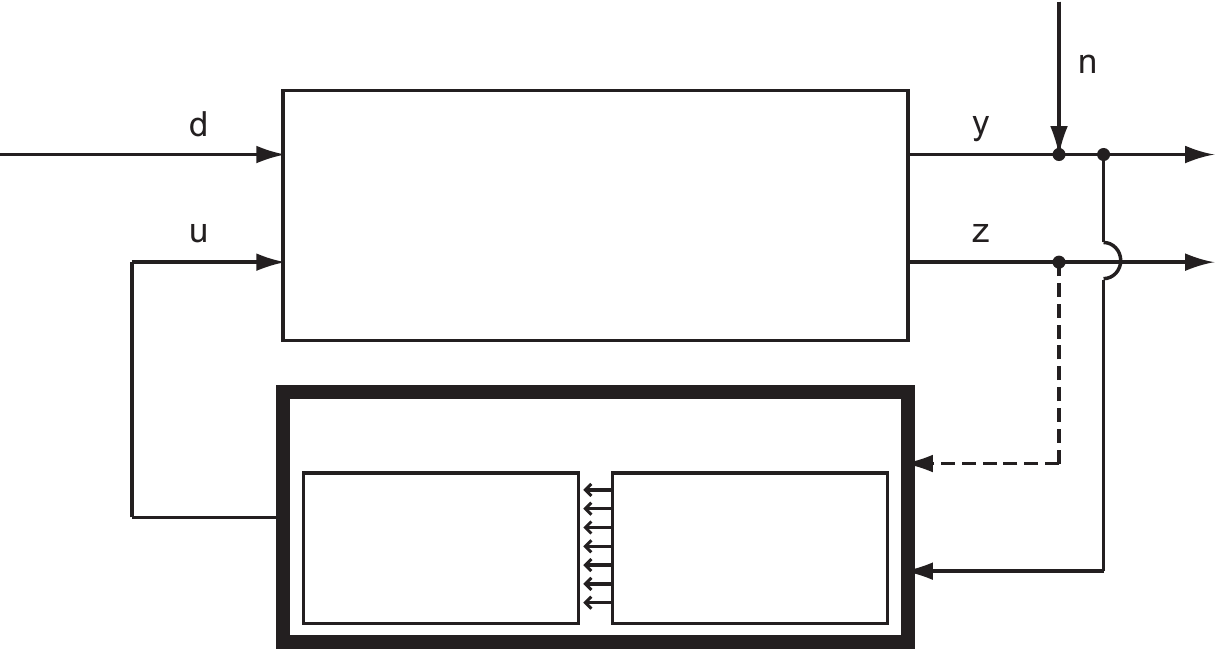}
 \setlength{\unitlength}{\textwidth} \begin{picture}(0,0) \put(-.260, .159){\scriptsize\sffamily system}
                                                          \put(-.328, .032){\scriptsize\sffamily controller}
                                                          \put(-.210, .032){\scriptsize\sffamily estimator}
                                                          \put(-.280, .075){\scriptsize\sffamily\textbf{compensator}}\end{picture} \hspace{-6pt}
 \caption{Block-diagram of the closed-loop system. The compensator, consisting of a controller coupled to an estimator, computes the control signal $\u(t)$ given the measurement $\y(t)$. The minimization of the measurement $\z(t)$ is the target parameter of the controller. Note that in a feedforward controller, the output $\z$ can be used to add robustness to the compensator (for instance, in adaptive filters, \S\ref{SEC:compensator-FXLMS}).}
 \label{FIG:compensator-scheme}
\end{figure}
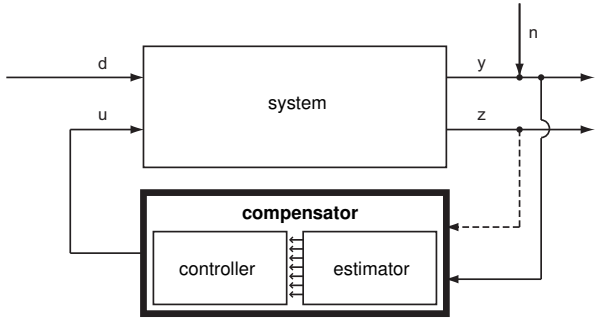

Using the theory developed in \refsec{control} and \refsec{estimation}, we are now ready to tackle the full control problem (\reffig{compensator-scheme}):  given the measurement $\y(\t)$, compute the modulation signal $\u(\t)$ in order to minimize a cost function based on $\z(\t)$. In the first part of this section we will focus on the LQG regulator, that couples a Kalman filter  to a LQR controller. Then we present a compensator based on adaptive algorithms using LMS techniques.

\subsection{Linear-quadratic Gaussian (LQG) regulator}\label{SEC:compensator-lqg}
By solving the control and estimation Riccati equations and the associated gains ($\L$ and $\K$),  we build a system that has as an input the measurement $\y(\t)$ and as an output the control signal $\u(\t)$:
\begin{align}
 \ddt{\hat{\q}}(\t) &= (\A+\B_{\u}\K+\L\C_{\y})\;\hat{\q}(\t) - \L\,\y(\t) \label{EQN:lqg-state} \\
             \u(\t) &= \K\;\hat{\q}(\t) \label{EQN:lqg-control-law}.
\end{align}
This linear system is referred to as the LQG compensator. The estimation and control problem, discussed in the previous sections,  are both optimal and guarantee stability as long as the system is observable and controllable \cite{2000CT-glad-ljung}. In particular, the disturbance $\d$ and the output $\z$ have to be placed respectively in the $\y$-observable and $\u$-controllable region (\reffig{gramians}). Under these conditions, a powerful theorem, known as the separation principle \cite{2000CT-glad-ljung}, states that optimality and stability transfer to the LQG compensator.
 
The closed-loop system obtained by connecting the compensator to the plant becomes
\begin{equation}
 \left[\begin{array}{c}  \dot{\q}(\t)\\ \dot{\hat{\q}}(\t)\end{array}\right] = 
 \left[\begin{array}{cc} \A         & \B_{\u}\K              \\
 						            -\L\C_{\y} & \A+\B_{\u}\K+\L\C_{\y} \end{array}\right] 
 \left[\begin{array}{c}  \q(\t)\\ \hat{\q}(\t)\end{array}\right] + 
 \left[\begin{array}{c}  \B_{\d}\\ \mat{0} \end{array}\right]\;\d(\t).
 \label{EQN:lqg-augmented}
\end{equation}
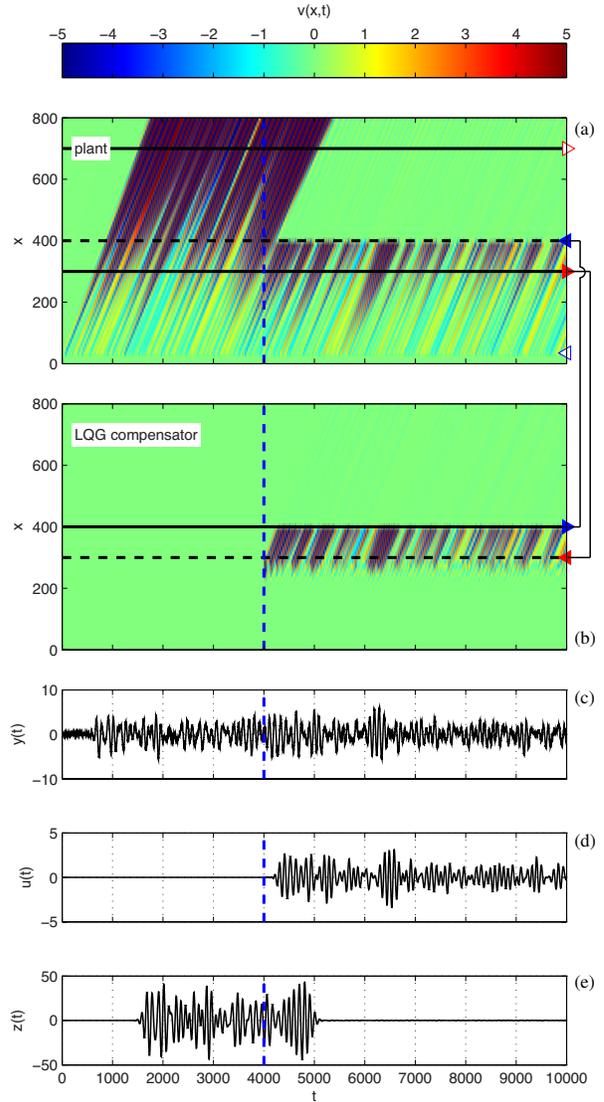
\begin{figure}
 \centering
 \includegraphics[width=.45\textwidth]{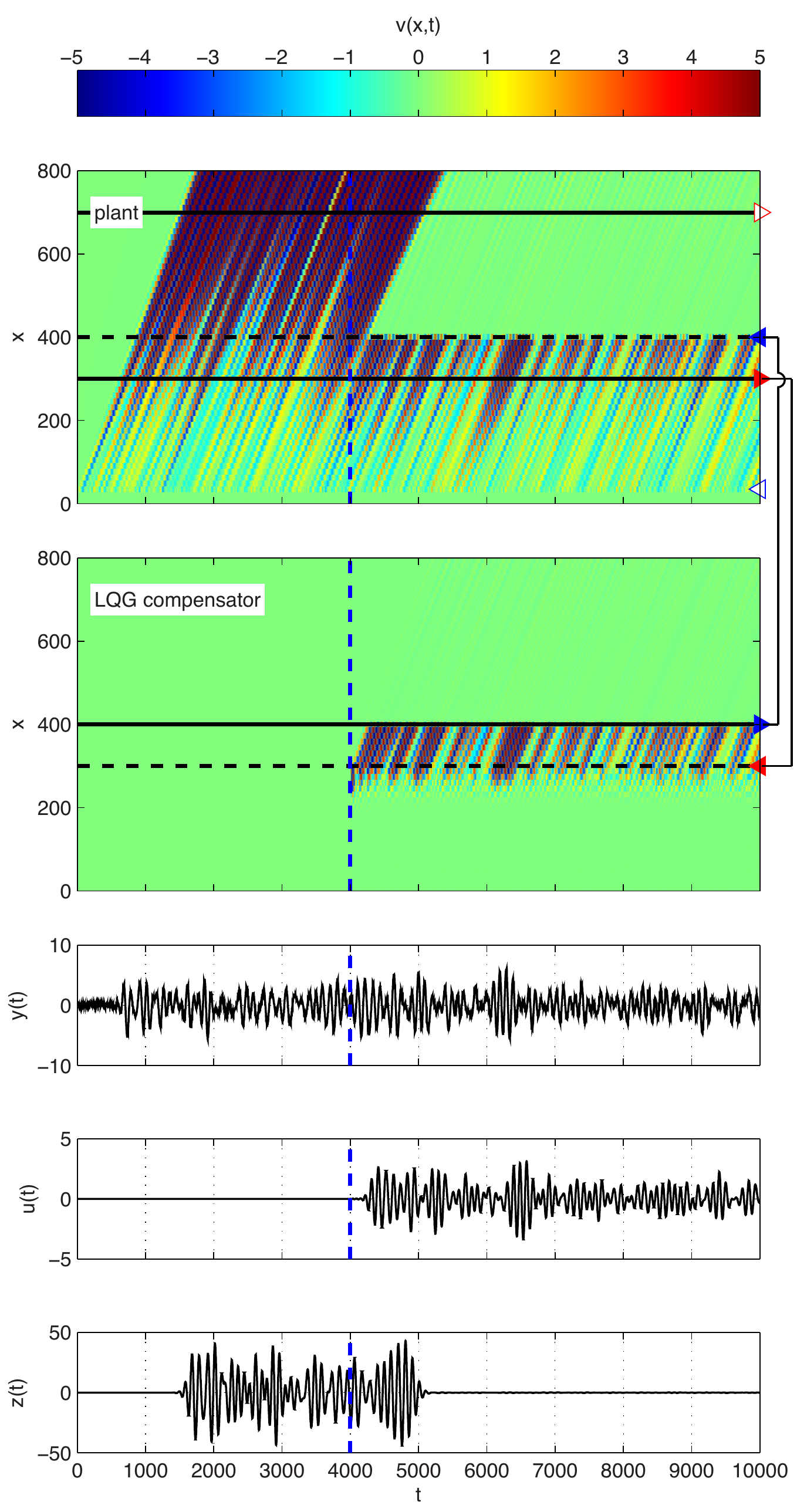}
 \setlength{\unitlength}{\textwidth} \begin{picture}(0,0) \put(-.025, .745){\scriptsize(a)}
                                                          \put(-.025, .355){\scriptsize(b)}
                                                          \put(-.025, .310){\scriptsize(c)}
                                                          \put(-.025, .200){\scriptsize(d)}
                                                          \put(-.025, .092){\scriptsize(e)}\end{picture} \hspace{-6pt}
 \caption{Spatio-temporal response in presence of a white noise input $\d(t)$ for the closed-loop system (a) and the compensator (b); the disturbance is shown as a function of the streamwise direction ($x$) and time ($t$). The measurement $\y(t)$, feeding the compensator, is shown in (c). At $t=4000$ ($\color{blue}-~-$), the compensator starts its action and after a short lag the actuator is fed with the computed control signal $\u(t)$. The perturbation is cancelled, as shown in the contours reported in (a) and the output $\z(t)$ minimized ($t>5000$). {\small\texttt{[script08.m]}}}
 \label{FIG:lqg-state}
\end{figure}
\reffig{lqg-state} shows the response of \refeqn{lqg-augmented} when a white random noise is considered as an input in $\d(t)$. The horizontal solid black line in the top frame depicts the location of $\y$ sensor: this signal is used to force the compensator at the location depicted in the lower frame with a black dashed line. The compensator then provides a signal to the actuator (dashed black line in the upper frame) to cancel the propagating wave-packet.  We let the two systems start to interact at $\t = 4000$, as depicted by the dashed blue line. As soon as the first wave-packet, that is reconstructed by the compensator, reaches the actuation area, the compensator starts to provide a non-zero actuation signal back to the plant. Recall that the state $\hat{\q}(\t)$ of the LQG compensator is an estimation of the state of the real plant $\q(\t)$. This can be seen by comparing \reffig{lqg-state}(a) and \reffig{lqg-state}(b); downstream of the sensor $\y$ the state of the compensator matches the controlled plant. 

Optimal controllers were applied to a large variety of flows, including oscillator flows, such as cavity and cylinder-wake flow, where the dynamic is characterized by self-sustained oscillations at well-defined frequencies, see \cite{sipp2010dynamics}. Note that $\q(\t)$ and $\hat{\q}(\t)$ have the same size: if complex systems are considered, a full-order compensator can be computationally demanding \cite{semeraro2013riccati
}; model reduction and compensator reduction enable to tackle these limitations and design low-order compensators, see \refsec{reduction}.

\subsection{Proportional controller with a time delay}
One may ask how a simple  proportional controller compares to the LQG for our configuration.
In a proportional compensator, the control signal $\u(\t)$ is simply obtained by multiplying the measurement signal $\y(t)$ by a constant $P$. Because of the strong time delays in our system,  one needs to introduce  also a time-delay $\tau$ between the measurement $\y(\t)$ and the control signal $\u(\t)$. The simplest control law for our system is 
\begin{equation}
 \u(t) = P\;y(t-\tau),
\end{equation}
where the ``best'' gain $P$ and the time-delay $\tau$ can be found via a trial-and-error basis (in our case, $\tau=250$ and $P=-0.5432$). This technique is also similar to opposition control \cite{choi94}, where blowing and suction is applied at the wall in opposition to the wall-normal fluid velocity, measured a small distance from the wall. 

\begin{figure}
 \centering
 \includegraphics[width=.45\textwidth]{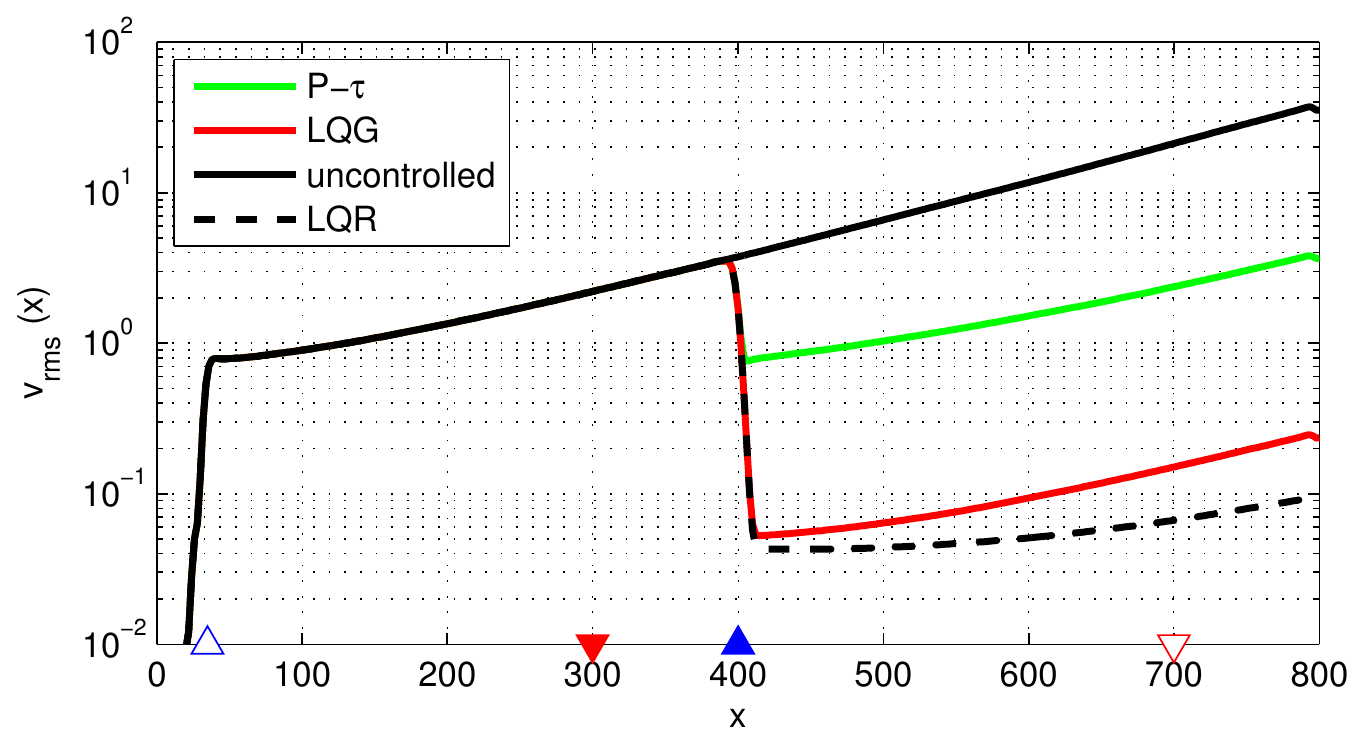}
 \caption{The $\textstyle{rms}$ velocity as a function of the streamwise location $x$ is shown for the uncontrolled case ($\color{black}-$), the LQG ($\color{red}-$), the LQR ($\color{black}{-~-}$) and the opposition controller $P-\tau$ ($\color{green}{-}$). {\small\texttt{[script08.m, script09.m]}}}
 \label{FIG:rms-compare-compensator}
\end{figure}
In \reffig{rms-compare-compensator},  we compare the velocity $\textstyle{rms}$ obtained with LQG compensator (red) and $P$-$\tau$ compensator (green). It can be observed that although both techniques reduce the perturbation amplitude downstream of the actuator position ($\x = 400$),  the performance of the {LQG} regulator is nearly an order of magnitude better than the proportional controller. This can be mainly attributed to the additional degrees of freedom given by the $n_\q\times n_\q$ LQG feedback gains, as opposed to the two-degree freedom $P-\tau$ controller. Indeed, the LQG gains are computed assuming an accurate knowledge of the state-space model. Also shown (dashed-solid line) is the full-information LQR control whose performance is comparable the partial-information LQG controller: the difference between the two is due to the difference between the estimated state $\hat{\q}(\t)$ and the real state $\q(\t)$, i.e. the estimation error $\e$. 

\subsection{Model uncertainties}
The LQG compensator is based on  coupling  an LQR controller and a Luenberger observer. Both of them are based on a model of the system and, as a consequence, their effectiveness is highly dependent on the quality of the model itself. Any difference between the model and the real plant can cause an abrupt reduction of the performances of the compensator \cite{ieee1978-doyle,pof2013-belson-semeraro-et-al}. Model error can be attributed  to, for example, nonlinearities due to the violation of the small perturbation hypothesis, nonlinearities of the actuator  or sensors/actuators shape and positioning. 

The robustness problem can be illustrated using a simple example. Suppose that one wants to cancel a  travelling wave with a localized actuator; what one should do is to generate a wave that is exactly counter-phase with respect to the original one. Suppose that exact location of the actuation action is difficult to model.  Shifting the actuator position slightly is equivalent to adding an error in the estimation of the phase of the original signal. This will in turn cause a mismatch between the wave that is meant to be cancelled and the wave created by the actuator, thus resulting in  an ineffective wave cancellation -- in the worst case, it may result in an amplification of the original wave.

\begin{figure}
 \centering
 \includegraphics[width=.45\textwidth]{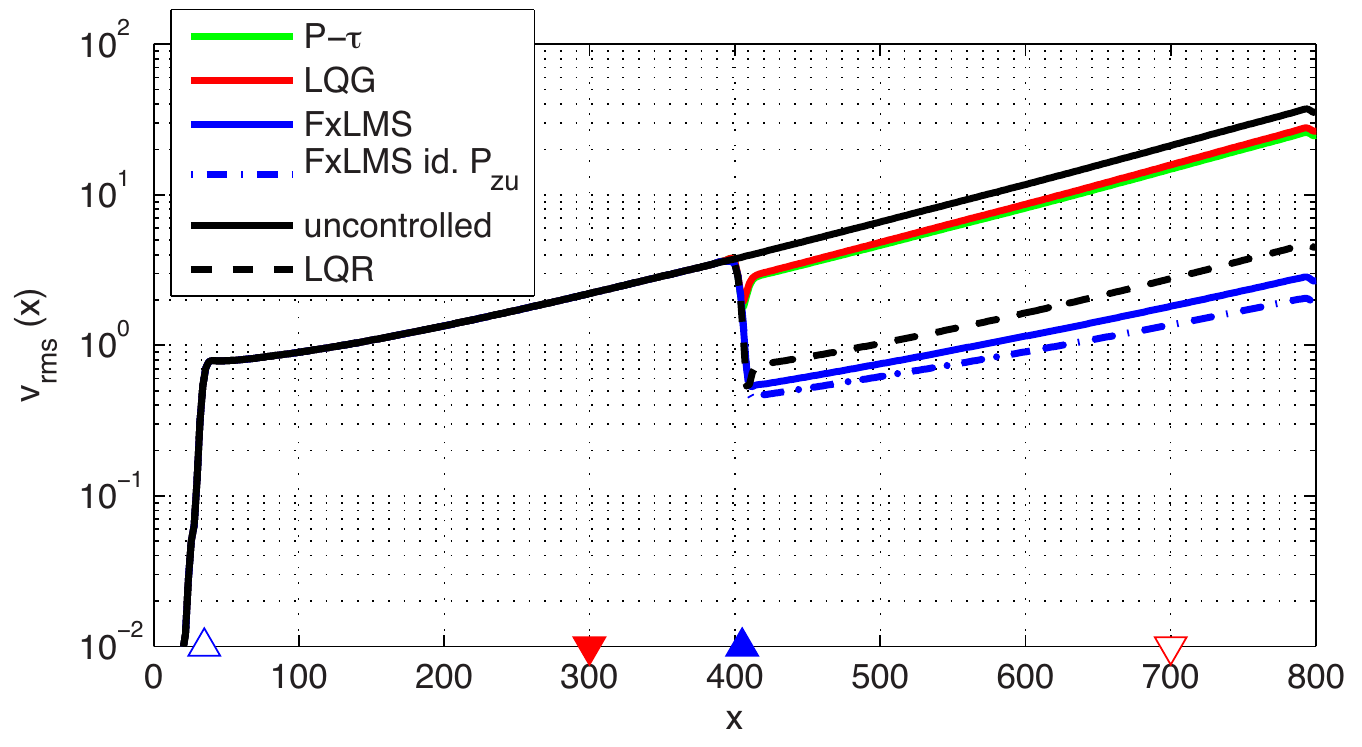}
 \caption{Robustness to uncertainties of the system: the actuator is dispaced of  5 length units from its nominal position.  The performance of the adaptive filter FXLMS ($\color{blue}-~-$ and $\color{blue}\cdot\,-$) are compared to the LQR ($\color{black}-~-$), LQG  ($\color{red}-$) and $P$-$\tau$ ($\color{green}-$) compensators; as a reference, the uncontrolled case is shown ($\color{black}-$). The $\textstyle{rms}$-velocity is shown as a function of the streamwise direction ($x$). The adaptive filter performs reasonably well in presence of un-modelled dynamics; the performances are enhanced by the use of a on-line identified $\pl_{\z\u}$ ($\color{blue}-~-$). The performances of the LQG ($\color{red}-$) and $P$-$\tau$ ($\color{green}-$) compensators are significantly reduced (compare with \reffig{rms-compare-compensator}).{\small\texttt{[script10.m]}}}
 \label{FIG:rms-u-robustness-compensator}
\end{figure}

As shown in \reffig{rms-u-robustness-compensator}, when we  displace the actuator further downstream by $5$ spatial units and apply the compensator designed for the nominal condition to this modified system,  the performance of the LQG regulator deteriorates.  Since, the compensator provides a control signal that is meant to be applied in the nominal position of the actuator the control signal is not able to cancel the upcoming disturbance. 
Essentially, we are suffering from the lack of robustness of the feedforward configuration, since the sensor cannot measure the consequence of the defective actuator signal.  There are different means to address this issue.

One can combine the feedforward configuration with a feedback action, in order to increase robustness. This can be accomplished using the second sensor $\z$ -- downstream of the actuator -- in combination with the estimation sensor $\y$ -- placed upstream of the actuator. The combination of feedback and feedforward is the underlying idea of the MPC controller applied to our configuration \cite{Goldin2013Laminar-flow-co}. However, there are some drawbacks due to the computational costs of the algorithm; indeed, the entries of the dynamic matrix \refeqn{dynamicmat-mpc} are computed during the prediction-step using time integration, whose domain increases with the time-delays of the system. Thus, the integration and the dimensions of the resulting matrices can represent a bottleneck for the on-line optimization. An alternative is the use of an adaptive algorithm, which adapts the compensator response according to the information given by $\z(\t)$, as shown in the next section.

\subsection{Filtered-X least-mean square (FXLMS)} \label{SEC:compensator-FXLMS}
\begin{figure}
 \centering
 \includegraphics[width=.45\textwidth]{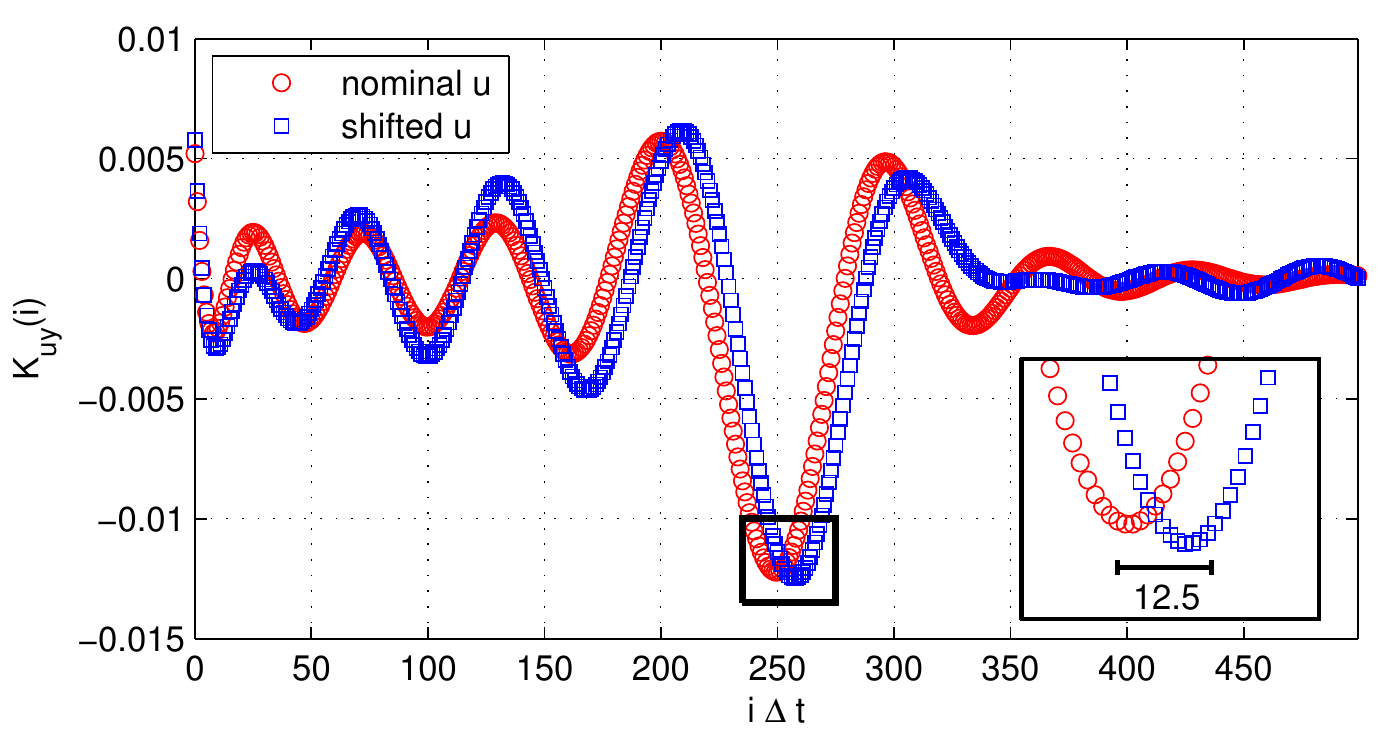}
 \caption{Robustness to uncertainties of the system: FXLMS control gain $\ct_{uy}(i)$ ($\color{blue}\square$) is shifted along the time-discrete coordinate if compared to the static LQG gain ($\color{red}\circ$) to compensate for the un-modelled shift in actuator position. {\small\texttt{[script10.m]}}
 \label{FIG:g-u-robustness-compensator}
 }
\end{figure}

The objective of FXLMS algorithm is to adapt the response of the compensator based on the information given by the downstream output $\z$. The first step of the design is to describe the compensator in a suitable way in order to modify its response. The FXLMS algorithm is based on a FIR description of the compensator. Recall again that the compensator is a linear system (input is the measurement $\y(t)$  and output is the control signal $\u(t)$), which in time-discrete form can be represented by,
\begin{equation}
 \u(k) =\sum_{j=1}^{\infty} \ct_{\u\y}(j)\;\y(k-j) \approx \sum_{j=1}^{N_{uy}} \ct_{\u\y}(j)\;\y(k-j), \label{EQN:compensator-IIR}
\end{equation}
where $ \ct_{\u\y}(j)$ is a time-discrete  kernel. Due to the stability of the system, we have  $ \ct_{\u\y}(j)\rightarrow 0$ as $t\rightarrow\infty$,  so that the sum can be truncated after $N_{\u\y}$ steps. In the case of LQG compensator $ \ct_{\u\y}$ has the form
\begin{equation*}
 \ct_{\u\y}(j) \triangleq \K\,\exp\left [\left(\A+\L\C_{\y}+\B_{\u}\K\right)\,\Delta\t\,(j-1)\right ]\,\L 
 \end{equation*}
for $i=1,2,\dots$ The kernel $\ct_{\u\y}(j)$ of the LQG controller is shown with red circles in \reffig{g-u-robustness-compensator}. In this case $N_{\u\y}=533$, which gives $\left|\ct_{\u\y}(j)\right|<10^{-2}$ for $j>N_{\u\y}$.

The FXLMS technique modifies on-line the  kernel $\ct_{\u\y}(j)$ in order to minimize the square of measurement $\z(\t)$ at each time step, \cite{ijhff2003-sturzebecher-nitsche}, i.e
\begin{equation}
 \min_{\ct_{\u\y}(j)} \z^2(k).
\end{equation}

The procedure is closely connected to the LMS filter discussed in \refsec{estimation-sign-proc-lms} for the estimation problem. The  kernel $\ct_{\u\y}(j)$ is updated at each time step by a steepest-descend method:
\begin{equation}
 \ct_{\u\y}(j|k+1) = \ct_{\u\y}(j|k) + \step(k)\,\lam(j|k)
\end{equation}
where $\step(k)$ is calculated from \refeqn{lms-ste} and $\lam(j|k)$ is the gradient of the cost function $\z(k)$ with respect of the control gains $\ct_{\u\y}(j)$. In order to obtain the update direction, consider the time-discrete convolution for $\z(k)$,
\begin{equation*}
\begin{split}
 \z(k) &= \sum_{i=0}^{\infty} \pl_{\z\d}(i)\;\d(k-i) + \sum_{i=0}^{\infty} \pl_{\z\u}(i)\;\u(k-i) = \\
       &= \sum_{i=0}^{\infty} \pl_{\z\d}(i)\;\d(k-i) + \sum_{i=0}^{\infty} \pl_{\z\u}(i)\;\sum_{j=0}^{N_{\u\y}} \ct_{\u\y}(j)\;\y(k-i-j) = \\
       &= \sum_{i=0}^{\infty} \pl_{\z\d}(i)\;\d(k-i) + \sum_{j=0}^{N_{\u\y}} \ct_{\u\y}(j)\;\sum_{i=0}^{\infty} \pl_{\z\u}(i)\;\y(k-j-i).
\end{split}
\end{equation*}
From this expression it is possible to obtain the gradient
\begin{equation}
 \lam(j|k) = -\dd{}{\z(k)^2}{\ct_{\u\y}(j)} = -2\:\z(k)\:\sum_{i=0}^{\infty} \pl_{\z\u}(i)\:\y(k-j-i), \label{EQN:FXLMS-lambda-iir}
\end{equation}
which can be  simplified by introducing the filtered signal $\y_f(k)$,
\begin{equation}
 \y_f(k) = \sum_{i=0}^{\infty} \pl_{\z\u}(i)\;\y(k-j-i) \approx \sum_{i=N_{i,\,\z\u}}^{N_{f,\,\z\u}} \pl_{\z\u}(i)\;\y(k-i)
\end{equation}
Note that a FIR approximation of $\pl_{\z\u}(i)$ has been used. Hence, the expression in \refeqn{FXLMS-lambda-iir} becomes,
\begin{equation}
 \lam(j|k) = -2\z(k)\;\y_f(k-j).
\end{equation}
In order to get the descend direction, the measurement $\y(t)$ is filtered by the plant transfer function $\pl_{\z\u}(i)$. 

Starting the on-line optimization from the compensator kernel $\ct_{\u\y}(j)$ given by the LQG solution, the algorithm is tested on our problem. In \reffig{rms-u-robustness-compensator} we observe that the algorithm is able to recover some of the lost performance of LQG (due to shift in actuator position) and it is comparable to the full-information control performed by the LQR controller with the nominal gain $\K$. This is possible because of the adaptation of the kernel $\ct_{uy}(j)$, to the new actuator location.
\reffig{g-u-robustness-compensator} shows how the convolution kernel has been modified by the algorithm; the  kernel is shifted in time in order to restore the correct phase shift between the control signal $\u(\t)$ and the measurement signal $\y(\t)$ in the modified system. The shift in time between the two peaks (visible in the inset figure) is exactly the time that it takes for the wave-packet to cover the additional  distance between the sensor and the actuator. Recalling  from \refsec{system}, that the wave-packet travels with a speed $\kubase=0.4$, it will take $\Delta\x_{\u}/\kubase = 5/{0.4} = 12.5$ time units to cover the extra space between $\u$ and $\y$.

From \refeqn{FXLMS-lambda-iir}, it can be noted that the FXLMS is not completely independent from a model of the system; in fact the convolution kernel $\pl_{\z\u}(i)$ is needed to compute the gradient $\lam(j|k)$ used by the algorithm. In the previous example, the nominal transfer function has been used, given by the model of the plant
\begin{equation}
 \pl_{\z\u}(i) = \C_{\z}\,e^{\A\:\Delta\t(i-1)}\,\B_{\u},\quad i = 1,2,...
 \label{EQN:xlms:p}
\end{equation}
One may obtained a kernel $\pl_{\z\u}(i)$ that is totally independent by the model -- thus without any assumption on placement/shape of both actuator and sensors  -- by using the LMS identification algorithm derived in \refsec{estimation-sign-proc-lms}. In \reffig{rms-u-robustness-compensator}, we compare $\pl_{\z\u}(i)$ obtained from \refeqn{xlms:p} using inaccurate state-space model (since actuator position has shifted) (solid blue) with $\pl_{\z\u}(i)$ obtained by model-free identification using LMS technique (dashed blue). We observe that when combining adaptiveness with a more accurate model-free identification of $\pl_{\z\u}(i)$, the performance is improved significantly.

Note  that this algorithm when applied to flows dominated by convection, and thus characterized by strong time-delays, results in a feedforward controller where the feedback information is recovered by the processing of the measurements in $\z$.  This method is known to as \emph{active noise cancellation} \cite{ijhff2003-sturzebecher-nitsche,phtr2012-erdmann-et-al}.  We can identify two time scales: a fast time-scale related to the estimation process and a slow time-scale related to the adaptive procedure \cite{gad2007flow}. For this reason, this method is suitable for static or slowly varying model discrepancies.

\section{Discussion}                \label{SEC:reduction}\label{SEC:discussion}
In this section, we discuss a few aspects that have not been addressed so far, but are important to apply the presented techniques to an actual flowing fluid. Many other important subjects such as choice of actuator and sensors, nonlinearities and receptivity  are not covered by this discussion.

\paragraph{Low-order control design.} The discretization of the Navier-Stokes system leads to high-dimensional systems that easily exceed  $10^5$  degrees of freedom. For instance, the full-order solution of Riccati equations for optimal control and Kalman filter problems cannot be obtained using standard algorithms  \cite{ieeecs2004-benn}.
\begin{figure}
 \centering
 \includegraphics[width=.45\textwidth]{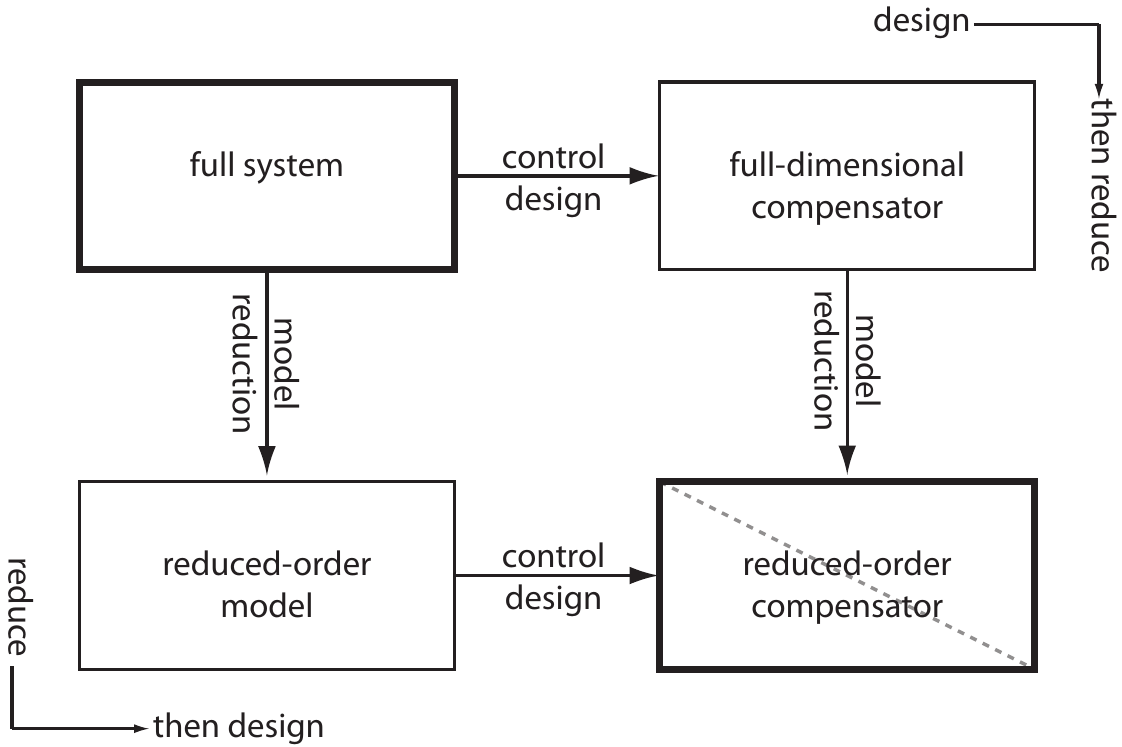}
 \caption{Two strategies are possible to compute a reduced-order compensator, reduce-then-design an design-then-reduce. In general, the two paths do not lead at the same results.}
 \label{FIG:reduction-scheme}
\end{figure}
One common strategy is to replace the high-dimensional system  with a low-order system able to reproduce the essential input-output dynamics of the original plant. This approach is referred to as {reduce-then-design} \cite{anderson:liu:89} (left part of \reffig{reduction-scheme}). First, a reduced-order model is identified using an appropriate model reduction or {system identification} technique;  then  the validated reduced-order model is used to design a low-order compensator. The dual approach is
called {design-then-reduce} or {compensator reduction} (right part of \reffig{reduction-scheme}). In this case, a high-order compensator is designed as first step (if possible). The second step is the reduction of the compensator to a low-order approximation.

Both the approaches lead to a low-order compensator that can be used to control the full-order plant, but they are not necessarily  equivalent \cite{anderson:liu:89}. I the reduce-then-design approach, we neglect a number of states during the model-order reduction of the open loop, that might become important for the dynamics of the closed-loop system. Despite these limitations, the reduce-then-design approach is the most common in flow control due to its computational advantages; indeed, the challenge of designing a high-dimensional compensator to be reduced strongly limits this alternative. 

\paragraph{Model reduction.} Following the reduce-then-design approach, the first step consists of identifying a reduced-order model, typically reproducing the I/O behaviour of the system. We can distinguish two classes of algorithms. The first category is based on a Petrov-Galerkin projection of the full-order system. In this case, the I/O  behaviour of the system is reconstructed starting from a low-order approximation of the state-vector $\q_r$, characterized by a number of degree of freedom $r\ll n$; the projection can be performed on global modes \cite{aakervik:hoepffner:ehrenstien:henning:07}, proper orthogonal modes (POD), obtained from the diagonalization of the controllability Gramian (see \refsec{gramians}), or balanced modes, for which the controllability and observability Gramians are equal and diagonal \cite{moore:81,rowley2005model,amr2009-bagheri-et-al}. This strategy has been widely used in the flow-control community in the past years for the identification of linear \cite{aakervik:hoepffner:ehrenstien:henning:07,ilak:rowley:08,bagheri2009input, barba:sipp:schmid:09,jfm2010-semeraro-et-al} and nonlinear models \cite{jfm2003-noack-afanasief-et-al,ref:siegel,Ilak2010Model-Reduction}. In particular, when  nonlinear effects are considered, it is necessary to take into account the effect that a finite disturbance in the flow has on the base-flow, as shown by \cite{jfm2003-noack-afanasief-et-al} for a cylinder wake flow. At low Reynolds numbers, a small number of modes are sufficient to reproduce the behaviour of oscillators such as the cylinder wake, while a larger number of modes is required to reproduce the I/O behaviour of convective unstable flows. This is mainly due to the presence of strong time-delays, \cite{2000CT-glad-ljung}, that characterize this type of systems, \refsec{io}.

The second approach stems from the I/O analysis of the formal solution carried out in \refsec{io}; we note that a low-order representation of the transfer function is enough to reconstruct the I/O behaviour of the system. The computation of this representation can be performed applying \emph{system identification} algorithms \cite{ljung1999system}. Once the transfer functions are identified, one constructs a reduced-order model in canonical form. These techniques were widely used for experimental investigations (see e.g. \cite{2007:jfm:lundell:control,2003:jfm:Rathnasingham:breuer}) and have been recently applied also in numerical studies \cite{2007:huang:kim:pof, jfm2012-herve-sipp-schmid-samuelides}. Indeed, for linear systems, it can be shown that projection-based techniques and system identification techniques can provide equivalent reduced-order models \cite{ma2011reduced}. We refer the reader to the reviews by  \cite{phtr2011-bagheri-henningson} and \cite{sipp2013closed} for a broader overview. 

\paragraph{Control of three-dimensional disturbances.}
\begin{figure}
 \centering
 \includegraphics[width=.45\textwidth]{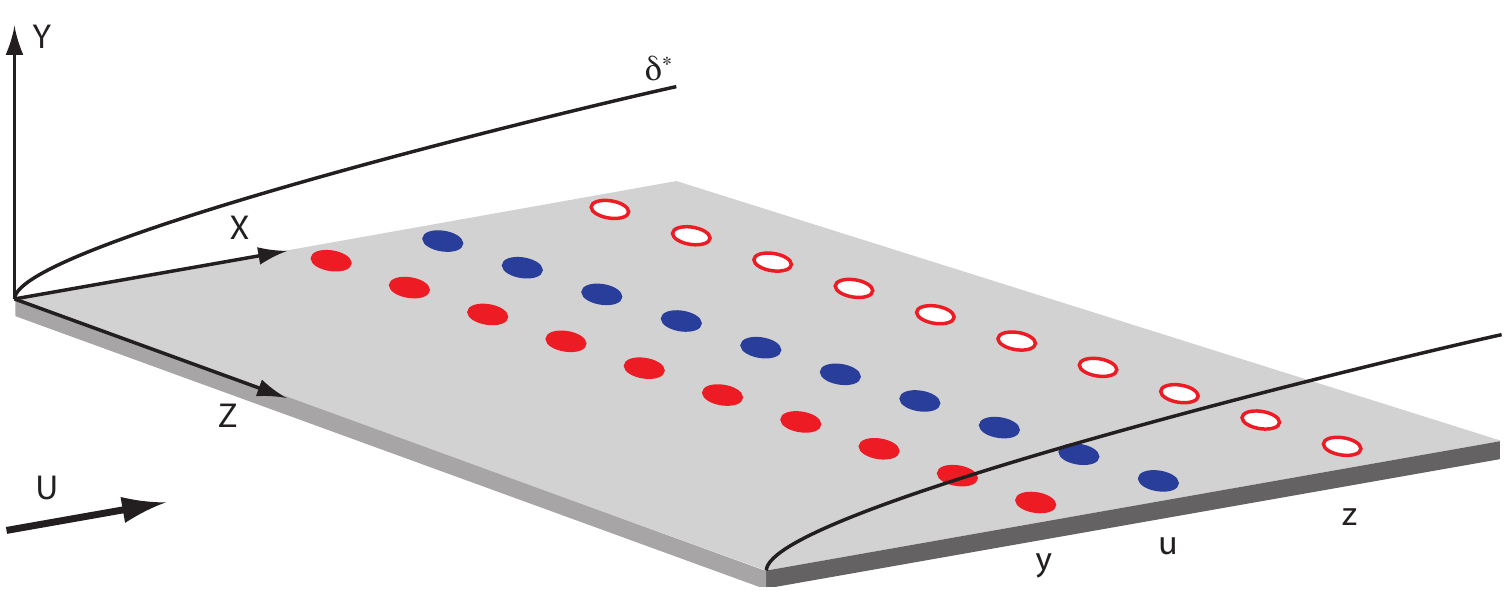}
 \setlength{\unitlength}{\textwidth} \begin{picture}(0,0) \put(-.443, .024){\tiny$\infty$}\end{picture} \hspace{-6pt}
 \caption{Control configuration for a three dimensional (3D) flow developing over a flat plate. A possible configuration consists of localized sensors and actuators placed along the spanwise direction.}
 \label{FIG:3D-setup}
\end{figure}
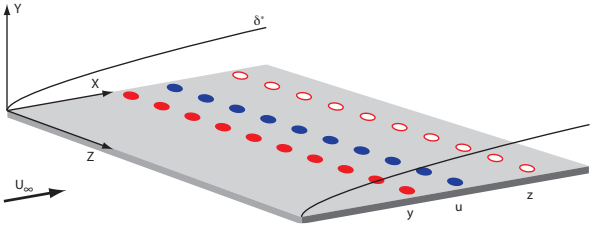
A sketch of the three-dimensional control setup of the flow over a flat-plate is shown \reffig{3D-setup}.  Compared to the 2D boundary-layer flow a single actuator $\u$, sensor $\y$ and output $z$ are now replaced by arrays of elements localized along the span-wise direction, resulting in a multi-input multi-output (MIMO) system. The localization (size and distance between elements) of sensors and actuators may significantly influence efficiency of the compensator \cite{jfm2010-semeraro-et-al} and \cite{semeraro2013transition}.  An important question one must address for MIMO systems  is how to connect inputs to outputs. A first approach consists of coupling one actuator with only one sensor (for instance, the one upstream); in this case, the number of single-input single-output (SISO) \emph{control units} equals the number of sensor/actuator pairs. This approach is called \emph{decentralized} control-design; despite its simplicity in practical implementations, the stability in closed loop is not guaranteed \cite{2000CT-glad-ljung}. The dual approach where only one control-unit is designed and all the sensors are coupled to all the available actuators is called \emph{centralized} control.  In \cite{jfm2010-semeraro-et-al}, the centralized-controller strategy was found necessary for the design of a stable {TS}-wave controller. The main drawback of a fully centralized-control approach is that the number of connections for a flat plate of large span quickly becomes impractical due to all the wiring. One may then introduce a \emph{semi-decentralized} controller \cite{dadfar2013centralised}, where small MIMO control-units are designed and connected to each other; in \cite{dadfar2013centralised}, it is shown that a number of control-units can efficiently replace a full centralized control with a limited lost of performance.

Another important aspect that has be accounted for in a MIMO setting, is the choice of the objective function $\z$. The minimization of a set of signals obtained from localized outputs with compact support does not necessarily correspond to a reduction of the actual perturbation amplitude in a global sense. For 1D and 2D flow systems any measurement taken locally, close to the solid wall and downstream in the computational domain, is sufficient for obtaining consistency between the perturbation and signal minimization \cite{bagheri2009input}; this is not the case for 3D systems. An optimal way for choosing the output $\C_{\z}$ is the \emph{output projection} suggested by \cite{rowley2005model}, where a projection on a POD basis is performed. The resulting signal $\z(t)$ corresponds to the amplitude coefficients of the {POD} modes, i.e.~the temporal behaviour of the most energetic coherent structure of the flow. This method can also provide useful guidelines for the location of output sensors.

\section{Summary and conclusions}               \label{SEC:conclusions}
This work provides a comprehensive review on standard model-based techniques (LQR, Kalman filter, LQG, MPC) and model-free techniques (LMS, X-filtered LMS) for the delay of the transition from laminar to turbulence. We have focussed on the control of perturbation evolving in convective flows, using the linearized Kuramoto-Sivashinsky equation as a model of the flow over the flat-plate to characterize and compare these techniques. Indeed, this model provides the two important traits of convectively unstable fluid systems, namely, the amplifying behaviour of a stable system and a very large time delay.

Much research have been performed on flow control using the very elegant techniques based on LQR and LQG, \cite{amr2009-bagheri-et-al,semeraro2013transition,jfm2013-julliet-schmid-huerre}. Although, these techniques may lead to the best possible performance and they have stability guarantees (under certain restrictions), their  implementation in experimental flow control settings raises a number obstacles: (1) The choice of actuator and sensor placement that yields a good performance of convectively unstable systems results in a feedforward system. We have highlighted the robustness issues arising from this configuration when using standard LQG-based techniques. (2) Disturbances, such as free-stream turbulence, and  actuators, such as plasma actuators, can be difficult to model under realistic conditions. (3) The requirement of solving two Riccati equations is a major computational hassle, although it has successfully been addressed by the community using model-order reduction techniques \cite{bagheri2009input} or iterative methods \cite{semeraro2013riccati}.

Model-free techniques based on classical system-identification methods or adaptive-noise-cancellation techniques can cope with the limitations of model-based methods, \cite{ijhff2003-sturzebecher-nitsche}. For example, we have presented algorithms that improve robustness by adapting to varying and un-modelled conditions. However, model-free techniques have their own limitations; (i) one may often encounter instabilities, which in contrast to LQR/LQG, cannot always be addressed in a straight-forward manner by using concepts such as controllability and observability. (ii) The number of free parameters (such as the limits of the sums appearing in  FIR filters) that need to be modelled are many and chosen in a somewhat {\it ad-hoc} manner.

The conclusion is that there does not exist one single method that is able to deal with all issues, and the final choice depends on the particular conditions that must be addressed. While a model-based technique may provide optimality and physical insight, it may lack the robustness to uncertainties that adaptive methods are able to provide. We believe that future research will head towards hybrid methods, where controllers  are partially  designed using numerical simulations and partially using adaptive experiment-based techniques.

\begin{acknowledgment}
 The authors acknowledge support the Swedish Research Council (VR-2012-4246, VR-2010-3910) and the Linn\`e  Flow Centre.
\end{acknowledgment}

\appendix

\section{Numerical method} \label{SEC:matlab}
Finite-difference (FD) schemes are used to approximate the spatial derivatives in \refeqn{LKS}. In particular, a centered scheme based on stencils of five-nodes are used for the second-order and fourth-order derivatives while a one-node-backward scheme is used for the first-order derivative. The latter is required due to the convective nature of the system: a de-centered scheme reduces the spurious, numerical oscillation of the approximated solution\cite{2009NMDP-quarteroni}.

The grid is equispaced $\x_i = i\,\frac{L}{n_\q}$, with $i = 1,2,...,n_\q$. Once the FD scheme is introduced, the time evolution at each of the internal node is solution of the ODE equation
\begin{equation}
\begin{split}
  \frac{d\kuper_i(\t)}{d\t} = - \kubase        \,&\sum_{j=-3}^{1}d^{b}_{1,j}\;\kuper_{i+j}(\t)
                              - \frac{\kp}{\kr}\, \sum_{l=-2}^{2}d^{c}_{2,l}\;\kuper_{i+l}(\t)\;+\\
                              - \frac{1}{\kr}  \,&\sum_{l=-2}^{2}d^{c}_{4,l}\;\kuper_{i+l}(\t)
                              + \kb_\d(\x_i)\;\d(\t) + \kb_\u(\x_i)\;\u(\t), \label{EQN:A-detailed}
\end{split}
\end{equation}
where $\kuper_i(\t) = \kuper(\x_i,\t)$ for $i = 1,2,...,n_{\q}$. The outflow boundary conditions in \refeqn{bc-L} on the right boundary of the domain lead to the linear system of equations,
\begin{align}
  \left.\dd{}{\kuper}{\x}\right|_{\x=L} = 0  \quad&\Rightarrow\quad \sum_{j=-3}^{1}d^{b}_{1,j}\;\kuper_{n_\q+j}(\t) = 0 \\
  \left.\dd{3}{\kuper}{\x}\right|_{\x=L} = 0 \quad&\Rightarrow\quad \sum_{j=-2}^{2}d^{c}_{3,j}\;\kuper_{n_\q+j}(\t) = 0
\end{align}
The solution of this system allows us to express the boundary nodes $i=n_\q+1,n_\q+2$ as a linear combination of the inner nodes.
Similarly, the left boundary condition in \refeqn{bc-0} leads to an expression for the nodes $i=0,-1$:
\begin{align}
  \left.\kuper\right|_{x=0} = 0            \quad&\Rightarrow\quad \kuper_0(\t) = 0 \\
  \left.\dd{}{\kuper}{\x}\right|_{x=0} = 0 \quad&\Rightarrow\quad \sum_{j=-1}^{3}d^{f}_{1,j}\;\kuper_{0+j}(\t) = 0
\end{align}
where a forward {FD} scheme is used for the first-order derivative approximation. Equation \refeqn{A-detailed} together with the boundary conditions can be rewritten in compact form as
\begin{equation*}
  \ddt{\q}(t) = \A    \,\q(t) + \B_{\d} \,\d(t) + \B_{\u} \,\u(t)
\end{equation*}
where $\B_{\d} = \{\kb_\d(\x_i)\}$, $\B_{\u} = \{\kb_\u(\x_i)\}$ and the matrix $\A\in\Real^{n_\q \times n_\q}$ is a banded matrix (see also \refeqn{space-state-detailed}).

The Crank-Nicolson method is used to march the system forward in time \refeqn{space-state-detailed}. Given a time step $\Delta\t$, the value of the state $\q(\t+\Delta\t)$ is given by the expression:
\begin{equation}
  \q(t+\Delta\t) = \mat{CN}_I^{-1} \left[\mat{CN}_E\,\q(t) + \Delta\t\,\left(\B_\d\,\d(\t) + \B_\u\,\u(\t)\right)\right]
\end{equation}
where $\mat{CN}_I = \mat{I} - \frac{\Delta\t}{2}\A$ and $\mat{CN}_E = \mat{I} + \frac{\Delta\t}{2}\A$. This is an implicit method, i.e. requires the solution of the linear system $\mat{CN}_I^{-1}$, and this operation can be numerically expensive.

\section{Numerical code}
A downloadable package of the MATLAB\copyright~routines used to produce the results presented in this paper can be found at \small{\url{http://www.mech.kth.se/~nicolo/ks/}}. The 11 scripts listed below cover all the methods that are presented in this work.

\paragraph{\texttt{script00.m}: Time evolution of a spatially localized initial condition.} The time response of the plant to a Gaussian-shaped initial condition is calculated: the generated wave-packet travels downstream while growing and is detected by the outputs $y$ and $z$. The spatio-temporal time evolution of $v(x,t)$ is plotted together with the output signals.

\paragraph{\texttt{script01.m}: Response to a white Gaussian disturbance $d(t)$.} A white noise signal is considered as input $d(t)$ and the time-response of the plant is calculated. The statistics of the velocity are computed and visualized for comparison with the controlled cases.

\paragraph{\texttt{script02.m}: External description.}  An alternative description of the system, based on the Input/Output behaviour of the system is calculated. In particular, the response of the system is calculated via a {FIR} filter and compared with the {LTI} system description, i.e. internal description.

\paragraph{\texttt{script03.m}: Controllability and observability Gramians.} The controllability and observability Gramians are computed solving the Lyapunov equations in \refeqns{obser-lyap}{contr-lyap}.

\paragraph{\texttt{script04.m}: Linear-Quadratic Regulator.} A {LQR} controller is applied to the plant and tested when the system is excited by a white Gaussian noise $d(t)$. The statistics of the velocity are computed and visualized in order to be compared to the other controlled cases.

\paragraph{\texttt{script05.m}: Model Predictive Control.} Constrained {MPC} is used in presence of saturation of the actuator. The system is excited by a white Gaussian noise $d(t)$. The statistics of the velocity are computed and visualized in order to be compared with the other controlled cases. 

\paragraph{\texttt{script06.m}: Kalman filter.} A Kalman filter is designed for the plant and used to estimate the system state when excited by a white Gaussian noise $d(t)$.

\paragraph{\texttt{script07.m}: Least-Mean Square filter} A {LMS} filter is used to identify the FIR-kernel $E_{zy}$. The resulting kernel is compared with the Kalman filter solution.

\paragraph{\texttt{script08.m}: Linear-Quadratic Gaussian compensator} A {LQG} compensator is designed coupling a {LQR} controller and a Kalman filter. The compensator is tested when the system is excited by a white Gaussian noise $d(t)$.

\paragraph{\texttt{script09.m}: $P-\tau$ compensator.} A simple opposition control is designed using explicitly the time-delay. The system is excited by a white Gaussian noise $d(t)$. The control gain has been obtained by a trial and error procedure.

\paragraph{\texttt{script10.m}: Filtered-X Least-Mean Square algorithm} {FXLMS} algorithm is implemented. The initial condition is provided by the impulse response of the corresponding {LQG} compensator; a robustness test is carried by displacing the actuator location.  

\bigskip
\noindent Following functions are required by the above scripts:

\paragraph{\texttt{[A,x,I] = KS\_init(nq)}} Given the number of degree of freedom $n_{\q}$, it provides the state matrix $\A$ obtained by a FD discretization of the spatial derivatives. Five grid-point stencil {FD} schemes are used: in particular, a one grid point de-centered scheme is used to enhance the stability of the numerical solution.

\paragraph{\texttt{d = fd\_coeff(n,dx)}} It provides the {FD} coefficients used by \texttt{KS\_init}.


\begin{thebibliography}{10}

\bibitem{Bushnell1991Drag-reduction-}
Bushnell, D.~M., and Moore, K.~J., 1991.
\newblock ``{Drag reduction in nature}''.
\newblock {\em Ann. Rev. Fluid Mech., {\bf 23}}, pp.~65--79.

\bibitem{arfm2007-kim-bewley}
Kim, J., and Bewley, T.~R., 2007.
\newblock ``{A Linear Systems Approach to Flow Control}''.
\newblock {\em Ann. Rev. Fluid Mech., {\bf 39}}, pp.~39--383.

\bibitem{schlichting}
Schlichting, H., and Gersten, K., 2000.
\newblock {\em {Boundary-Layer Theory}}.
\newblock Springer Verlag, Heidelberg.

\bibitem{arfm2002-saric-et-al}
Saric, W.~S., Reed, H.~L., and Kerschen, E.~J., 2002.
\newblock ``{Boundary-layer receptivity to freestream disturbances}''.
\newblock {\em Ann. Rev. Fluid Mech., {\bf 34}}(1), pp.~291--319.

\bibitem{2001STSF-schmid-henningson}
Schmid, P.~J., and Henningson, D.~S., 2001.
\newblock {\em {Stability and Transition in Shear Flows}}.
\newblock No.~v. 142 in {Applied Mathematical Sciences}. Springer-Verlag.

\bibitem{jovanovic}
Jovanovic, M.~R., and Bamieh, B., 2005.
\newblock ``{Componentwise Energy Amplification in Channel Flows}''.
\newblock {\em J. Fluid Mech., {\bf 534}}, pp.~145--183.

\bibitem{arfm2007-schmid}
Schmid, P.~J., 2007.
\newblock ``{Nonmodal Stability Theory}''.
\newblock {\em Ann. Rev. Fluid Mech., {\bf 39}}, pp.~129--62.

\bibitem{2000CT-glad-ljung}
Glad, T., and Ljung, L., 2000.
\newblock {\em {Control Theory}}.
\newblock Taylor \& Francis, London.

\bibitem{arfm1990-huerre-monkewitz}
Huerre, P., and Monkewitz, P.~A., 1990.
\newblock ``{Local and Global Instabilities in Spatially Developing Flows}''.
\newblock {\em Ann. Rev. Fluid Mech., {\bf 22}}, pp.~473--537.

\bibitem{joshi}
Joshi, S.~S., Speyer, J.~L., and Kim, J., 1997.
\newblock ``{A Systems Theory Approach to the Feedback Stabilization of
  Infinitesimal and Finite-amplitude Disturbances in Plane Poiseuille Flow}''.
\newblock {\em J. Fluid Mech., {\bf 332}}, pp.~157--184.

\bibitem{bewley98}
Bewley, T.~R., and Liu, S., 1998.
\newblock ``{Optimal and Robust Control and Estimation of Linear Paths to
  Transition}''.
\newblock {\em J. Fluid Mech., {\bf 365}}, pp.~305--349.

\bibitem{cortelezzi98}
Cortelezzi, L., Speyer, J.~L., Lee, K.~H., and Kim, J., 1998.
\newblock ``{Robust reduced-order control of turbulent channel flows via
  distributed sensors and actuators}''.
\newblock {\em IEEE 37th Conf. on Decision and Control}, pp.~1906--1911.

\bibitem{hogberg:bewley:henning:03:a}
H{\"o}gberg, M., Bewley, T.~R., and Henningson, D.~S., 2003.
\newblock ``{Linear Feedback Control and Estimation of Transition in Plane
  Channel Flow}''.
\newblock {\em J. Fluid Mech., {\bf 481}}, pp.~149--175.

\bibitem{chevalier:hoepffner:akervik:henning:07}
Chevalier, M., H{\oe}pffner, J., Akervik, E., and Henningson, D.~S., 2007.
\newblock ``{Linear Feedback Control and Estimation Applied to Instabilities in
  Spatially Developing Boundary Layers}''.
\newblock {\em J. Fluid Mech., {\bf 588}}, pp.~163--187.

\bibitem{monokrousos2008dns}
Monokrousos, A., Brandt, L., Schlatter, P., and Henningson, D.~S., 2008.
\newblock ``{{DNS} and {LES} of estimation and control of transition in
  boundary layers subject to free-stream turbulence}''.
\newblock {\em Intl J. Heat and Fluid Flow, {\bf 29}}(3), pp.~841--855.

\bibitem{lee01}
Lee, K.~H., Cortelezzi, L., Kim, J., and Speyer, J., 2001.
\newblock ``{Application of Reduced-order Controller to Turbulent Flow for Drag
  Reduction}''.
\newblock {\em Phys. Fluids, {\bf 13}}, pp.~1321--1330.

\bibitem{hogberg:bewley:henning:03:b}
H{\"o}gberg, M., Bewley, T.~R., and Henningson, D.~S., 2003.
\newblock ``{Relaminarization of {$Re_{\tau} =100$} Turbulence Using Gain
  Scheduling and Linear State-feedback Control Flow}''.
\newblock {\em Phys. Fluids, {\bf 15}}, pp.~3572--3575.

\bibitem{chevalier:hoepffner:bewley:henning:06}
Chevalier, M., H{\oe}pffner, J., Bewley, T.~R., and Henningson, D.~S., 2006.
\newblock ``{State Estimation in Wall-bounded Flow Systems. Part 2: Turbulent
  Flows}''.
\newblock {\em J. Fluid Mech., {\bf 552}}, pp.~167--187.

\bibitem{ljung1999system}
Ljung, L., 1999.
\newblock {\em {System identification}}.
\newblock Wiley Online Library.

\bibitem{Elliott1993Active-noise-co}
Elliott, S., and Nelson, P., 1993.
\newblock ``{Active noise control}''.
\newblock {\em Signal Processing Magazine, IEEE, {\bf 10}}(4), pp.~12--35.

\bibitem{milling1981tollmien}
Milling, R.~W., 1981.
\newblock ``{{T}ollmien--{S}chlichting wave cancellation}''.
\newblock {\em Phys. Fluids, {\bf 24}}, p.~979.

\bibitem{98:jacobson:reynolds}
Jacobson, S.~A., and Reynolds, W.~C., 1998.
\newblock ``{Active control of streamwise vortices and streaks in boundary
  layers}''.
\newblock {\em J. Fluid Mech., {\bf 360}}, pp.~179--211.

\bibitem{ijhff2003-sturzebecher-nitsche}
Sturzebecher, D., and Nitsche, W., 2003.
\newblock ``{Active cancellation of Tollmien--Schlichting instabilities on a
  wing using multi-channel sensor actuator systems }''.
\newblock {\em Intl J. Heat and Fluid Flow, {\bf 24}}, pp.~572--583.

\bibitem{2003:jfm:Rathnasingham:breuer}
Rathnasingham, R., and Breuer, K.~S., 2003.
\newblock ``{Active control of turbulent boundary layers}''.
\newblock {\em J. Fluid Mech., {\bf 495}}, pp.~209--233.

\bibitem{2007:jfm:lundell:control}
Lundell, F., 2007.
\newblock ``{Reactive control of transition induced by free-stream turbulence:
  an experimental demonstration}''.
\newblock {\em J. Fluid Mech., {\bf 585}}, pp.~41--71.

\bibitem{McKeon2013Experimental-ma}
McKeon, B.~J., Sharma, A.~S., and Jacobi, I., 2013.
\newblock ``{Experimental manipulation of wall turbulence: A systems
  approacha)}''.
\newblock {\em Phys. Fluids, {\bf 25}}(3), pp.~--.

\bibitem{Goldin2013Laminar-flow-co}
Goldin, N., King, R., P{\"a}tzold, A., Nitsche, W., Haller, D., and Woias, P.,
  2013.
\newblock ``{Laminar flow control with distributed surface actuation: damping
  Tollmien-Schlichting waves with active surface displacement}''.
\newblock {\em Exp. Fluids, {\bf 54}}(3), pp.~1--11.

\bibitem{sipp2010dynamics}
Sipp, D., Marquet, O., Meliga, P., and Barbagallo, A., 2010.
\newblock ``{Dynamics and control of global instabilities in open-flows: a
  linearized approach}''.
\newblock {\em Appl. Mech. Rev., {\bf 63}}(3), p.~30801.

\bibitem{phtr2011-bagheri-henningson}
Bagheri, S., and Henningson, D.~S., 2011.
\newblock ``{Transition delay using control theory}''.
\newblock {\em Philos. Trans. R. Soc., {\bf 369}}, pp.~1365--1381.

\bibitem{amr2009-bagheri-et-al}
Bagheri, S., H{\oe}pffner, J., Schmid, P.~J., and Henningson, D.~S., 2009.
\newblock ``{Input-Output analysis and control design applied to a linear model
  of spatially developing flows}''.
\newblock {\em Appl. Mech. Rev., {\bf 62}}.

\bibitem{sipp2013closed}
Sipp, D., and Schmid, P.~J., 2013.
\newblock {Closed-Loop Control of Fluid Flow: a Review of Linear Approaches and
  Tools for the Stabilization of Transitional Flows}.
\newblock AerospaceLab Journal.

\bibitem{gad96}
el~Hak, M.~G., 1996.
\newblock ``{Modern Developments in Flow Control}''.
\newblock {\em Appl. Mech. Rev., {\bf 49}}, pp.~365--379.

\bibitem{Bewley01}
Bewley, T.~R., 2001.
\newblock ``{Flow Control: New Challenges for a New Renaissance}''.
\newblock {\em Progr. Aerospace. Sci., {\bf 37}}, pp.~21--58.

\bibitem{Collis2004Issues-in-activ}
Collis, S.~S., Joslin, R.~D., Seifert, A., and Theofilis, V., 2004.
\newblock ``{Issues in active flow control: theory, control, simulation, and
  experiment}''.
\newblock {\em Progress in Aerospace Sciences, {\bf 40}}(4), pp.~237--289.

\bibitem{bagheri2009input}
Bagheri, S., Brandt, L., and Henningson, D.~S., 2009.
\newblock ``{Input--output analysis, model reduction and control of the
  flat-plate boundary layer}''.
\newblock {\em J. Fluid Mech., {\bf 620}}(1), pp.~263--298.

\bibitem{2007SIMSON-chavalier}
Chevalier, M., Schlatter, P., Lundbladh, A., and Henningson, D.~S., 2007.
\newblock {A pseudo-spectral solver for incompressible boundary layer flows}.
\newblock Tech. Rep. TRITA-MEK 2007:07, KTH Mechanics, Stockholm, Sweden.

\bibitem{ef2008-grundmann-tropea}
Grundmann, S., and Tropea, C., 2008.
\newblock ``{Active cancellation of artificially introduced
  {T}ollmien--{S}chlichting waves using plasma actuators}''.
\newblock {\em Exp. Fluids, {\bf 44}}(5), pp.~795--806.

\bibitem{ptp1976-kuramoto-tsuzuki}
Kuramoto, Y., and Tsuzuki, T., 1976.
\newblock ``{ Persistent Propagation of Concentration Waves in Dissipative
  Media Far from Thermal Equilibrium}''.
\newblock {\em Progress of Theoretical Physics, {\bf 55}}(2), pp.~356--369.

\bibitem{aa1977-sivashinsky}
Sivashinsky, G.~I., 1977.
\newblock ``{Nonlinear analysis of hydrodynamic instability in laminar flames
  -- I.Derivation of basic equations}''.
\newblock {\em Acta Astronautica, {\bf 4}}, pp.~1177--1206.

\bibitem{manneville1995dissipative}
Manneville, P., 1995.
\newblock {\em {Dissipative structures and weak turbulence}}.
\newblock Springer.

\bibitem{2012CCQ4-cvitanovic-et-al}
Cvitanovi{\'c}, P., Artuso, R., Mainieri, R., Tanner, G., and Vattay, G., 2012.
\newblock ``{Turbulence?}''.
\newblock In {\em {Chaos: Classical and Quantum}}. Niels Bohr Institute,
  Copenhagen, ch.~4.
\newblock \href{http://ChaosBook.org/version14}{ChaosBook.org/version14}.

\bibitem{2011HI-charru}
Charru, F., 2011.
\newblock {\em {Hydrodynamic Instabilities}}, first~ed.
\newblock Cambridge.

\bibitem{skogestad05}
Skogestad, S., and Postlethwaite, I., 2005.
\newblock {\em {Multivariable Feedback Control, Analysis to Design}}, 2nd~ed.
\newblock Wiley.

\bibitem{1995AC-astrom-wittenmark}
Astr{\"o}m, K.~J., and Wittenmark, B., 1995.
\newblock {\em {Adaptive Control}}, second~ed.
\newblock Addison Wesley.

\bibitem{doyle89}
Doyle, J.~C., Glover, K., Khargonekar, P.~P., and Francis, B.~A., 1989.
\newblock ``{State-space solutions to standard ${H}_2$ and ${H}_\infty$ control
  problems}''.
\newblock {\em IEEE Trans. Autom. Control, {\bf 34}}, pp.~831--847.

\bibitem{zhou:doyle:glover:02}
Zhou, K., Doyle, J.~C., and Glover, K., 2002.
\newblock {\em {Robust and Optimal Control}}.
\newblock Prentice Hall, New Jersey.

\bibitem{gad2007flow}
el~Hak, M.~G., 2007.
\newblock {\em {Flow control: passive, active, and reactive flow management}}.
\newblock Cambridge University Press.

\bibitem{jfm2013-julliet-schmid-huerre}
Julliet, F., Schmid, P.~J., and Huerre, P., 2013.
\newblock ``{Control of amplifier flows using subspace identification
  techniques}''.
\newblock {\em J. Fluid Mech., {\bf 725}}, pp.~522--565.

\bibitem{pof2013-belson-semeraro-et-al}
Belson, B.~A., Semeraro, O., Rowley, C.~W., and Henningson, D.~S., 2013.
\newblock ``{Feedback control of instabilities in the two-dimensional Blasius
  boundary layer: The role of sensors and actuators}''.
\newblock {\em Phys. Fluids, {\bf 25}}.

\bibitem{1995OC-lewis}
Lewis, F.~L., and Syrmos, L.~V., 1995.
\newblock {\em {Optimal Control}}.
\newblock John Wiley \& Sons, New York.

\bibitem{2001-bewley-moin-temam}
Bewley, T.~R., Moin, P., and Temam, R., 2001.
\newblock ``{DNS-based predictive control of turbulence: an optimal benchmark
  for feedback algorithms}''.
\newblock {\em J. Fluid Mech., {\bf 447}}(1), pp.~179--225.

\bibitem{gunzburger2003perspectives}
Gunzburger, M., 2003.
\newblock {\em {Perspectives in Flow Control and Optimization}}.
\newblock SIAM.

\bibitem{boyd2004convex}
Boyd, S., and Vandenberghe, L., 2004.
\newblock {\em {Convex optimization}}.
\newblock Cambridge university press.

\bibitem{2007NR-press-et-al}
Press, W.~H., Teukolsky, S.~A., Vetterling, W.~T., and Flannery, B.~P., 2007.
\newblock {\em {Numerical Recipes 3rd Edition: The Art of Scientific
  Computing}}, 3~ed.
\newblock Cambridge University Press.

\bibitem{corbett2001optimal}
Corbett, P., and Bottaro, A., 2001.
\newblock ``{Optimal control of nonmodal disturbances in boundary layers}''.
\newblock {\em Theoretical and Computational Fluid Dynamics, {\bf 15}}(2),
  pp.~65--81.

\bibitem{arnold1984riccati}
Arnold, W.~I., and Laub, A., 1984.
\newblock ``{Generalized eigenproblem algorithms and software for algebraic
  {R}iccati equations}''.
\newblock {\em Proceedings of the IEEE, {\bf 72}}(12), pp.~1746--1754.

\bibitem{nlaa2008-benner-li-penz}
Benner, P., Li, J., and Penzl, T., 2008.
\newblock ``{Numerical solution of large-scale {L}yapunov equations, {R}iccati
  equations, and linear-quadratic optimal control problems}''.
\newblock {\em Numer. Linear Algebra Appl., {\bf 15}}, pp.~755--777.

\bibitem{siamjco1991-bank-ito}
Banks, H.~T., and Ito, K., 1991.
\newblock ``{A numerical algorithm for optimal feedback gains in high
  dimensional linear quadratic regulator problems}''.
\newblock {\em SIAM J. Control and Optimization, {\bf 29}}(3), pp.~499--515.

\bibitem{ieeecs2004-benn}
Benner, P., 2004.
\newblock ``{Solving large-scale control problems}''.
\newblock {\em Control Systems IEEE, {\bf 24}}(1), pp.~44--59.

\bibitem{bamieh2002}
Bamieh, B., Paganini, F., and Dahleh, M., 2002.
\newblock ``{Distributed control of spatially invariant systems}''.
\newblock {\em IEEE Trans. Autom. Control, {\bf 47}}(7), pp.~1091--1107.

\bibitem{2000:hogberg:bewley}
H{\"o}gberg, M., and Bewley, T.~R., 2000.
\newblock ``{Spatially localized convolution kernels for feedback control of
  transitional flows}''.
\newblock {\em IEEE 39th Conf. on Decision and Control, {\bf 3278-3283}}.

\bibitem{2010:akht:borg:miro:ziet}
Akhtar, I., Borggaard, J., Stoyanov, M., and Zietsman, L., 2010.
\newblock ``{{O}n {C}ommutation of {R}eduction and {C}ontrol: {L}inear
  {F}eedback {C}ontrol of a von {K}{\'a}rm{\'a}n {S}treet}''.
\newblock In {5th {F}low {C}ontrol {C}onference}, American Institute of
  Aeronautics and Astronautics, pp.~1--14.

\bibitem{maartensson2011synthesis}
Martensson, K., and Rantzer, A., 2011.
\newblock ``Synthesis of structured controllers for large-scale systems''.
\newblock {\em IEEE Trans. Autom. Control}.

\bibitem{iutam2009-pralits-luchini}
Pralits, J.~O., and Luchini, P., 2010.
\newblock ``{Riccati-less optimal control of bluff-body wakes}''.
\newblock In {Seventh IUTAM Symposium on Laminar-Turbulent Transition},
  P.~Schlatter and D.~S. Henningson, eds., Vol.~18, Springer.

\bibitem{semeraro2013riccati}
Semeraro, O., Pralits, J.~O., Rowley, C.~W., and Henningson, D.~S., 2013.
\newblock ``{Riccati-less approach for optimal control and estimation: an
  application to two-dimensional boundary layers}''.
\newblock {\em J. Fluid Mech., {\bf 731}}, pp.~394--417.

\bibitem{garcia1989model}
Garcia, C.~E., Prett, D.~M., and Morari, M., 1989.
\newblock ``{Model predictive control: theory and practice -- a survey}''.
\newblock {\em Automatica, {\bf 25}}(3), pp.~335--348.

\bibitem{qin2003survey}
Qin, S.~J., and Badgwell, T.~A., 2003.
\newblock ``{A survey of industrial model predictive control technology}''.
\newblock {\em Control engineering practice, {\bf 11}}(7), pp.~733--764.

\bibitem{noack2011reduced}
Noack, B.~R., Morzynski, M., and Tadmor, G., 2011.
\newblock {\em {Reduced-Order Modelling for Flow Control}}, Vol.~528.
\newblock Springer Verlag.

\bibitem{siamjo1999-bryd-et-al}
Bryd, R.~H., Hribar, M.~E., and Nocedal, J., 1999.
\newblock ``{An Interior Point Algorithm for Large-Scale Nonlinear
  Programming}''.
\newblock {\em SIAM Journal on Optimization}.

\bibitem{corke2010}
Corke, T.~C., Enloe, C.~L., and Wilkinson, S.~P., 2010.
\newblock ``{Dielectric Barrier Discharge Plasma Actuators for Flow Control}''.
\newblock {\em Ann. Rev. Fluid Mech., {\bf 42}}(1), pp.~505--529.

\bibitem{suzen2005numerical}
Suzen, Y., Huang, P., Jacob, J., and Ashpis, D., 2005.
\newblock ``{Numerical simulations of plasma based flow control
  applications}''.
\newblock {\em AIAA paper, {\bf 4633}}, p.~2005.

\bibitem{2011:Kriegseis}
Kriegseis, J., 2011.
\newblock ``{Performance Characterization and Quantification of Dielectric
  Barrier Discharge Plasma Actuators}''.
\newblock PhD thesis, TU Darmstadt.

\bibitem{siamjo1996-coleman-li}
Coleman, T.~F., and Li, Y., 1996.
\newblock ``{A reflective Newton method for minimizing a quadratic function
  subject to bounds on some of the variables}''.
\newblock {\em SIAM Journal on Optimization, {\bf 6}}(4), pp.~1040--1058.

\bibitem{anderson:moore:90}
Anderson, B., and Moore, J., 1990.
\newblock {\em {Optimal control: Linear Quadratic Methods}}.
\newblock Prentice Hall, New York.

\bibitem{mpcps1955-penrose}
Penrose, R., 1955.
\newblock ``{A generalized inverse for matrices}''.
\newblock {\em Mathematical Proceedings of the Cambridge Philosophical Society,
  {\bf 51}}, pp.~406--413.

\bibitem{1979IDS-luenberger}
Luenberger, D.~G., 1979.
\newblock {\em {Introduction to Dynamic System}}.
\newblock John Wiley \& Sons, New York.

\bibitem{jfm2012-herve-sipp-schmid-samuelides}
Herv{\'e}, A., Sipp, D., Schmid, P.~J., and Samuelides, M., 2012.
\newblock ``{A physics-based approach to flow control using system
  identification }''.
\newblock {\em J. Fluid Mech., {\bf 702}}, pp.~26--58.

\bibitem{1986AFT-haykin}
Haykin, S., 1986.
\newblock {\em {Adaptive Filter Theory}}.
\newblock Prentice-Hall.

\bibitem{choi94}
Choi, H., Moin, P., and Kim, J., 1994.
\newblock ``{Active Turbulence Control for Drag Reduction in Wall-bounded
  Flows}''.
\newblock {\em J. Fluid Mech., {\bf 262}}, pp.~75--110.

\bibitem{ieee1978-doyle}
Doyle, J.~C., 1978.
\newblock ``{Guaranteed Margins for {LQG} Regulators}''.
\newblock {\em IEEE Trans. Autom. Control, {\bf AC-23}}(4), pp.~756--757.

\bibitem{phtr2012-erdmann-et-al}
Erdmann, R., P{\"a}tzold, A., Engert, M., Peltzer, I., and Nitsche, W., 2012.
\newblock ``{On active control of laminar-turbulent transition on
  two-dimensional wings}''.
\newblock {\em Philos. Trans. R. Soc., {\bf 369}}, pp.~1382--1395.

\bibitem{anderson:liu:89}
Anderson, B., and Liu, Y., 1989.
\newblock ``{Controller Reduction: Concepts and Approaches}''.
\newblock {\em IEEE Trans. Autom. Control, {\bf 34}}, pp.~802--812.

\bibitem{aakervik:hoepffner:ehrenstien:henning:07}
Akervik, E., H{\oe}pffner, J., Ehrenstein, U., and Henningson, D.~S., 2007.
\newblock ``{Optimal Growth, Model Reduction and Control in a Separated
  Boundary-layer Flow Using Global Eigenmodes}''.
\newblock {\em J. Fluid Mech., {\bf 579}}, pp.~305--314.

\bibitem{moore:81}
Moore, B., 1981.
\newblock ``{Principal component analysis in linear systems: Controllability,
  observability, and model reduction}''.
\newblock {\em IEEE Trans. Autom. Control, {\bf 26}}(1), pp.~17--32.

\bibitem{rowley2005model}
Rowley, C.~W., 2005.
\newblock ``{Model reduction for fluids, using balanced proper orthogonal
  decomposition}''.
\newblock {\em Intl J. of Bifurcation and Chaos, {\bf 15}}(03), pp.~997--1013.

\bibitem{ilak:rowley:08}
Ilak, M., and Rowley, C.~W., 2008.
\newblock ``{Modeling of Transitional Channel Flow Using Balanced Proper
  Orthogonal Decomposition}''.
\newblock {\em Phys. Fluids, {\bf 20}}, p.~034103.

\bibitem{barba:sipp:schmid:09}
Barbagallo, A., Sipp, D., and Schmid, P.~J., 2009.
\newblock ``{Closed-loop control of an open cavity flow using reduced order
  models}''.
\newblock {\em J. Fluid Mech., {\bf 641}}, pp.~1--50.

\bibitem{jfm2010-semeraro-et-al}
Semeraro, O., Bagheri, S., Brandt, L., and Henningson, D.~S., 2011.
\newblock ``{Feedback control of three-dimensional optimal disturbances using
  reduced-order models}''.
\newblock {\em J. Fluid Mech., {\bf 677}}, pp.~63--102.

\bibitem{jfm2003-noack-afanasief-et-al}
Noack, B.~R., Afanasief, K., Morzynski, M., Tadmor, G., and Thiele, F., 2003.
\newblock ``{A hierarchy of low-dimensional models for the transient and
  post-transient cylinder wake }''.
\newblock {\em J. Fluid Mech., {\bf 497}}, pp.~335--363.

\bibitem{ref:siegel}
Siegel, S.~G., Siegel, J., Fagley, C., Luchtenburg, D.~M., Cohen, K., and
  McLaughlin, T., 2008.
\newblock ``{Low Dimensional Modelling of a Transient Cylinder Wake Using
  Double Proper Orthogonal Decomposition}''.
\newblock {\em J. Fluid Mech., {\bf 610}}, pp.~1--42.

\bibitem{Ilak2010Model-Reduction}
Ilak, M., Bagheri, S., Brandt, L., Rowley, C.~W., and Henningson, D.~S., 2010.
\newblock ``{Model Reduction of the Nonlinear Complex {G}inzburg-{L}andau
  Equation}''.
\newblock {\em SIAM J. Appl.Dyn. Sys., {\bf 9}}(4), pp.~1284--1302.

\bibitem{2007:huang:kim:pof}
Huang, S., and Kim, J., 2008.
\newblock ``{Control and system identification of separated flow}''.
\newblock {\em Phys. Fluids, {\bf 20}}, p.~101509.

\bibitem{ma2011reduced}
Ma, Z., Ahuja, S., and Rowley, C.~W., 2011.
\newblock ``{Reduced-order models for control of fluids using the eigensystem
  realization algorithm}''.
\newblock {\em Theoretical and Computational Fluid Dynamics, {\bf 25}}(1-4),
  pp.~233--247.

\bibitem{semeraro2013transition}
Semeraro, O., Bagheri, S., Brandt, L., and Henningson, D.~S., 2013.
\newblock ``Transition delay in a boundary layer flow using active control''.
\newblock {\em J. Fluid Mech., {\bf 731}}(9), pp.~288--311.

\bibitem{dadfar2013centralised}
Dadfar, R., Fabbiane, N., Bagheri, S., and Henningson, D.~S., 2014.
\newblock ``{Centralised versus decentralised active control of boundary layer
  instabilities}''.
\newblock {\em Flow, Turb. and Comb.}(Under considaration).

\bibitem{2009NMDP-quarteroni}
Quarteroni, A., 2009.
\newblock {\em {Numerical Models for Differential Problems}}.
\newblock Springer.

\end{thebibliography}
\end{document}